\pdfoutput=1
\documentclass[a4paper,11pt]{article}
\usepackage{jheppub} 
\usepackage{lineno}

\usepackage{tikz} 
\usepackage{tikzscale} 
\usetikzlibrary{patterns} 
\usepackage{caption} 
\usepackage{microtype} 

\usepackage{calrsfs}
    	\DeclareMathAlphabet{\mathcal}{OMS}{zplm}{m}{n} 
\usepackage{amsmath}
\usepackage{amssymb}

\usepackage{mathtools}
\usepackage{physics}
\usepackage{upgreek}
\usepackage{tensor} 
\usepackage{url} 
\usepackage{cancel} 
\usepackage{cases} 
\usepackage{anyfontsize}
\usepackage{wrapfig}
\usepackage{verbatim}
\usepackage{amsfonts}
\usepackage{bm} 
\usepackage{bold-extra} 
\usepackage{multicol} 
\usepackage{comment}
\usepackage{pgfplots} 
\usepackage{standalone} 

\DeclareMathAlphabet{\mathbbold}{U}{bbold}{m}{n}

\tikzset{every picture/.style={line width=0.75pt}}

\DeclareRobustCommand{\corr}[1]{#1}
\DeclareRobustCommand{\corrIV}[1]{#1}
\DeclareRobustCommand{\chg}[1]{#1}
\DeclareRobustCommand{\rev}[1]{#1}
\DeclareRobustCommand{\corrIX}[1]{#1}

\title{\boldmath Dirichlet walls beyond JT gravity and their quantum group interpretation}

\author[a]{Luca Griguolo,}
\author[b]{Jacopo Papalini,}
\author[c]{Lorenzo Russo,}
\author[c]{Domenico Seminara}
\author[a]{and Alex Tarana}

\affiliation[a]{Dipartimento SMFI, Universit\`a di Parma and INFN Gruppo Collegato di Parma,\\
Viale G. P. Usberti 7/A, 43100 Parma, Italy}

\affiliation[b]{Department of Physics and Astronomy,\\
Ghent University, Krijgslaan 281-S9, 9000 Gent, Belgium}

\affiliation[c]{Dipartimento di Fisica, Universit\`a di Firenze and INFN Sezione di Firenze,\\
Via G. Sansone 1, 50019 Sesto Fiorentino, Italy}

\emailAdd{luca.griguolo@unipr.it}
\emailAdd{jacopo.papalini@ugent.be}
\emailAdd{lorenzo.russo@unifi.it}
\emailAdd{domenico.seminara@unifi.it}
\emailAdd{alex.tarana@unipr.it}

\abstract{We canonically quantize \corrIV{sinh dilaton gravity} with finite Dirichlet boundary conditions. The reduced phase space is formulated in terms of the length of a two-sided spacelike geodesic in an effective AdS$_2$ geometry and its conjugate momentum. \corrIV{Canonical quantization} leads to a finite-difference Schrödinger equation whose wavefunctions of the universe are given by $q$-hypergeometric functions. These furnish a quantum deformation of the hypergeometric wavefunctions of JT gravity. We further show that the finite cutoff Hamiltonian derived from the bulk analysis admits a natural interpretation as the reduced Casimir of \rev{$U_q(\mathfrak{sl}(2,\mathbb{R}))$} acting on hyperbolic matrix elements. The two scattering wave solutions are hyperbolic matrix elements of the modular double $U_q(\mathfrak{sl}(2,\mathbb R))\otimes U_{\tilde q}(\mathfrak{sl}(2,\mathbb R))$, and the wavefunction of the universe is their unique combination regular at vanishing length.}

\begin{document}
\maketitle
\flushbottom

\section{Introduction}

The holographic principle is one of the most concrete frameworks we have for studying quantum gravity. In its standard formulation, the holographic dictionary is set up with the boundary at asymptotic infinity. A natural question is how this picture changes when the boundary is moved to a finite radial position. Finite Dirichlet walls give a controlled deformation of holography. They interpolate between asymptotic observables and local bulk physics, and they probe the radial realization of the UV/IR correspondence \corr{\cite{McGough:2016lol}}.
 
From the bulk viewpoint, finite cutoff gravity is appealing because it gives access to observables that the asymptotic description cannot reach. A finite cutoff regulates the short-distance singularities of boundary correlators, and it naturally encodes irrelevant deformations of the dual theory, such as $T\bar T$ \corr{\cite{Zamolodchikov:2004ce,Smirnov:2016lqw,Cavaglia:2016oda,McGough:2016lol,Gross:2019ach,Iliesiu:2020zld}}. The cutoff surface can also be seen as a dynamical ingredient of the gravitational system, rather than just a choice of boundary conditions. This view makes it possible to describe localized observers in quantum gravity \corr{\cite{Abdalla:2025gzn,Blommaert:2026lvp}}, and it may offer new perspectives on the black-hole interior and on the emergence of local bulk physics \corr{\cite{AliAhmad:2025kki}}.
 
These questions make sense in any dimension, but they become especially tractable in low dimensions, where analytic control is often possible. Two-dimensional dilaton gravity is a remarkably rich laboratory in this respect. Jackiw--Teitelboim (JT) gravity plays a distinguished role, because it is dual to the Schwarzian theory of the boundary reparametrization mode \corr{\cite{Teitelboim:1983ux,Jackiw:1984je,Almheiri:2014cka,Maldacena:2016upp,Jensen:2016pah,Engelsoy:2016xyb}}, which describes the low-energy limit of SYK model \corr{\cite{Sachdev:1992fk,Maldacena:2016hyu,Kitaev:2017awl}}. It also captures the universal near-horizon dynamics of near-extremal black holes
\cite{Nayak:2018qej,Iliesiu:2020qvm,Castro:2022cuo}, and its quantum theory is exactly
solvable \cite{Stanford:2017thb,Mertens:2017mtv,Mertens:2018fds}; see \cite{Mertens:2022irh} for a review. In this framework, a finite cutoff has a transparent meaning: the physical boundary clock is moved from asymptotic infinity into the bulk, and one can follow how both the gravitational and the holographic descriptions change under this displacement \cite{Iliesiu:2020zld,Stanford:2020qhm,Goel:2020yxl,Griguolo:2021wgy,Ferrari:2024kpz,Chaudhuri:2024yau,Aguilar-Gutierrez:2024nst}.  Considerable progress has recently been made on this problem directly from the bulk, through the canonical quantization of finite cutoff JT gravity \corr{\cite{Griguolo:2025kpi,Griguolo:2026eoi}}.\footnote{A complementary first-order (BF) derivation of the finite cutoff phase space, which also includes
end-of-the-world branes, was given in \cite{Belaey:2026xvh}.} In that approach the reduced phase space is described by the length $L$ of the two-sided Einstein--Rosen bridge (ERB) and its conjugate momentum $P$. The finite cutoff Hamiltonian is a function of a P\"oschl--Teller operator, and the wavefunctions of the universe are hypergeometric functions, with a precise group-theoretic meaning. They are hyperbolic matrix elements of $SL(2,\mathbb R)$, with both hyperbolic labels fixed by the cutoff \corr{\cite{Griguolo:2026eoi}}.
 
The purpose of this work is to extend this program beyond JT gravity, to dilaton potentials that deform the linear JT potential \corr{\cite{Witten:2020ert,Turiaci:2020fjj,Mertens:2020hbs,Blommaert:2024ymv,Griguolo:2025kpi,Kruthoff:2024gxc}}.
Among the possible deformations, an interesting one is given by sinh dilaton gravity, which admits a reformulation as non-critical string theory, i.e. Liouville gravity \corr{\cite{Mertens:2020hbs,Fan:2021bwt,Collier:2023cyw}}. Its sine counterpart admits an analogous reformulation as complex Liouville gravity \cite{Collier:2024kmo,Blommaert:2025eps}.
This embeds it in a well-developed framework that connects two-dimensional gravity with conformal field theory, matrix models and higher-genus amplitudes, and that offers powerful tools for their non-perturbative quantum dynamics \corr{\cite{Saad:2019lba,Stanford:2019vob,Turiaci:2020fjj,Mertens:2020hbs,Blommaert:2025avl}}.
 
Second, it exhibits an underlying quantum group structure \corr{\cite{Fan:2021bwt,Mertens:2022aou,Blommaert:2023wad,Blommaert:2023opb,Berkooz:2022mfk}}.
This theory is therefore a natural arena for $q$-deformed holography \corr{\cite{Berkooz:2022mfk,Lin:2022rbf,Lin:2023trc,Blommaert:2023opb,Almheiri:2024ayc,Belaey:2025ijg,vanderHeijden:2025zkr,Schouten:2025tvn,Belaey:2025kiu}}. A finite cutoff adds a second, independent deformation of the holographic dictionary on top of the quantum group one.
Third, these theories are motivated by the holographic description of the SYK model.
In particular, sine dilaton gravity has recently emerged as the gravitational counterpart of double-scaled SYK (DSSYK) \corr{\cite{Berkooz:2018qkz,Berkooz:2018jqr}}, a regime of SYK that is exactly solvable at all energies. Recent work has also linked this theory and three-dimensional de Sitter gravity \corr{\cite{Verlinde:2024znh,Tietto:2025oxn,Blommaert:2026ofx,Marini:2026zjk}}, opening a new way to study quantum aspects of de Sitter space anchored in a microscopic model.
 
In this paper we develop the canonical quantization of sinh dilaton gravity with finite Dirichlet boundary conditions. Building on the bulk quantization of finite cutoff JT gravity \corr{\cite{Griguolo:2026eoi}}, we show that the exact solvability of the theory survives the introduction of a finite boundary. We also show that the associated quantum group structure carries over directly to the finite cutoff setting.
 
The theory depends on a parameter $b$ that controls how far the dilaton potential deviates from its linear approximation, and JT is recovered as $b\to0$ \corr{\cite{Mertens:2020hbs}}. This parameter is related to the central charge of the Liouville description by $c=1+6\left(b+b^{-1}\right)^2$, and to the quantum group deformation parameter by $q=e^{i\pi b^2}$.
 
\paragraph{Reduced phase space and Hamiltonian.}
Following the finite cutoff JT construction, we describe the reduced phase space with two diffeomorphism-invariant variables: the length $L$ of a two-sided spacelike slice and its conjugate momentum $P$. The key difference from JT gravity is that $L$ is not measured in the physical metric. It is measured in an auxiliary AdS$_2$ geometry, obtained from the physical one by the Weyl rescaling $\dd s^2_{\rm AdS}=e^{2\pi b^2\Phi}\dd s^2$. This choice is motivated by the asymptotic formulation of sinh and sine dilaton gravity, where the geodesic length in this effective geometry is the natural canonical variable \corr{\cite{Blommaert:2024ymv,Blommaert:2024whf}}. In that setting it led to identifying the gravitational Hamiltonian with the transfer matrix of DSSYK for the sine model \corr{\cite{Blommaert:2024ymv}}, and with the Hamiltonian of a particle on the quantum group $SL_q(2,\mathbb R)$ for the sinh model \corr{\cite{Blommaert:2023wad}}. A new feature at finite cutoff is that the image of the physical boundary in the auxiliary geometry depends on the black-hole parameter.
 
The two-sided Brown--York energy, written in terms of the boundary value of the dilaton $\Phi_b$ and its horizon value $\Phi_h$, is
\begin{equation}
H=\frac{1}{2\pi b^2}\left[\sqrt{2\cosh(2\pi b^2\Phi_b)-2}-\sqrt{2\cosh(2\pi b^2\Phi_b)-2\cosh(2\pi b^2\Phi_h)}\,\right].
\label{intro:H}
\end{equation}
All the dynamics is encoded in the operator $\cosh(2\pi b^2\Phi_h)$. With the canonical transformation
\begin{equation}
\Lambda\equiv 2\log\sinh(L/2),
\qquad
\Pi\equiv P\tanh(L/2),
\label{intro:canonical}
\end{equation}
it takes the compact form
\begin{equation}
\cosh(2\pi b^2\Phi_h)=
\sqrt{\left(1+e^{4\pi b^2\Phi_b}e^{-\Lambda}\right)\left(1+e^{-\Lambda}\right)}\,
\cosh(2\pi b^2\Pi)
-e^{2\pi b^2\Phi_b}e^{-\Lambda}.
\label{intro:H0}
\end{equation}
Equation~\eqref{intro:H0} is a \emph{double deformation} of the Liouville quantum mechanics of asymptotic JT gravity, $H\propto P^2+e^{-L_{\rm ren}}$ \corr{\cite{Bagrets:2016cdf,Harlow:2018tqv,Kitaev:2018wpr,Yang:2018gdb,Mertens:2017mtv}}. The finite value of $\Phi_b$ encodes the finite cutoff deformation, already present in JT gravity \corr{\cite{Iliesiu:2020zld,Griguolo:2026eoi}}. The parameter $b$ encodes the quantum group deformation characteristic of sinh dilaton gravity, through the hyperbolic dependence on the momentum. The two limits that undo each deformation are both under control, and they commute. For $b\to0$, \eqref{intro:H0} reduces to the P\"oschl--Teller Hamiltonian of finite cutoff JT gravity \corr{\cite{Griguolo:2026eoi}}. For $\Phi_b\to\infty$ at fixed renormalized length, it reduces to the Hamiltonian of asymptotic Liouville gravity \corr{\cite{Mertens:2020hbs,Blommaert:2023wad}}.
 
\paragraph{Wavefunctions of the universe.}
Upon quantization, $e^{\pm2\pi b^2\hat\Pi}$ acts as a finite shift of $\Lambda$ by $\mp2\pi i b^2$ (with $\hbar=1$), so the spectral problem of the gravitational Hamiltonian becomes a finite-difference equation. We fix the operator ordering by \corrIV{a symmetric prescription that} reproduces, as $b\to0$, the ordering of the P\"oschl--Teller operator singled out by the $\mathfrak{sl}(2,\mathbb R)$ Casimir in finite cutoff JT gravity\corrIV{. The $b\to0$ limit alone does not determine the ordering; what singles it out is that, at finite $b$, it reproduces exactly the radially reduced Casimir of $U_q(\mathfrak{sl}(2,\mathbb R))$, see below}. Setting $s\equiv\Phi_h$ and $\nu\equiv\Phi_b$ in units of $\hbar$, the difference equation can be solved exactly in Fourier space. The resulting wavefunctions of the universe are given by $b$-deformed ($q$-) hypergeometric functions \corr{\cite{NishizawaUeno1996,Ponsot:1999uf,PonsotTeschner2001,Ip2013QuantumPlane}}:
\begin{equation}
\psi_s(\Lambda)=
\sqrt{\frac{\phi_b\!\left(-\frac{\Lambda}{2\pi b}\right)}{\phi_b\!\left(\frac{-\Lambda+4\pi b^2\nu}{2\pi b}\right)}}\;
e^{(\Lambda-2\pi b^2\nu)\left(\frac12+i\nu\right)}\,
F_b\!\left(\tfrac b2+ib\nu+ibs,\ \tfrac b2+ib\nu-ibs,\ b;\ -e^{\Lambda}\right),
\label{intro:wavefunction}
\end{equation}
where $\phi_b$ is Faddeev's quantum dilogarithm. These wavefunctions reduce to the hypergeometric wavefunctions of finite cutoff JT gravity as $b\to0$ \corr{\cite{Griguolo:2026eoi}}, and to the $q$-deformed Bessel-$K$ functions of asymptotic Liouville gravity as $\nu\to\infty$ \corr{\cite{Mertens:2020hbs,Fan:2021bwt}}. From their asymptotic behavior we extract the finite cutoff spectral density
\begin{equation}
\rho(s)=\frac{1}{\pi b^2}\,
\frac{\sinh(2\pi b^2 s)\,\sinh(2\pi s)}{\cosh(2\pi s)+\cosh(2\pi\nu)},
\label{intro:density}
\end{equation}
which interpolates between the finite cutoff JT density ($b\to0$) and the Liouville gravity density $\sinh(2\pi b^2s)\sinh(2\pi s)$ ($\nu\to\infty$). As in JT gravity, the finite cutoff affects the disk partition function in two distinct ways: a perturbative deformation of the energy, \eqref{intro:H}, and a non-perturbative deformation of the measure, \eqref{intro:density}.
 
\paragraph{Quantum group interpretation.}
In finite cutoff JT gravity, the bulk Hamiltonian coincides with the $\mathfrak{sl}(2,\mathbb R)$ Casimir acting on hyperbolic matrix elements, with both hyperbolic labels fixed by the cutoff \corr{\cite{Griguolo:2026eoi}}. We show that the same holds for sinh dilaton gravity once $\mathfrak{sl}(2,\mathbb R)$ is replaced by its quantum deformation $U_q(\mathfrak{sl}(2,\mathbb R))$. We reduce the Casimir acting on functions of the non-commuting Gauss coordinates $(\gamma,\phi,\beta)$ of the quantum group, restricted to fixed left and right Cartan charges. The result is a second-order difference equation in the single invariant $Z=\gamma e^{2\phi}\beta$. With the charges set to $h_L=h_R=\nu$ and the identification $e^{\Lambda}=e^{2\pi b^2\nu}Z$, this equation is \emph{exactly} the eigenvalue equation of the gravitational Hamiltonian. Schematically,
\begin{equation}
\bra{\nu}\,g\,\mathcal C\,\ket{\nu}
\quad\Longleftrightarrow\quad
\cosh(2\pi b^2\hat\Phi_h)\,\psi_s=\cosh(2\pi b^2 s)\,\psi_s .
\end{equation}
In this language, the canonical variable $\Lambda$ is the logarithm of the quantum group radial coordinate, the cutoff $\nu$ is the common hyperbolic charge and $s$ labels the principal-series representation.
 
\corr{The difference equation alone determines the wavefunction only up to multiplication by an $i$-periodic function of $\Lambda/(2\pi b^2)$. This ambiguity is invisible in both the JT and the large cutoff limits. We show that it is removed by the conditions that make the gravitational Hamiltonian a well-defined operator: analyticity of the wavefunction in a strip whose width is set by the imaginary shifts, tempered growth at large length, and regularity at $L=0$, the finite-$b$ counterpart of the Friedrichs condition of finite cutoff JT gravity. These conditions select the $q$-hypergeometric wavefunction \eqref{intro:wavefunction} uniquely\corrIV{, for $0<b<1$,} without input from representation theory.}

\paragraph{Modular double and the regular wavefunction.}
\corr{The group-theoretic interpretation extends from the Hamiltonian to the wavefunctions once $U_q(\mathfrak{sl}(2,\mathbb R))$ is completed to the modular double $U_q(\mathfrak{sl}(2,\mathbb R))\otimes U_{\tilde q}(\mathfrak{sl}(2,\mathbb R))$, $\tilde q=e^{i\pi/b^2}$, which underlies Liouville gravity \cite{Ponsot:1999uf,Fan:2021bwt,Mertens:2022aou}. In finite cutoff JT gravity the two solutions with plane-wave behavior at large $L$ are the hyperbolic matrix elements $K_{++}$ and $K_{--}$ of $SL(2,\mathbb R)$, and the wavefunction of the universe is their combination that is regular at $L=0$. We show that the same holds at finite $b$. Using a $q$-deformed Gauss--Euler decomposition of the group element built from Faddeev's quantum dilogarithm \cite{Mertens:2022aou}, we compute hyperbolic matrix elements of the modular double with labels $\frac b2\pm ibs$. Their radial parts are the two scattering solutions of the gravitational equation at $L\rightarrow +\infty$. Each of them also solves the dual Casimir equation, a difference equation with shifts $\Lambda\to\Lambda\pm2\pi i$, and the two Casimir equations fix it without periodic ambiguity\corrIV{ for irrational $b^2$}. The wavefunction \eqref{intro:wavefunction} is their unique combination regular at $L=0$, and the coefficients are the asymptotic amplitudes that determine the spectral density \eqref{intro:density}. }
 
\paragraph{Outline.}
The paper is organized as follows. In Section~\ref{sec:JT_review} we review the reduced phase space and canonical quantization of finite cutoff JT gravity, emphasizing the ingredients needed for its deformation. In Section~\corr{\ref{sec:sinh_box}} we construct the reduced phase space of sinh dilaton gravity at finite cutoff, quantize it, and solve the resulting difference equation. We also study its JT and large cutoff limits, \corr{show that the wavefunction is uniquely determined by the conditions that make the Hamiltonian well defined,} and compute the spectral density and the disk partition function. In Section~\corr{\ref{sec:casimir}} we show that the bulk difference equation is the radially reduced Casimir of $U_q(\mathfrak{sl}(2,\mathbb R))$. \corr{In Section~\ref{sec:wave} we construct the hyperbolic matrix elements of the modular double, and show that the wavefunction is their unique regular combination.} We conclude in Section~\corr{\ref{sec:conclusions}} with a discussion and an outlook. Technical details are collected in the appendices.

\section{Review: canonical quantization of JT gravity}
\label{sec:JT_review}

In this section we recall the finite cutoff quantization of JT gravity presented in \corr{\cite{Griguolo:2026eoi}}, pointing out the ingredients that will be essential in the extension to sinh dilaton gravity.

The Lorentzian action of JT gravity is given by
\begin{equation}
S_{\rm JT}[g,\Phi]
=
\frac12
\int_{\mathcal M}\! \dd^2 x\,\sqrt{-g}\,\Phi(R+2)
+
\int_{\partial\mathcal M}\! \dd y\,\sqrt{|h|}\,\Phi(K-1),
\label{eq:JT_action_review}
\end{equation}
and we impose Dirichlet boundary conditions on both the dilaton and induced metric on the boundary
\begin{equation}
\Phi\big|_{\partial\mathcal M}=\Phi_b,
\qquad
\dd s^2\big|_{\partial\mathcal M} = -\dd\tau^2.
\label{eq:JT_Dirichlet_review}
\end{equation}
The classical field configurations in the Schwarzschild gauge read
\begin{equation}
\dd s^2=-(r^2-\Phi_h^2)\,\dd t^2+\frac{\dd r^2}{r^2-\Phi_h^2},
\qquad
\Phi=r,
\qquad
\Phi_h<r<\Phi_b,
\label{eq:JT_black_hole_review}
\end{equation}
which correspond to an AdS$_2$ black hole with a horizon located at $r = \Phi_h$ and a timelike boundary at $r = \Phi_b$.
The goal is to perform a reduction of the gravitational phase space of the theory and quantize the resulting classical system.
The solutions \eqref{eq:JT_black_hole_review} can be extended to a two-sided strip topology, and the proper times along the corresponding boundaries are given by
\begin{equation}
\dd\tau_{L/R}=\sqrt{\Phi_b^2-\Phi_h^2}\,\dd t_{L/R},
\qquad
\tau\equiv\tau_L+\tau_R .
\label{eq:JT_boundary_times_review}
\end{equation}
The contraction of the Brown--York stress tensor with the Killing vector generating translations in $\tau_{L/R}$ gives equal left and right quasi-local charges, and we interpret their average as the Hamiltonian of the reduced two-sided system, i.e.
\begin{equation}
H(\Phi_h)=\frac{H_L+H_R}{2}=\Phi_b-\sqrt{\Phi_b^2-\Phi_h^2} .
\label{eq:JT_energy_phi_review}
\end{equation}
Next, we must specify the classical state of the system in terms of a pair of canonically conjugate variables. We foliate the maximally extended solution by horizontal boundary-to-boundary spacelike geodesics and define $L$ to be their length.
For a horizontal slice $t_L=t_R=t$, and one can compute its length to be
\begin{equation}
L(\tau,\Phi_h)
=
2\sinh^{-1}\!\left[
\frac{\sqrt{\Phi_b^2-\Phi_h^2}}{\Phi_h}
\cosh\!\left(
\frac{\Phi_h\tau}{2\sqrt{\Phi_b^2-\Phi_h^2}}
\right)
\right] .
\label{eq:JT_length_tau_review}
\end{equation}
In the spirit of \corr{\cite{Harlow:2018tqv}}, we then define the momentum $P$ conjugate to $L$ by requiring the symplectic form to be preserved $\omega=\dd\tau\wedge\dd H=\dd L\wedge\dd P$.
Consequently, the finite cutoff Hamiltonian \eqref{eq:JT_energy_phi_review} can be written as
\begin{equation}
H(L,P)
=
\Phi_b-
\sqrt{
\Phi_b^2-
\left(
P^2+\frac{\Phi_b^2}{\cosh^2(L/2)}
\right)
},
\label{eq:hamiltonian_LP}
\end{equation}
where the quantity under the square root is a P\"oschl--Teller Hamiltonian. 

We recall in passing that ordinary JT gravity at large cutoff is recovered by sending $\Phi_b\to\infty$ while keeping the renormalized length 
\begin{equation}\label{eq:geodesic_length_renormalized}
L_{\rm ren}=L-2\log(2\Phi_b)
\end{equation}
fixed.
In this limit, the Hamiltonian \eqref{eq:hamiltonian_LP} reduces to
\begin{equation}
H(L_{\rm ren},P)
=
\frac{1}{2\Phi_b}
\left(P^2+e^{-L_{\rm ren}}\right),
\label{eq:JT_Liouville_H_review}
\end{equation}
which is the familiar Liouville quantum mechanics describing asymptotic JT gravity \corr{\cite{Bagrets:2016cdf,Harlow:2018tqv,Kitaev:2018wpr,Yang:2018gdb}}.

We now proceed by quantizing the Hamiltonian system \eqref{eq:hamiltonian_LP} by promoting the canonical pair $(L,P)$ to operators satisfying $[\hat{L},\hat{P}]=i\hbar$. Since the Brown--York Hamiltonian is purely a function of
\begin{equation}\label{eq:Phi_h^2_LP}
    \Phi_h^2(L,P) = P^2 + \frac{\Phi_b^2}{\cosh^2(L/2)},
\end{equation}
it is sufficient to quantize and diagonalize the seed operator $\hat{\Phi}_h^2$.

In the following, we adopt units where $\hbar=1$ and rename the parameters as $s \equiv \Phi_h$, $\nu \equiv \Phi_b$ to streamline the notation. Having done so, the Schr\"odinger equation corresponding to \eqref{eq:Phi_h^2_LP} yields\footnote{As discussed in detail in \cite{Griguolo:2026eoi}, we are adopting a non-trivial operator ordering for the kinetic term
\begin{equation}
    P^2 \to \frac{1}{2} \left( \frac{1}{\sqrt{\sinh \hat{L}}} \hat{P} \sqrt{\sinh\hat{L}} \hat{P} + \corrIX{\hat{P}\sqrt{\sinh \hat{L}} \hat{P} \frac{1}{\sqrt{\sinh\hat{L}}}} \right).
\end{equation}}
\begin{equation}\label{eq:Schroedinger}
\left[
-\frac{\dd^2}{\dd L^2}
+\frac{\nu^2}{\cosh^2(L/2)}
-\frac{1}{4 \sinh^2(L)}
\right]\psi^\text{JT}_s(L)
=s^2 \, \psi^\text{JT}_s(L).
\end{equation}
This differential equation can be brought to hypergeometric form, and we choose the solution to be \corr{\cite{Griguolo:2026eoi}}
\begin{equation}\label{eq:waveJT}
\psi^\text{JT}_s(L)
=\sinh(L/2)^{1/2} \cosh^{1/2+2i\nu}(L/2) \,
{}_2F_1\!\left(
\frac12+i\nu+is,
\frac12+i\nu-is, 1; -\sinh^2(L/2)
\right).
\end{equation}
This wavefunction is the solution of \eqref{eq:Schroedinger} that is compatible with our choice of self-adjoint extension, i.e. it vanishes as $\sqrt{L/2}$ near the origin as required by the interpretation of $L$ as a geodesic length satisfying $L\geq0$.
It is worth emphasizing that any other linearly independent solution would generate a branch-cut singularity at the origin.

Exploiting the connection formula for hypergeometric functions, one can decompose the solution \eqref{eq:waveJT} \corr{on} the basis of independent solutions $\psi_{s}^{\pm}(L)$ as
\begin{equation}\label{eq:connection}
\begin{split}
\psi^\text{JT}_s(L)
=\frac{\Gamma(-2is)}{\Gamma(\frac12 +i \nu-is)\Gamma(\frac12 -i\nu-is)}\psi_{s}^{-}(L)
+\frac{\Gamma(2is)}{\Gamma(\frac12 +i \nu+is)\Gamma(\frac12 -i\nu+is)}\psi_{s}^{+}(L),
\end{split}
\end{equation}
where $\psi_{s}^{\pm}(L)$ are given by
\begin{equation}\label{eq:sol_infty}
\psi_{s}^{\pm}(L)=\frac{\sinh(L/2)^{\pm 2i s}}{\tanh(L/2)^{1/2+2i\nu}} \,{}_2F_1\!\left(\frac12 +i \nu\mp i s,\frac12 +i \nu\mp i s;1\mp 2is;-\frac{1}{\sinh^2(L/2)}\right).
\end{equation}
Importantly, the wavefunctions \eqref{eq:waveJT} are real and obey the orthogonality condition
\begin{equation}\label{eq:JTspectral}
    \int_0^\infty
    \dd L\,
    \corrIV{\psi^\text{JT}_{s'}(L)^* \, \psi^\text{JT}_{s}(L)}
    =
    \frac{\delta(s-s')}{\rho(s)} \qquad \rho(s) = 
    \frac{\corrIV{2}s\sinh(2\pi s)}
    {\cosh(2\pi s)+\cosh(2\pi\nu)}.
\end{equation}
Taking the large cutoff limit at the level of the wavefunction gives
\begin{equation}
\psi^\text{JT}_s(L_{\rm ren}) = \frac{e^{\pi \nu}}{\pi}
K_{2is}\!\left(2e^{-L_{\rm ren}/2}\right) ,
\label{eq:JT_Bessel_limit_review}
\end{equation}
which is the Liouville scattering wavefunction of ordinary asymptotic JT gravity.

The disk partition function of the model can then be obtained as an Euclidean transition amplitude between states of vanishing geodesic length, i.e.
\begin{equation}
Z(\beta)=\bra{L=0}e^{-\beta\hat H}\ket{L=0} = 
e^{2 \pi \nu} \int_0^\infty \dd s\,
\rev{\rho}(s)
\,e^{-\beta (\nu - \sqrt{\nu^2 - s^2})}.
\label{eq:JT_partition_spectral_review} 
\end{equation}
This is the finite cutoff version of the usual asymptotic prescription in which one projects onto the state $\ket{L_{\rm ren}=-\infty}$.
The resulting partition function displays two logically distinct finite cutoff effects.
The first is the deformation of the energy in the Boltzmann factor, which is the effect that is also captured by $T\bar T$-inspired descriptions of finite cutoff JT gravity \corr{\cite{Iliesiu:2020zld}}.
The second is the nonperturbative deformation of the spectral density \eqref{eq:JTspectral}.

The operator ordering adopted in \eqref{eq:Schroedinger} is selected by a group-theoretic realization of the finite cutoff problem.
Specifically, the gravitational Hamiltonian can be related to the $\mathfrak{sl}(2,\mathbb{R})$ Casimir in the left and right regular representations.
In \corr{\cite{Griguolo:2026eoi}} we adopted the conventions
\begin{equation}
    [H,E] = E, \qquad [H,F] = -F, \qquad [E,F] = 2H
\end{equation}
for the generators, and considered the following local decomposition of an $SL(2,\mathbb R)$ element 
\begin{equation}
g=e^{aX}e^{L H}e^{bX}
\qquad \text{where} \qquad
X=\frac{E+F}{2}.
\label{eq:JT_hyperbolic_decomposition_review}
\end{equation}
Diagonalizing the Casimir together with the left and right action of $X$ with eigenvalue $i\nu$, one obtains precisely the operator \eqref{eq:Schroedinger}. 

In the present paper, we will exploit another useful hyperbolic decomposition of the group element, namely
\begin{equation}\label{eq:hyperbolicdecomp2}
g=e^{u H}e^{\frac{L}{2}(E+F)}e^{v H}
\end{equation}
which is related to \eqref{eq:JT_hyperbolic_decomposition_review} by conjugation \corr{\cite{Vilenkin1968,vilenkin2012representation}}. One can check that reducing the $\mathfrak{sl}(2,\mathbb R)$ Casimir on the hyperbolic decomposition \eqref{eq:hyperbolicdecomp2} yields the same differential operator \eqref{eq:Schroedinger} \corr{\cite{Kitaev:2017hnr}}. Moreover, by equating the parameterization above with the standard Gauss-Euler parameterization used in the large cutoff limit of JT gravity \corr{\cite{Blommaert:2018oro,Blommaert:2018iqz}}, namely
\begin{equation}
g=e^{\gamma F}e^{2\phi H}e^{\beta E},
\end{equation}
and working in the fundamental representation, \corr{one finds} the following mapping between the finite cutoff chart $\left(L,u,v\right)$ and the Gauss-Euler one $\left(\phi,\beta,\gamma\right)$:
\begin{equation}\label{eq:map}
\corrIV{\beta=e^{-v}\tanh(L/2) \qquad \gamma=e^{-u}\tanh(L/2) \qquad e^{2\phi}=\cosh^2(L/2)\,e^{u}e^{v},}
\end{equation}
\corrIV{where $H$ is normalized as $[H,E]=E$, i.e.\ $H=\frac12\mathrm{diag}(1,-1)$ in the fundamental representation,} which will be useful in the following. An important quantity is given by the following combination:
\begin{equation}
Z=\beta e^{2\phi} \gamma=\sinh^2 (L/2) 
\end{equation}
which allows to express the finite cutoff bulk length $L$ in terms of Gauss-Euler coordinates. This ingredient will be essential in Section~\corr{\ref{sec:casimir}}, \corr{where the corresponding invariant of the quantum group is identified with the canonical variable of sinh dilaton gravity.}

In particular, the JT gravity wavefunctions presented above, i.e. Equations \eqref{eq:waveJT} and \eqref{eq:sol_infty}, can be shown to have a precise group theoretic origin as matrix elements associated to the hyperbolic decomposition \eqref{eq:hyperbolicdecomp2}. As reviewed in Appendix \ref{app:JTsteps}, $SL(2,\mathbb R)$ hyperbolic states have in general a branched structure, with labels $+$ and $-$ labeling the two different branches of the hyperbolic basis. Therefore, hyperbolic representation matrix elements are characterized by a $2\times 2$ matrix structure \cite{Vilenkin1968}, whose entries are usually denoted as $K_{++}$, $K_{--}$, \corr{$K_{+-}$} \rev{and $K_{-+}$}, whose explicit form is computed in \corr{\cite{Vilenkin1968,vilenkin2012representation}}.

Indeed, by setting \corrIV{both hyperbolic labels of \cite{Vilenkin1968} equal to $\frac12+i \nu+is$} (with $j=-\frac12+is$ the continuous series representation label), and \corrIV{the hyperbolic angle equal to $L/2$} in the hyperbolic matrix elements found in \cite{Vilenkin1968}, one finds the following expressions for $K_{++}$ and $K_{--}$ sectors:
\begin{equation}\label{eq:K++JT}
\begin{split}
K_{++}&=\frac{\Gamma\left(\frac12 +is\pm i\nu\right)}{\Gamma(1+2is)}\left(\sinh(L/2)\right)^{-1} \left(\tanh (L/2)\right)^{-2i\nu} \sinh^{-2is}\left(L/2\right) \times 
\\
&\times {}_2F_1\!\left(
\frac12+i\nu+ is,
\frac12+i\nu+ is;
1+2is;
-\frac{1}{\sinh^2(L/2)}
\right)
\end{split}
\end{equation}
\begin{equation}
\begin{split}
K_{--}&=\frac{\Gamma\left(\frac12 -is\pm i\nu\right)}{\Gamma(1-2is)}\left(\sinh(L/2)\right)^{-1} \left(\tanh (L/2)\right)^{-2i\nu} \sinh^{2is}\left(L/2\right) \times 
\\
&\times {}_2F_1\!\left(
\frac12+i\nu- is,
\frac12+i\nu- is;
1-2is;
-\frac{1}{\sinh^2(L/2)}
\right)
\end{split}
\end{equation}
On the other hand, the $K_{+-}$ component takes the form \cite{Vilenkin1968}: \footnote{Choosing $\epsilon=0$ in the notation of \cite{Vilenkin1968}.}
\begin{equation}\label{eq:K+-JT}
\begin{split}
K_{+-}&=\pi \left(\frac{\cosh(\pi s)\cosh(\pi \nu)}{\cosh(2\pi s)+\cosh(2\pi \nu)}\right) \left(\cosh(L/2)\right)^{2i\nu}\times 
\\
&\times {}_2F_1\!\left(
\frac12+i\nu+is,
\frac12+i\nu- is;
1;
-\sinh^2(L/2)\right)
\end{split}
\end{equation}
We provide an independent derivation of \corr{$K_{+-}$} in Appendix \ref{app:JTsteps}.
Apart from the normalization, one finds the following identification between the gravitational wavefunctions and the hyperbolic representation matrix elements displayed above:
\begin{equation}
\begin{gathered}
\corrIV{\psi_{s}(L)\propto \mu(L)\,K_{+-}, \qquad \psi_{s}^{+}(L)\propto \mu(L)\,\corrIX{K_{--}}, \qquad \psi_{s}^{-}(L)\propto \mu(L)\,\corrIX{K_{++}},}\\
\corrIV{\mu(L)\equiv\sqrt{\sinh(L/2)\cosh(L/2)}.}
\end{gathered}
\end{equation}
\corrIV{The common factor $\mu(L)$ converts matrix elements, which are functions on the group, into wavefunctions normalized with the flat measure $\dd L$; it is the JT analogue of the factor $Z^{-1/2}$ in \eqref{eq:Phi-chi}.}
In other words, $K_{+-}$ represents the regular solution we picked at $L=0$ and chose as the wavefunction of the universe, while $K_{++}$, $K_{--}$ provide the basis of solutions at $L=\infty$. Individually and not in the combination \eqref{eq:connection}, both $K_{++}$ and $K_{--}$ feature a branch cut singularity at $L=0$ and therefore do not satisfy the Friedrichs' self adjoint extension chosen in \cite{Griguolo:2026eoi}. 

This precise interplay between gravitational wavefunctions and group hyperbolic matrix elements will be a key benchmark also in the $b$-deformed setup of sinh dilaton gravity, \corr{where we will recover its Casimir side in Section~\ref{sec:casimir}, and the matrix elements $K_{\pm\pm}$ together with their regular combination in Section~\ref{sec:wave}.}

\section{Sinh dilaton gravity in a box}\label{sec:sinh_box}
After reviewing the case of JT gravity, we now shift our focus to the finite cutoff formulation of another relevant two-dimensional dilaton gravity theory that has received considerable attention in recent years, namely Liouville gravity \corr{\cite{Mertens:2020hbs,Fan:2021bwt,Blommaert:2023wad}}. This model can be thought of as a one-parameter deformation of JT gravity, with the parameter being denoted here as $b$.
The Lorentzian action of the theory is given by\footnote{We have set $\kappa=\frac{1}{8 \pi G_{N}}=1$ in order to streamline the notation.}
\begin{equation}\label{eq:sinh_action}
S[g,\Phi] = \frac{1}{2} \int_{\mathcal{M}} \dd^2x \sqrt{-g} \left(\Phi R + \frac{\sinh(2\pi b^2 \Phi)}{\pi b^2}\right) + \int_{\partial \mathcal{M}} \dd y \sqrt{\abs{h}}\,\bigg(\Phi K-\sqrt{\frac{\cosh(2\pi b^2 \Phi)-1}{2\pi^2 b^4}}\bigg).
\end{equation}
In the limit $b \to 0$, both the bulk and boundary terms in \eqref{eq:sinh_action} reduce to those of JT gravity \eqref{eq:JT_action_review}. For this reason, throughout the rest of this work we will refer to the $b \to 0$ limit of the theory as its ``JT limit''.

We supplement the action \eqref{eq:sinh_action} with Dirichlet boundary conditions on the dilaton and on the induced metric,
\begin{equation}\label{eq:boundary_conditions}
\Phi\big|_{\partial\mathcal M}
=
\Phi_b,
\qquad
\mathrm{d}s^2\big|_{\partial\mathcal M}
=
-\mathrm{d}\tau^2,
\end{equation}
where $\tau$ denotes proper time along the timelike cutoff surface.
The purpose of this section is to construct a Hamiltonian description of the theory by following the same general strategy adopted for finite cutoff JT gravity in \corr{\cite{Griguolo:2026eoi}}. In order to do so, it is first useful to review the classical solutions and the corresponding quasi-local energy. The equations of motion obtained by varying \eqref{eq:sinh_action} are
\begin{equation}
R+2\cosh(2\pi b^2\Phi)=0, \qquad
\nabla_\mu\nabla_\nu\Phi
-
g_{\mu\nu}\nabla^2\Phi
+
\frac{g_{\mu\nu}}{2\pi b^2}
\sinh(2\pi b^2\Phi)
=
0,
\end{equation}
and the static solutions can be written in Schwarzschild gauge\footnote{The boundary condition on the metric \eqref{eq:boundary_conditions} in the Schwarzschild gauge we adopt here reads
\begin{equation}\label{eq:boundary_condition_metric}
    g \big\lvert_{\partial\mathcal{M}} = - \frac{\cosh(2 \pi b^2 \Phi_b) - \cosh(2 \pi b^2 \Phi_h)}{2 \pi^2 b^4} \, \dd t^2,
\end{equation}
where $\Phi_h$ is some real parameter that we recognize to be the location of the black hole horizon. Here we are implicitly assuming that the boundary data $(\Phi_h,\Phi_b)$ are such that $\Phi_h < \Phi_b$. Nevertheless, in what follows we will be able to relax this assumption.}
\begin{equation}\label{geom}
    \dd s^2 = - \frac{\cosh(2\pi b^2 r) - \cosh(2 \pi b^2 \Phi_h)}{2\pi^2b^4} \dd t^2 + \frac{2\pi^2b^4}{ \cosh(2\pi b^2 r) - \cosh(2 \pi b^2 \Phi_h)} \dd r^2,
    \quad
    \Phi = r.
\end{equation}
The geometry \eqref{geom} describes a spacetime with a black hole horizon located at $r=\Phi_h$, a boundary at $r = \Phi_b$ and it reduces to the AdS$_2$ \eqref{eq:JT_black_hole_review} of the JT gravity black hole in the $b \rightarrow 0$ limit. What changes with respect to the ordinary treatment of sinh dilaton gravity \corr{\cite{Mertens:2020hbs,Blommaert:2023wad}} is that, in our case, the boundary is located at finite distance $\Phi_b < \infty$.

\rev{In the following we will also need} the inverse Hawking temperature of the black hole described by the on-shell metric \eqref{geom}, which reads\footnote{One must not confuse the black hole inverse temperature $\beta_\text{BH}$ with the inverse Tolman temperature $\beta$. The former only depends on local data at the black hole horizon, whereas the latter is interpreted as the inverse temperature felt by an observer at the boundary $r = \Phi_b$ accounting for redshift.}
\begin{equation}\label{eq:Hawkingsinh}
    \beta_\text{BH} = \frac{4 \pi^2 b^2}{\sinh(2\pi b^2 \Phi_h)}.
\end{equation}

\subsection{Quasi-local Hamiltonian}
Having discussed the classical solutions of the theory, we now turn to the dynamics of sinh dilaton gravity at finite cutoff. The main goal of this section is to reformulate the theory as a one-dimensional Hamiltonian system, following the same general strategy adopted for JT gravity.

As in the JT analysis, we regard $\Phi_b$ as a fixed datum of the theory, whereas $\Phi_h$ labels the family of on-shell geometries described by \eqref{geom}. We will establish a correspondence between each of these geometries and a classical trajectory of an auxiliary Hamiltonian system. At the classical level, this correspondence allows us to replace the gravitational description, which contains gauge redundancies, with an ordinary one-dimensional mechanics that is more suitable for quantization.

The Schwarzschild coordinates appearing in \eqref{geom} cover only the exterior region of the black hole, where the coordinate $r$ is bounded between $\Phi_h < r < \Phi_b$. In principle, one might introduce a set of Kruskal-like coordinates allowing to maximally extend the geometry \eqref{geom} to a two-sided strip just like what is usually done for JT gravity.
We will not need the explicit form of the maximally extended geometry, and it will be sufficient to label points on its left- and right-boundaries by the corresponding proper times
\begin{equation}\label{eq:proper_times}
\tau_{\text{L}/\text{R}}
=
\sqrt{
\frac{
\cosh(2\pi b^2\Phi_b)
-
\cosh(2\pi b^2\Phi_h)
}{
2\pi^2b^4
}
}
t_{\text{L}/\text{R}}.
\end{equation}
Guided by diffeomorphism invariance, we choose as the time parameter on the auxiliary phase space the sum
\begin{equation}\label{eq:tau_definition}
\tau
\equiv
\tau_{\mathrm L}
+
\tau_{\mathrm R}.
\end{equation}
The Hamiltonian generating evolution along $\tau$ is then naturally identified with the quantity that is conserved by virtue of the existence of the Killing vector fields $\partial_{\tau_{\text{L}/\text{R}}}$ in the geometries \eqref{geom}.
Following an identical procedure to the one discussed in Section 2.2 of \corr{\cite{Griguolo:2026eoi}}, we find that the latter is obtained by contracting the Brown--York tensor on each boundary\footnote{For a timelike boundary, the Brown--York boundary stress-energy tensor is defined as
\begin{equation*}
    T^{\text{L}/\text{R}}_{\alpha\beta}
    \equiv
    - \frac{2 }{\sqrt{\abs{h}}}
    \frac{\delta S}{\delta h^{\alpha\beta}} \bigg\lvert_{\partial\mathcal{M}_{\text{L}/\text{R}}}
    =
    \left( n^\mu \nabla_\mu \Phi - \corrIV{\sqrt{\frac{\cosh(2\pi b^2\Phi)-1}{2\pi^2b^4}}} \right)\bigg\lvert_{\partial\mathcal{M}_{\text{L}/\text{R}}}
    h_{\alpha\beta}
    =
    - H_{\text{L}/\text{R}} h_{\alpha\beta}.
\end{equation*}
} with the Killing vector field $\partial_{\tau_{\text{L}/\text{R}}}$ twice. \rev{Averaging} the contributions from the two boundaries, we obtain the following expression for the quasi-local energy
\begin{equation}\label{quasilocal}
    H(\Phi_h) \equiv \frac{H_\text{L} + H_\text{R}}{2} = \corrIV{\frac{1}{2 \pi b^2} \left[\sqrt{2\cosh(2 \pi b^2 \Phi_b)-2} - \sqrt{2\cosh(2 \pi b^2 \Phi_b)-2\cosh(2 \pi b^2 \Phi_h)} \right]}.
\end{equation}
We notice in passing that, as $\Phi_b \to \infty$, the Hamiltonian \eqref{quasilocal} reduces to
\begin{equation}
    H(\Phi_h) \overset{\Phi_b \to \infty}{\sim} \frac{1}{2 \pi b^2} \frac{\cosh(2\pi b^2 \Phi_h)\corrIV{-1}}{\sqrt{2 \cosh(2 \pi b^2 \Phi_b)}},
\end{equation}
which reproduces the ADM energy of asymptotic Liouville gravity up to a cutoff-dependent normalization\corrIV{, while for $b\to0$ \eqref{quasilocal} reduces to the JT energy \eqref{eq:JT_energy_phi_review}}. This fact provides a first consistency check of the current construction.

Before turning to the quantization of the theory, we must replace the pair $(\tau,H)$ with a more convenient set of phase space coordinates. On the gravitational side, this procedure corresponds to choosing a foliation of the maximally extended geometries with spacelike Cauchy slices labeled by the boundary proper time $\tau$ of their endpoints. On each slice, we then specify the state of the system in terms \corr{of} a pair of canonically conjugate variables. This is possible since the phase space of any dilaton gravity is known to be two-dimensional \corr{\cite{Louis-Martinez:1993bge}}.

\subsection{Auxiliary AdS$_2$ geometry and reduced phase space}\label{sec:auxiliary_phase_space}
Before specifying which set of spacelike slices we find convenient for the foliation, we \rev{review} the construction of the auxiliary AdS$_2$ spacetime introduced in \cite{Blommaert:2024ymv}. This is the approach that has led to the identification of the gravitational Hamiltonian with the transfer matrix of DSSYK \corr{\cite{Blommaert:2024ymv}}, and with the reduced Hamiltonian of a particle on a quantum group \cite{Blommaert:2023wad}, for the sine and sinh dilaton cases respectively, thereby leading to the establishment of the corresponding infinite cutoff holographic dualities.
Thus, we recall it here and adapt it to our finite cutoff setup.

In two spacetime dimensions, every metric can be related to any other via a suitable Weyl rescaling. In particular, we seek a Weyl factor $\omega(r)$ and a radial coordinate transformation $\rho(r)$ that map the on-shell metric \eqref{geom} to the Schwarzschild patch of AdS$_2$,
\begin{equation}\label{eq:auxiliary_AdS}
\mathrm{d}s_{\mathrm{AdS}}^2
=
-
\left(
\rho^2-\rho_h^2
\right)
\mathrm{d}t^2
+
\frac{
\mathrm{d}\rho^2
}{
\rho^2-\rho_h^2
},
\qquad
\rho_h<\rho<\rho_b,
\end{equation}
according to
\begin{equation}\label{eq:Weyl_rescaling}
\corrIV{\mathrm{d}s_{\mathrm{AdS}}^2(t,\rho)
=
\omega^2(r)\,
\mathrm{d}s^2(t,r)},
\qquad
\rho=\rho(r),
\end{equation}
while keeping the Schwarzschild time coordinate $t$ unchanged. This leads to the differential equation
\begin{equation}
    \dv{\rho}{r} = \frac{2 \pi^2 b^4}{\cosh(2 \pi b^2 r) - \cosh(2 \pi b^2 \Phi_h)} (\rho^2 - \rho_h^2),
\end{equation}
with the Weyl factor given in turn by $\omega^2(r) = \dv{\rho}{r}$.
The most general solution to this equation, subject to the requirement that the horizon of the physical spacetime is mapped to the corresponding AdS$_2$ horizon, $r(\rho_h)=\Phi(\rho_h)=\Phi_h$, and that the Hawking temperatures of the two geometries coincide, is
\begin{equation}\label{eq:changegeneral}
    r(\rho) = \frac{1}{2\pi b^2} \log\!\left[\frac{\rho_h \cosh(\pi b^2 \Phi_h-\delta)+\rho \sinh \left(\pi b^2 \Phi_h-\corr{\delta}\right)}{\rho_h \cosh(\pi b^2 \Phi_h+\delta)-\rho \sinh \left(\pi b^2 \Phi_h+\corr{\delta}\right)}\right] \qquad \rho_h=\frac{\sinh(2\pi b^2 \Phi_h)}{2 \pi b^2},
\end{equation}
where $\delta$ parametrizes the residual freedom associated with the choice of boundary conditions.\footnote{Notice indeed that the inverse Hawking temperature of the auxiliary geometry $\beta_{\mathrm{AdS}}=2\pi/\rho_h$ matches with the inverse Hawking temperature of the  sinh dilaton black hole \eqref{eq:Hawkingsinh}. On the other hand, the mapping of the black hole horizons preserves the Bekenstein–Hawking entropy.} 
We now further require the asymptotic boundary of the physical spacetime, located at $r\to\infty$, to be mapped to the asymptotic boundary of AdS$_2$, $\rho\to\infty$. This condition fixes the boundary parameter in \eqref{eq:changegeneral} to $\delta=-\pi b^2\Phi_h$. We will comment on the interpretation of this specific Weyl rescaling in terms of the Liouville formulation of sinh dilaton gravity in the concluding Section \corr{\ref{sec:conclusions}}.

The Weyl factor and radial coordinate transformation then reduce to
\begin{equation}\label{eq:change}
\omega^2(r)
=
e^{2\pi b^2r},
\qquad
\rho(r)
=
\frac{
e^{2\pi b^2r}
-
\cosh(2\pi b^2\Phi_h)
}{
2\pi b^2
}.
\end{equation}
With the choice \eqref{eq:change}, the sinh dilaton metric \eqref{geom} is mapped precisely to the AdS$_2$ black hole geometry \eqref{eq:auxiliary_AdS}, with the auxiliary horizon and boundary located at
\begin{equation}\label{eq:rho_h_rho_b}
\rho_h
\equiv
\rho(\Phi_h)
=
\frac{
\sinh(2\pi b^2\Phi_h)
}{
2\pi b^2
},
\qquad
\rho_b
\equiv
\rho(\Phi_b)
=
\frac{
e^{2\pi b^2\Phi_b}
-
\cosh(2\pi b^2\Phi_h)
}{
2\pi b^2
}.
\end{equation}

An important feature of \eqref{eq:rho_h_rho_b} is that the location $\rho_b$ of the AdS$_2$ boundary depends explicitly on the black hole parameter $\Phi_h$. Thus, while the physical boundary is fixed by the Dirichlet condition $\Phi=\Phi_b$, its image under the Weyl transformation varies for each geometry labeled by $\Phi_h$. This dependence has no analogue in the finite cutoff JT construction and must be taken into account in the following.

The auxiliary geometry \eqref{eq:auxiliary_AdS} allows us to introduce a convenient foliation of the maximally extended spacetime. In \corr{\cite{Harlow:2018tqv,Griguolo:2026eoi}} the AdS$_2$ spacetime was foliated using horizontal boundary-to-boundary geodesics as Cauchy slices; here, we can choose to foliate the sinh dilaton geometry with the image of these curves under the map \eqref{eq:change}. These slices are not geodesics in the sinh dilaton geometry \eqref{geom}, but the positivity of the Weyl factor preserves their spacelike character.

We choose the first phase space variable to be \corr{the} length of these curves measured in the auxiliary AdS$_2$ metric, and we denote this quantity as $L$. For a symmetric slice whose endpoints satisfy $t_\text{L}=t_\text{R}=t$, the length is given by
\begin{equation}\label{eq:length_computation}
    L = \int_\text{slice} \dd s_\text{AdS} = \int_\text{slice} \omega \, \dd s = 2 \sinh^{-1}\!\left[ \frac{\sqrt{\rho_b^2 - \rho_h^2}}{\rho_h} \cosh(\rho_h \, t_{\text{L}/\text{R}}) \right].
\end{equation}
Expression \eqref{eq:length_computation} is found by first maximally extending the AdS$_2$ metric \eqref{eq:auxiliary_AdS} and then integrating over the aforementioned horizontal geodesics, see Equation 2.15 of \corr{\cite{Griguolo:2026eoi}}.
Using the relation \eqref{eq:proper_times} between the Schwarzschild time and the proper times of the two boundaries, this quantity can be written entirely in terms of $\tau$ and the boundary data $\Phi_h$, $\Phi_b$. Doing so results in
\begin{equation}\label{eq:length_sinh}
\begin{split}
    L(\tau,\Phi_h)
    = 2 \sinh^{-1}\!\Bigg[ &\sqrt{\left(\frac{e^{2 \pi b^2 \Phi_b} - \cosh(2 \pi b^2 \Phi_h)}{\sinh(2 \pi b^2 \Phi_h)}\right)^2 - 1} \times \\
    &\times \cosh\!\left( \frac{\sinh(2 \pi b^2 \Phi_h)}{\sqrt{2 \cosh(2 \pi b^2 \Phi_b) - 2 \cosh(2 \pi b^2 \Phi_h)}} \frac{\tau}{2} \right) \Bigg].
\end{split}
\end{equation}
It remains to determine the momentum $P$ conjugate to $L$. This can be done by inverting Hamilton's equation
\begin{equation}
\frac{\mathrm{d}L}{\mathrm{d}\tau}
=
\frac{\partial H}{\partial P}
=
\frac{\partial H}{\partial\Phi_h}
\frac{\partial\Phi_h}{\partial P},
\end{equation}
which results in\footnote{To find $P$ in \eqref{eq:phase_space_LP}, one has to compute the integral \begin{equation}
    P = \int \dd \Phi_h \,
    \left( 1 - \frac{\rho_b(\Phi_h)^2}{\rho_h(\Phi_h)^2} \frac{1}{\cosh^2(L/2)} \right)^{-1/2}.
\end{equation}
}
\begin{equation}\label{eq:phase_space_LP}
\cosh\!\left(
2\pi b^2P\tanh\frac{L}{2}
\right)
=
\frac{
e^{2\pi b^2\Phi_b}
+
\cosh(2\pi b^2\Phi_h)\sinh^2(L/2)
}{
\cosh(L/2)
\sqrt{
e^{4\pi b^2\Phi_b}
+
\sinh^2(L/2)
}
}.
\end{equation}

For later purposes, it is useful to invert \eqref{eq:phase_space_LP} and express the parameter $\Phi_h$ directly in terms of the new variables $(L,P)$. This gives
\begin{equation}\label{eq:cosh_hamiltonian}
\cosh(2\pi b^2\Phi_h)
=
\sqrt{
1
+
\frac{
e^{4\pi b^2\Phi_b}
}{
\sinh^2(L/2)
}
}
\frac{
\cosh\!\left(
2\pi b^2P\tanh(L/2)
\right)
}{
\tanh(L/2)
}
-
\frac{
e^{2\pi b^2\Phi_b}
}{
\sinh^2(L/2)
}.
\end{equation}
Substituting \eqref{eq:cosh_hamiltonian} into the quasi-local energy \eqref{quasilocal} yields the complete Hamiltonian of the two-sided theory as a function of the canonical variables on the slice $H=H(L,P)$. In practice, however, it will be more convenient to quantize the simpler quantity appearing on the left-hand side of \eqref{eq:cosh_hamiltonian} and to recover the full energy spectrum through \eqref{quasilocal} only after the eigenvalue problem has been solved.

\begin{figure}[t]
    \centering\includegraphics[width=\linewidth]{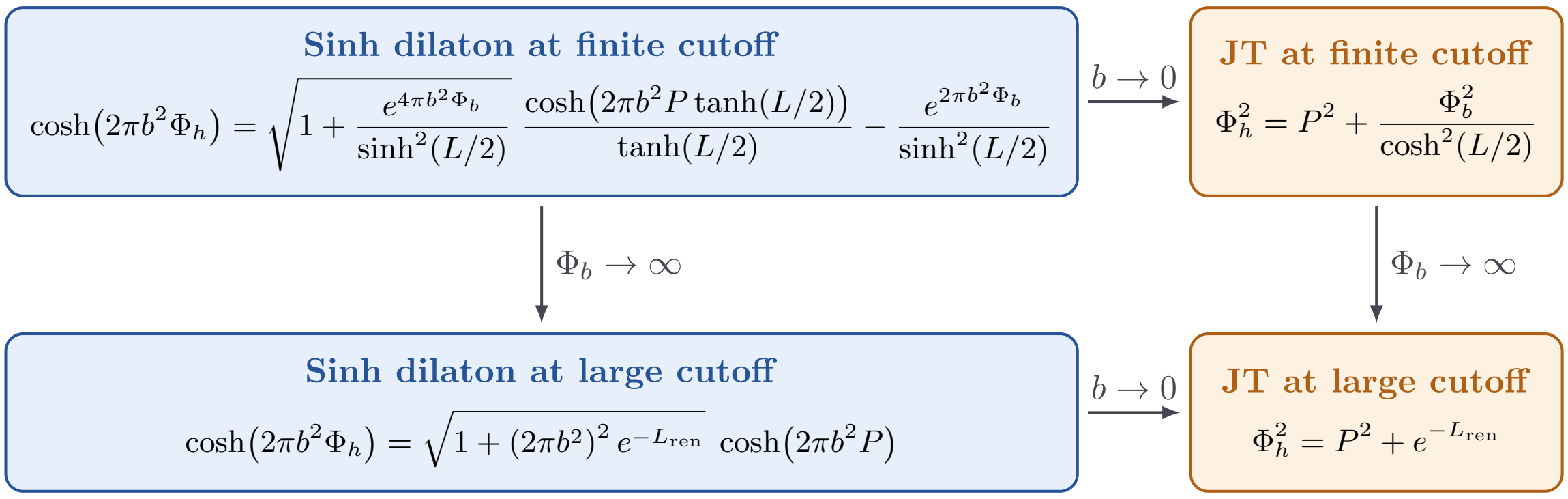}
    \caption{Classical Hamiltonians of sinh dilaton gravity and of JT gravity, at finite and at large cutoff. The JT limit $b\to0$ and the large cutoff limit $\Phi_b\to\infty$ commute.}
    \label{fig:sinh_limits}
\end{figure}

It is useful to pause and examine the structure of \eqref{eq:cosh_hamiltonian}. The expression incorporates two independent deformations of the Liouville Hamiltonian \eqref{eq:JT_Liouville_H_review} that governs asymptotic JT gravity. The parameter $b$ controls the sinh deformation and introduces a hyperbolic dependence on the canonical momentum, which will produce finite-translation operators upon quantization. The finite value of $\Phi_b$, on the other hand, encodes the effect of placing the theory in a box. The finite cutoff JT and asymptotic Liouville limits provide two immediate checks of this interpretation.

To recover the finite cutoff JT theory, we expand \eqref{eq:cosh_hamiltonian} at small $b$. At leading nontrivial order, one finds
\begin{equation}\label{eq:classical_JT_limit}
\begin{split}
1
+
\frac{1}{2}
(2\pi b^2)^2\Phi_h^2
+
O(b^6)
=
1
+
\frac{1}{2}
(2\pi b^2)^2
\left[
P^2
+
\frac{
\Phi_b^2
}{
\cosh^2(L/2)
}
\right]
+
O(b^6),
\end{split}
\end{equation}
and therefore we identify
\begin{equation}
\Phi_h^2
=
P^2
+
\frac{
\Phi_b^2
}{
\cosh^2(L/2)
}.
\end{equation}
This is precisely the classical P\"oschl--Teller Hamiltonian obtained in the finite cutoff formulation of JT gravity \cite{Griguolo:2026eoi}.

The asymptotic Liouville gravity theory at finite $b$ is instead recovered by sending $\Phi_b \to \infty$ while keeping fixed the renormalized length
\begin{equation}\label{eq:renormalized_length}
L_{\mathrm{ren}}
\equiv
L
-
2\log\!\left[
\frac{2}{2\pi b^2}
e^{2\pi b^2\Phi_b}
\right].
\end{equation}
This choice of renormalization can be motivated starting from the expression \eqref{eq:length_sinh} of the Weyl-rescaled geodesic length in sinh dilaton gravity.
When the boundary value of the dilaton is large $\Phi_b \to \infty$, this expression simplifies to (recall that $\sinh^{-1}(x) \overset{x \to \infty}{\longrightarrow} \log(2x)$):
\begin{equation}\label{eq:sinh_length_large_cutoff}
\begin{split}
    L &\overset{\Phi_b \to \infty}{\longrightarrow} 2 \log \left[ 2 \frac{e^{2\pi b^2 \Phi_b} - \cosh(2 \pi b^2 \Phi_h)}{\sinh(2\pi b^2 \Phi_h)} \cosh\!\left( \frac{\sinh(2\pi b^2 \Phi_h)}{2\pi b^2}  t\right) \right] \\
    &= \underbrace{2 \log \left[ \frac{2\pi b^2}{\sinh(2\pi b^2 \Phi_h)} \cosh\!\left( \frac{\sinh(2\pi b^2 \Phi_h)}{2\pi b^2}  t\right) \right]}_{\displaystyle{= L_\text{ren}}} + \underbrace{2 \log \left[ 2 \frac{e^{2\pi b^2 \Phi_b} - \cosh(2\pi b^2 \Phi_h)}{2\pi b^2} \right]}_{\text{counterterm}},
\end{split}
\end{equation}
where in the second step we reorganized the expression into the standard renormalized geodesic length in the auxiliary AdS$_2$ geometry \eqref{eq:auxiliary_AdS} and its associated counterterm. 
Notice that this split is such that the two terms individually approach the renormalized length and the counterterm \eqref{eq:geodesic_length_renormalized} found in JT gravity, respectively, as $b \to 0$. Indeed, as $b\rightarrow 0$, both the Weyl rescaling and the radial coordinate change in \eqref{eq:change} become trivial and the auxiliary AdS$_2$ geometry of sinh dilaton gravity reduces to the AdS$_2$ black hole of JT gravity accordingly.

On the other hand, when we take $\Phi_b$ large prior to taking $b \rightarrow 0$, the more general counterterm in \eqref{eq:sinh_length_large_cutoff} reduces to the renormalization performed in \eqref{eq:renormalized_length}.

In the limit \eqref{eq:renormalized_length}, equation \eqref{eq:cosh_hamiltonian} becomes
\begin{equation}\label{eq:classical_Liouville_limit}
\cosh(2\pi b^2\Phi_h)
\overset{\Phi_b\to\infty}{\sim}
\sqrt{
1
+
(2\pi b^2)^2e^{-L_{\mathrm{ren}}}
}
\,
\cosh(2\pi b^2P),
\end{equation}
which indeed reproduces the Liouville gravity Hamiltonian \cite{Mertens:2020hbs}, up to a canonical transformation. Furthermore, one can check that the two limits commute, as summarized in Figure~\ref{fig:sinh_limits}.
These two limits will provide useful consistency checks for the quantum theory constructed below.

\subsection{Canonical quantization}\label{subsec:sinh_quantization}
We now turn to the quantization of the reduced Hamiltonian \eqref{eq:cosh_hamiltonian} obtained in the previous section. The canonical variables $L$ and $P$ are promoted to operators satisfying canonical commutation relations
\begin{equation}\label{eq:canonical_commutation}
[\hat L,\hat P]
=
i\hbar.
\end{equation}

As in the finite cutoff formulation of JT gravity, the classical expression \eqref{eq:cosh_hamiltonian} does not determine a unique quantum operator, since functions of $\hat{L}$ do not commute with terms depending on $\hat{P}$. An ordering prescription must therefore be specified. We propose to adopt a formally symmetric ordering that reproduces the P\"oschl--Teller Hamiltonian \eqref{eq:Schroedinger} as $b \to 0$, as we will explicitly check later.
An operator ordering satisfying all these properties is given by
\begin{equation}\label{eq:sinh_quantum_hamiltonian2}
\begin{split}
\cosh(2\pi b^2\hat\Phi_h)
=
{}&
\frac{1}{2}
\left(
e^{\pi b^2\hat\Pi} \,
f(\hat\Lambda) \,
e^{\pi b^2\hat\Pi}
+
e^{-\pi b^2\hat\Pi} \,
f(\hat\Lambda) \,
e^{-\pi b^2\hat\Pi}
\right)
-
e^{2\pi b^2\Phi_b} \, e^{-\hat\Lambda},
\end{split}
\end{equation}
where we introduced the pair of operators
\begin{equation}\label{eq:quantum_canonical_variables}
    \hat{\Lambda} \equiv 2\log\!\left( \sinh\!{(\hat{L}/2)} \right),
    \qquad
    \hat{\Pi} \equiv \frac{1}{2} \left( \hat{P} \tanh\!{(\hat{L}/2)} + \tanh\!{(\hat{L}/2)} \hat{P} \right)
\end{equation}
and the function
\begin{equation}\label{eq:sinh_f_function}
    f(\hat{\Lambda})
    \equiv \sqrt{ \left(1 + e^{4\pi b^2 \Phi_b} \, e^{-\hat{\Lambda}} \right) \left(1 + e^{-\hat{\Lambda}} \right)
    }
    = \sqrt{1 + \frac{e^{4 \pi b^2 \Phi_b}}{\sinh^2(\hat{L}/2)}} \frac{1}{\tanh\!{(\hat{L}/2)}}.
\end{equation}

Remarkably, the two operators \eqref{eq:quantum_canonical_variables} are themselves a canonical pair, as one can check that
\begin{equation}
    [\hat{\Lambda}, \hat{\Pi}] = i\hbar.
\end{equation}
Consequently, $\hat{\Pi}$ acts on functions of $\Lambda$ as the differential operator $\hat{\Pi} \to -i \hbar \partial_\Lambda$ and its exponential $e^{\pm \pi b^2 \hat{\Pi}}$ generates finite shifts. Therefore, the eigenvalue problem associated with \eqref{eq:sinh_quantum_hamiltonian2} reads\footnote{
At this stage we must relate the $L$-eigenstates to the $\Lambda$-eigenstates. In order to preserve the resolution of the identity $1 = \int_0^\infty \dd L \ket{L} \bra{L} = \int_{-\infty}^{+\infty} \dd\Lambda \ket{\Lambda} \bra{\Lambda}$, we choose to define $\ket{\Lambda} \equiv \sqrt{\dv{L}{\Lambda}} \ket{L(\Lambda)}$. Furthermore, we introduce the wavefunction $\psi_s(\Lambda) \equiv \bra{\Lambda}\ket{s}$, which is related to $\psi_s(L) \equiv \bra{L}\ket{s}$ as $\psi_s(\Lambda(L)) = \sqrt{\dv{L}{\Lambda}} \psi_s(L)$.}
\begin{equation}\label{eq:Schroedinger_equation}
\begin{split}
    0 &= \frac{1}{2} \left(
    f(\Lambda - i \pi b^2) \, \psi_s(\Lambda - 2 i \pi b^2) +
    f(\Lambda + i \pi b^2) \, \psi_s(\Lambda + 2 i \pi b^2)
    \right)
    - \\
    &\quad - \left( e^{2 \pi b^2 \nu} \, e^{-\Lambda} + \cosh\!{(2 \pi b^2 s)} \right) \psi_s(\Lambda),
\end{split}
\end{equation}
where we streamlined the notation by rescaling the parameters as
\begin{equation}
    \Phi_h \to \hbar s, \qquad \Phi_b \to \hbar \nu, \qquad b^2 \to b^2/\hbar.
\end{equation}

Although equation \eqref{eq:Schroedinger_equation} follows directly from the symmetric ordering \eqref{eq:sinh_quantum_hamiltonian2}, the square roots contained in $f(\Lambda)$ make it inconvenient for constructing explicit solutions. It is therefore useful to perform a similarity transformation involving the Faddeev's quantum dilogarithm function, see Appendix \corr{\ref{app:faddeev}}. To avoid confusing this function with the boundary value $\Phi_b$ of the dilaton, we denote it by $\phi_b(z)$ and adopt the convention where it satisfies the shift relation
\begin{equation}\label{eq:dilog_shift}
\phi_b(z\pm ib)
=
\left(
1+e^{2\pi bz}e^{\pm i\pi b^2}
\right)^{\pm1}
\phi_b(z).
\end{equation}
Making an ansatz of the form
\begin{equation}\label{eq:factor}
\psi_s(\Lambda)
=
\sqrt{
\frac{
\phi_b\!\left(
-\frac{\Lambda}{2\pi b}
\right)
}{
\phi_b\!\left(
\frac{-\Lambda+4\pi b^2\nu}{2\pi b}
\right)
}
}
\,
\chi_s(\Lambda),
\end{equation}
we bring the difference equation \eqref{eq:Schroedinger_equation} to the simpler form
\begin{equation}\label{eq:sinh_difference_equation_chi}
\begin{split}
\cosh(2\pi b^2s)\chi_s(\Lambda)
={}&
\frac{1}{2}
\left(
1+e^{-\Lambda+i\pi b^2}
\right)
\chi_s(\Lambda-2\pi ib^2) +
\\
&+
\frac{1}{2}
\left(
1+e^{4\pi b^2\nu}
e^{-\Lambda-i\pi b^2}
\right)
\chi_s(\Lambda+2\pi ib^2) -
\\
&-
e^{2\pi b^2\nu}e^{-\Lambda}
\chi_s(\Lambda).
\end{split}
\end{equation}
The transformed operator is no longer manifestly symmetric with respect to the flat measure $\mathrm{d}\Lambda$. Nevertheless, it is related to the symmetric operator \eqref{eq:sinh_quantum_hamiltonian2} by the similarity transformation \eqref{eq:factor}, and its asymmetric form is considerably more convenient for identifying explicit eigenfunctions.

The eigenvalue equation \eqref{eq:sinh_difference_equation_chi} is now polynomial in $e^{-\Lambda}$, meaning that it can be further simplified via a Fourier ansatz
\begin{equation}\label{eq:sol}
\chi_s(\Lambda)
=
\int_{\mathcal C_{\Pi}}
\mathrm{d}\Pi\,
e^{-i\Lambda\Pi}
\widetilde{\chi}_s(\Pi).
\end{equation}
The contour $\mathcal C_{\Pi}$ may initially be taken to coincide with the real axis, with subsequent deformations understood whenever required by analytic continuation. In deriving the momentum space equation, we assume that the contour can be shifted without crossing poles and that the integrand decays sufficiently rapidly at its endpoints. Under these assumptions, the finite translations of $\Lambda$ in \eqref{eq:sinh_difference_equation_chi} become multiplicative factors in momentum space, whereas multiplication by $e^{-\Lambda}$ shifts $\Pi$ by $i$. The resulting equation is the first order recurrence relation
\begin{equation}\label{eq:sinh_recurrence_relation}
\frac{
\widetilde{\chi}_s(\Pi+i)
}{
\widetilde{\chi}_s(\Pi)
}
=
e^{-2\pi b^2\nu}
\frac{
\sinh\!\left[
\pi b^2(s+\Pi)
\right]
\sinh\!\left[
\pi b^2(s-\Pi)
\right]
}{
\sinh^2\!\left[
\pi b^2(\Pi+\nu+i/2)
\right]
}.
\end{equation}

The structure of \eqref{eq:sinh_recurrence_relation} suggests expressing its solution in terms of the double sine function $S_b(z)$, whose shift property reads
\begin{equation}\label{eq:double_sine_shift}
    S_b(z+b) = 2\sin(\pi bz)\, S_b(z) \qquad S_b(z-b) = \frac{S_b(z)}{2 \sin(\pi b (z-b))}.
\end{equation}
In what follows, we use the shorthand $S_b(x\pm y) \equiv S_b(x+y)S_b(x-y)$.
A direct application of \eqref{eq:double_sine_shift} shows that
\begin{equation}\label{eq:fourier_solution}
\widetilde{\chi}_s(\Pi)
=
\frac{\corrIV{b^2}}{S_b(\frac{b}{2} + ib\nu \pm ibs)} \times
\frac{
S_b(-ib\Pi\pm ibs)\,
S_b\!\left(
\frac{b}{2}+ib\Pi+ib\nu
\right)
}{
S_b\!\left(
\frac{b}{2}-ib\Pi-ib\nu
\right)
}
e^{2\pi ib^2\nu\Pi}
\end{equation}
satisfies \eqref{eq:sinh_recurrence_relation}, where we also introduced a convenient normalization prefactor. Plugging \eqref{eq:fourier_solution} back into the ansatz \eqref{eq:sol} and performing the change of integration variable $t = -i\Pi - i\nu - \frac{1}{2}$, we obtain an explicit solution in integral form
\begin{equation}\label{eq:chi_eigenfunction}
    \chi_s(\Lambda) =
    e^{(\Lambda - 2 \pi b^2 \nu)(\frac{1}{2}+i\nu)} \frac{2 \pi b^2}{S_b(\frac{b}{2} + ib\nu \pm ibs)} \int_{\mathcal{C}_t} \frac{\dd t}{2 \pi i} e^{(\Lambda - 2 \pi b^2 \nu)t} \frac{S_b(\frac{b}{2} + ib\nu \pm ibs + b t) \, S_b(-bt)}{S_b(b+bt)}.
\end{equation}
Here $\mathcal C_t$ is a Mellin--Barnes contour separating the ascending and descending sequences of poles of the integrand.

The integral appearing in \eqref{eq:chi_eigenfunction} can be recognized as a $q$-deformation ($b$-deformation) of the Mellin--Barnes integral representation of the ordinary hypergeometric function. Indeed the $q$-deformed ($b$-deformed) hypergeometric function can be defined as \cite{Ponsot:1999uf,NishizawaUeno1996,PonsotTeschner2001,Ip2013QuantumPlane}
\begin{equation}\label{eq:b_hypergeometric}
\begin{split}
F_b(\alpha,\beta,\gamma;z)
=
\frac{
2\pi b\,S_b(\gamma)
}{
S_b(\alpha)S_b(\beta)
}
\int_{\mathcal C_t}
\frac{\mathrm{d}t}{2\pi i}\,
(-z)^t
e^{i\pi b t(\alpha+\beta-\gamma)}
\frac{
S_b(\alpha+bt)
S_b(\beta+bt)
S_b(-bt)
}{
S_b(\gamma+bt)
}.
\end{split}
\end{equation}

Collecting everything together, we conclude that a solution of the original Schr\"odinger equation \eqref{eq:Schroedinger_equation} is given by
\begin{equation}\label{eq:sinh_eigenfunction}
\begin{split}
\psi_s(\Lambda)
=
\sqrt{
\frac{
\phi_b\!\left(
-\frac{\Lambda}{2\pi b}
\right)
}{
\phi_b\!\left(
\frac{-\Lambda+4\pi b^2\nu}{2\pi b}
\right)
}
}
e^{
(\Lambda-2\pi b^2\nu)
(\frac{1}{2}+i\nu)
}
F_b\!\left(
\frac{b}{2}+ib\nu+ibs,\,
\frac{b}{2}+ib\nu-ibs,\,
b;\,
-e^\Lambda
\right).
\end{split}
\end{equation}
The identification of the wavefunction $\psi_{s}(\Lambda)$ as a $q$-deformed ($b$-deformed) hypergeometric function is remarkable as it shows that our procedure of constructing the finite cutoff Hamiltonian of sinh dilaton gravity is consistent with a quantum deformation of the JT gravity framework, where the eigenfunctions are given by ordinary hypergeometric functions \eqref{eq:waveJT}.
In Sections~\ref{sec:casimir} and \ref{sec:wave}, we will substantiate this aspect further in the following by identifying the Hamiltonian \eqref{eq:sinh_difference_equation_chi} with the reduced Casimir of \rev{$U_q(\mathfrak{sl}(2,\mathbb{R}))$}, and of the wavefunction \eqref{eq:sinh_eigenfunction} as a combination of hyperbolic matrix elements of the modular double. 

There is, however, an ambiguity at the level of the Schr\"odinger equation \eqref{eq:Schroedinger_equation} that we need to address. Indeed, starting from the solution \eqref{eq:sinh_eigenfunction} and multiplying it by some $2 \pi i b^2$-periodic function
\begin{equation}\label{eq:periodic_ambiguity}
    \psi_s(\Lambda) \to g(\Lambda) \psi_s(\Lambda) \qquad \text{with} \qquad g(\Lambda + 2 \pi i b^2)=g(\Lambda),
\end{equation}
we obtain a new and independent solution. \corr{We will argue in Section~\ref{subsec:uniqueness} that this ambiguity can be removed
by requiring the conditions that make \eqref{eq:sinh_quantum_hamiltonian2} a well-defined operator.} Before that, we discuss some important features of the wavefunctions obtained above.

\subsection{Properties of the eigenfunction}
In this Section, we discuss some properties of the eigenfunction \eqref{eq:sinh_eigenfunction} that provide consistency checks and will be useful for computing physically relevant quantities in the next Sections.

\paragraph{\bfseries The JT limit.}
A first consistency check for the discussion presented above is provided by the limit $b \to 0$ at fixed $s$, $\nu$, and $L$. In this regime, sinh dilaton gravity reduces to JT gravity both at the level of the classical Hamiltonian \eqref{eq:classical_JT_limit}
and of the corresponding spectral problem \eqref{eq:sinh_quantum_hamiltonian2}. Here we consider the $b \to 0$ limit of the eigenfunction \eqref{eq:sinh_eigenfunction}. The ratio of quantum dilogarithms appearing there simplifies in this limit using the asymptotics \eqref{eq:dilog_b=0}
\begin{equation}
\begin{split}
   e^{\frac{1}{2} \left( \log{\phi_b(\frac{-\Lambda}{2 \pi b})} - \log{\phi_b(\frac{-\Lambda+4 \pi b^2 \nu}{2 \pi b})} \right)}
    \overset{b \to 0}{\sim} e^{i \nu \frac{\dd}{\dd \Lambda} \text{Li}_2(-e^{-\Lambda})}
    = (1 + e^{-\Lambda})^{i \nu}
    = \tanh\!{(L/2)}^{-2i\nu}.
\end{split}
\end{equation}
Next, we recall that double sine functions reduce to ordinary Gamma functions as $b \to 0$, see Equation \eqref{eq:double_sine_b=0}. Consequently, the Mellin--Barnes integral defining the $b$-hypergeometric function \eqref{eq:b_hypergeometric} reduces to the corresponding integral representation of the ordinary hypergeometric function
\begin{equation}
    F_b(bA, bB, bC; z) \overset{b \to 0}{\sim} \frac{\Gamma(C)}{\Gamma(A)\Gamma(B)} \int_{\mathcal{C}_t} \frac{\dd t}{2 \pi i} (-z)^t \frac{\Gamma(A+t) \, \Gamma(B+t) \, \Gamma(-t)}{\Gamma(C+t)} = {}_2F_1(A,B,C;z).
\end{equation}
Thus, putting everything together and recalling that $e^{\Lambda/2}=\sinh\!{(L/2)}$, we find that the $b \to 0$ limit of the eigenfunction \eqref{eq:sinh_eigenfunction} is given by
\begin{equation}\label{eq:JT_limit}
\begin{split}
\psi_s(\Lambda)
\overset{b \to 0}{\sim} \sinh\!{(L/2)} \cosh\!{(L/2)}^{2i\nu} \,
{}_2F_1\!\left(
\frac{1}{2}+i\nu+is,\,
\frac{1}{2}+i\nu-is;\,
1;\,
-\sinh^2\!\frac{L}{2}
\right).
\end{split}
\end{equation}
If we now recall that $\psi_s(\Lambda(L)) = \sqrt{\dv{L}{\Lambda}} \psi_s(L) = \sqrt{\tanh(L/2)} \psi_s(L)$, we can conclude that
\begin{equation}
    \psi_s(L) \overset{b \to 0}{\sim} \psi_s^\text{JT}(L),
\end{equation}
where $\psi_s^\text{JT}(L)$ is the wavefunction of JT gravity at finite cutoff \eqref{eq:waveJT}.

This is perfectly consistent with the small $b$ expansion of the Hamiltonian of sinh dilaton gravity: starting from \eqref{eq:sinh_quantum_hamiltonian2}, we can expand
\begin{equation}
\cosh\!{(2\pi b^2 \hat{\Phi}_h)} = 1 + \frac{(2\pi b^2)^2}{2} \hat{\Phi}_h^2 + O(b^8)
\end{equation}
and, recalling the relation between the canonical variables \eqref{eq:quantum_canonical_variables}, we precisely recover the JT gravity Schr\"odinger equation \eqref{eq:Schroedinger} with the non-trivial operator ordering.

\paragraph{\bfseries The large cutoff limit.}
A second consistency check is obtained by taking the large cutoff limit $\nu \to \infty$ while keeping the parameter $b$ fixed.
Combining the renormalization prescription \eqref{eq:renormalized_length} for the geodesic length with the definition of $\Lambda$ \eqref{eq:quantum_canonical_variables}, we learn that the latter is renormalized as
\begin{equation}
    \Lambda \overset{\nu \to \infty}{\sim} L_\text{ren} + 2 \log\!\left[ \frac{e^{2 \pi b^2 \nu}}{2 \pi b^2} \right].
\end{equation}
First, using the asymptotics \eqref{eq:dilog_asymptotics} one finds that the \corrIX{numerator of the} dilogarithm prefactor in \eqref{eq:sinh_eigenfunction} goes to $1$.
Moreover, introducing $Q=b+b^{-1}$, we recall that the double sine function has the asymptotic behavior
\begin{equation}\label{eq:double_sine_asymptotics}
S_b(x)
\sim
\exp\!\left[
\pm \frac{i\pi}{2} \left( x(Q-x) - \frac{Q^2+1}{6}
\right) \right]
\qquad
\text{for}
\qquad
\text{Im}(x)\to\pm\infty.
\end{equation}
Applying this asymptotics to our solution yields
\begin{equation}\label{eq:Liouville_limit2}
\begin{split}
\psi_s(\Lambda)
\overset{\nu\to\infty}{\sim}
{}&
\corrIX{\frac{e^{\pi\nu} \, 2 \pi b^2}{\sqrt{\phi_b\!\left(-\frac{L_\text{ren}-2\log(2\pi b^2)}{2\pi b}\right)}}}
\int_{\mathcal{C}_t}
\frac{\dd t}{2 \pi i} \left[  (2 \pi b^2)^2 \, e^{-L_\text{ren}}\right]^{-t} S_b(bt + ibs) \, S_b(bt - ibs) \, e^{-i \pi b^2 (t^2 + s^2)},
\end{split}
\end{equation}
where the contour $\mathcal{C}_t$ runs upward with all the poles of the integrand on its left.
\corrIX{The dilogarithm prefactor in front of the integral is a pure phase.} The remaining Mellin-Barnes integral in \eqref{eq:Liouville_limit2} is precisely the $q$-deformed Bessel--$K$ wavefunction that solves the spectral problem of asymptotic Liouville gravity \cite{Mertens:2020hbs,Fan:2021bwt}.
Furthermore, it is clear that \eqref{eq:Liouville_limit2} reduces to the Mellin--Barnes representation of a Bessel-$K$ for small $b$\footnote{See Appendix C.1 of \cite{Griguolo:2026eoi} for a detailed discussion.}, yielding
\begin{equation}
    \psi_s(\Lambda) \overset{\substack{\nu \to \infty \\ b \to 0}}{\sim} \frac{e^{\pi\nu}}{\pi} K_{2is}(2 e^{-L_\text{ren}/2}),
\end{equation}
which are precisely the eigenfunctions of JT gravity at large cutoff.

The same large cutoff limit may also be formulated directly in terms of the Schr\"odinger equation \eqref{eq:sinh_quantum_hamiltonian2}, although it first requires to conjugate the symmetric Hamiltonian by the multiplicative factor introduced in \eqref{eq:factor}, thereby mapping it to the asymmetric representative \eqref{eq:sinh_difference_equation_chi}, and only then take the limit $\nu\to\infty$. This is discussed in detail in Appendix~\ref{appendix:large_cutoff}, and the resulting equations match those governing Liouville gravity \corr{\cite{Mertens:2020hbs,Blommaert:2023wad}}.
Therefore, we conclude that the scheme proposed at the end of Section \ref{sec:auxiliary_phase_space} holds at the quantum level for both the eigenvalue equations and the corresponding solutions.

\paragraph{\bfseries Asymptotics and overlap.}
In the following sections, we will propose a prescription that allows for the computation of physical observables within the canonical quantization framework at hand. As we will explain later, this prescription requires us to have computed further properties of the eigenfunctions \eqref{eq:sinh_eigenfunction} obtained in the current section. Specifically, we need the explicit form of the density of states $\rho(s)$ given by the orthogonality relation for the eigenfunctions \eqref{eq:sinh_eigenfunction} with respect to the measure $\dd\Lambda$, i.e.
\begin{equation}\label{eq:orthogonality}
    \bra{s'}\ket{s} = \int_{-\infty}^{+\infty} \dd\Lambda \, \psi^*_{s'}(\Lambda) \psi_s(\Lambda) = \frac{\delta(s-s')}{\rho(s)}.
\end{equation}
Indeed, once $\rho(s)$ is known, we are able to write a resolution of the identity in terms of energy eigenstates
\begin{equation}
    1 = \int_0^\infty \dd s \, \rho(s) \ket{s} \bra{s}.
\end{equation}
Preliminarily to the computation of the overlap \eqref{eq:orthogonality}, we discuss the asymptotics of the eigenfunction \eqref{eq:sinh_eigenfunction} for $\Lambda \to -\infty$ and $\Lambda \to +\infty$. 

A closed form expression for the eigenfunction \eqref{eq:sinh_eigenfunction} in these regimes can be obtained by closing the Mellin--Barnes contour $\mathcal{C}_t$ to the right and to the left, respectively.
For this reason, recalling the location of the poles of the double sine function given in \eqref{eq:poleszeros}, we infer the integrand appearing in \eqref{eq:chi_eigenfunction} has \corrIV{the following} poles in the $t$-complex plane:
\begin{figure}[h]
    \centering
    \includegraphics[width=0.9\linewidth]{images/poles2F1.tikz}
    \label{fig:poles2F1}
\end{figure}

\corrIV{The poles $t''_\times$ with label $(n,m)$ coincide with the poles $t'_\times$ with label $(n,m+1)$, so the integrand has simple poles at $t=n$ and double poles at $t=n+mb^{-2}$, $m\geq1$.}
First, let us consider the regime $\Lambda \to -\infty$. Since $e^{\Lambda} < 1$ here, we are allowed to close the contour to the right. Picking up the residues associated with $t'_\times$ and $t''_\times$, the integral would yield an infinite sum of the form
\begin{equation}
    \int_{\mathcal{C}_t} = \sum_{n \geq 0} c_{n}\,e^{\Lambda n} \corrIV{+\sum_{n\geq0,\,m\geq1}\big(c'_{n,m}+c''_{n,m}\,\Lambda\big)\,e^{\Lambda (n+m b^{-2})}}.
\end{equation}
In the $\Lambda \to -\infty$ limit, the main contribution thus comes from the \corrIV{simple} pole $t'_\times(n=m=0)$, whose residue is given by
\begin{equation}
    \int_{\mathcal{C}_t} \overset{\Lambda \to -\infty}{\sim} \frac{1}{2 \pi b^2} S_b\!\left(\frac{b}{2} + ib\nu \pm ibs\right).
\end{equation}
Furthermore, using the asymptotic behavior of the quantum dilogarithm \eqref{eq:dilog_asymptotics}, the prefactor in \eqref{eq:sinh_eigenfunction} reduces to
\begin{equation}
    \sqrt{
    \frac{\phi_b\!\left(-\frac{\Lambda}{2 \pi b}\right)}{\phi_b\!\left(-\frac{\Lambda-4 \pi b^2 \nu}{2 \pi b}\right)}
    }
    \overset{\Lambda \to -\infty}{\sim}
    \corrIV{e^{-i \nu (\Lambda - 2 \pi b^2 \nu)}}.
\end{equation}
Putting everything together, we conclude that the behavior of the eigenfunction \eqref{eq:sinh_eigenfunction} as $\Lambda \to -\infty$ is given by
\begin{equation}\label{eq:eigenfunction_L=0}
    \psi_s(\Lambda) \overset{\Lambda \to -\infty}{\sim}
    e^{\frac{1}{2}(\Lambda - 2 \pi b^2 \nu)},
\end{equation}
\corrIV{which is real and independent of $s$.}

Next, we consider the regime $\Lambda \to \infty$. Since $e^{\Lambda} > 1$ here, we are allowed to close the contour to the left. This means that now we should pick up the residues associated with the poles $t^\pm_\times$, yielding a structure of the form
\begin{equation}
    \int_{\mathcal{C}_t} = \sum_{n,m \geq 0} e^{-\Lambda (\frac{1}{2}+n+mb^{-2})} \left( e^{-i \Lambda (\nu - s)} d^{+}_{n,m} + e^{-i \Lambda (\nu + s)} d^{-}_{n,m} \right).
\end{equation}
Hence, if $\Lambda \to \infty$, the main contribution comes from the two poles $t^\pm_\times(n=m=0)$:
\begin{equation}
    \begin{split}
    \int_{\mathcal{C}_t} ={}& \frac{e^{-\corr{(\Lambda-2\pi b^2\nu)}(\frac{1}{2}+i\nu)}}{2 \pi b} \bigg( \frac{S_b(2ibs) \, S_b(\frac{b}{2}+ib\nu-ibs)}{S_b(\frac{b}{2} -ib\nu + ibs)} e^{is\corr{(\Lambda-2\pi b^2\nu)}}\\
    &+ \frac{S_b(-2ibs) \, S_b(\frac{b}{2} + ib\nu + ibs)}{S_b(\frac{b}{2} -ib\nu - ibs)} e^{-is\corr{(\Lambda-2\pi b^2\nu)}} \bigg).
    \end{split}
\end{equation}
The Faddeev dilogarithm prefactor reduces to $1$  in this limit, thus we conclude that the eigenfunctions \eqref{eq:sinh_eigenfunction} behave as plane waves at $\Lambda \to \infty$
\begin{equation}\label{eq:eigenfunction_L=infty}
    \psi_s(\Lambda) \overset{\Lambda \to \infty}{\sim} C_s \, e^{is\corr{(\Lambda-2\pi b^2\nu)}} + C^*_s \, e^{-is\corr{(\Lambda-2\pi b^2\nu)}},
\end{equation}
where we introduced the quantity
\begin{equation}
    C_s \equiv \frac{b \, S_b(2ibs)}{S_b(\frac{b}{2} + ib\nu + ibs) \, S_b(\frac{b}{2} - ib\nu + ibs)}.
\end{equation}

These asymptotic behaviors \eqref{eq:eigenfunction_L=0} and \eqref{eq:eigenfunction_L=infty} are the only two ingredients we need in order to compute the overlap \eqref{eq:orthogonality}. A \corrIV{derivation} of the overlap \corrIV{based on a formal expansion of the shift operators} is provided in Appendix \ref{appendix:overlap}.

Here we present \corr{a} heuristic argument that allows to extract the same final result in a quicker way. The idea is that for our purposes it is sufficient to focus on the asymptotic scattering behavior of the wavefunctions in the $\Lambda \to \infty$ region. This corresponds to the regime where the wavefunctions reduce asymptotically to plane waves and the delta-function normalization originates precisely from this asymptotic volume divergence:
\begin{equation}\label{eq:delta}
\delta(s-s')
\sim
\int \dd \Lambda \,
e^{- i \Lambda (s-s')} \, .
\end{equation}
Therefore, the spectral density can be extracted directly from the asymptotic normalization of the scattering states in \eqref{eq:eigenfunction_L=infty}.
Hence, we conclude that the overlap between two eigenfunctions reads
\begin{equation}
    \int_{-\infty}^{+\infty} \dd\Lambda \, \psi^*_{s'}(\Lambda) \psi_s(\Lambda) = 2 \pi \abs{C_s}^2 \delta(s-s').
\end{equation}
Finally, using the expression of $C_s$ in \eqref{eq:eigenfunction_L=infty} together with Equations \eqref{eq:double_sine_norm}, we can read off the spectral density
\begin{equation}\label{eq:spectral_density_s}
    \rho(s) = \frac{1}{2 \pi \abs{C_s}^2} = \frac{1}{\pi b^2} \frac{\sinh\!{(2 \pi b^2 s)} \, \sinh(2 \pi s)}{\cosh(2 \pi \nu) + \cosh(2 \pi s)}.
\end{equation}

Notice that in the limit $b \rightarrow 0$, \eqref{eq:spectral_density_s} reduces to the finite cutoff JT gravity spectral density \eqref{eq:JTspectral}. On the other hand, as $\nu \to \infty$, it reduces to
\begin{equation}
\rho(s) = \frac{2 \, e^{-2 \pi \nu}}{\pi b^2} \, \sinh\!{(2 \pi b^2 s)} \, \sinh(2\pi s),
\end{equation}
which is the standard spectral density of Liouville gravity \corr{\cite{Mertens:2020hbs}} up to an unimportant multiplicative factor.\footnote{Our notations are directly suitable for taking the JT limit $b \rightarrow 0$, while in the standard Liouville gravity literature one finds $\rho(s)
\propto \sinh(2 \pi b s)\,\sinh(2\pi s/b)$ and needs to rescale the energy variable $s\rightarrow b s$ before taking the limit.}

\paragraph{\bfseries Modular duality.}
So far, we have overlooked the multiplicative ambiguity \eqref{eq:periodic_ambiguity}. In the large cutoff setup for $b^2 \notin \mathbb{Q}$, this is resolved by requiring the physical wavefunction to simultaneously solve \eqref{eq:Schroedinger_equation} and a dual equation obtained by sending $b \to 1/b$ \cite{Mertens:2020hbs}.
In our finite cutoff setting, solving \eqref{eq:Schroedinger_equation} together with its $b \to 1/b$ image would yield an eigenfunction with no definite $b \to 0$ limit, hence spoiling its connection with JT gravity. Since we want to regard sinh dilaton gravity as a deformation of JT, we would like to retain the $b \to 0$ limit as a physical feature of the theory.
As a result, we will not impose the $b \to 1/b$ symmetry here.

Nevertheless, we mention that $\chi_s$
also solves a dual equation with $\tilde q=e^{i\pi/b^2}$, i.e.
\begin{equation}\label{eq:dual_difference_equation}
\begin{split}
\cosh(2\pi s)\,\chi_s(\Lambda)={}&\tfrac12\big(1+\tilde q\,e^{-\Lambda/b^2}\big) \, \chi_s(\Lambda-2\pi i)
+\tfrac12\big(1+\tilde q^{-1}e^{4\pi\nu}e^{-\Lambda/b^2}\big) \, \chi_s(\Lambda+2\pi i) +\\
& + \cos\!{(\pi/b^2)} \, e^{2\pi\nu}e^{-\Lambda/b^2} \, \chi_s(\Lambda),
\end{split}
\end{equation}
which can be checked to hold using the identity $S_b(x+b^{-1})=2\sin(\pi x/b)S_b(x)$ in \eqref{eq:fourier_solution}.
Of course, this is not the
$b\to1/b$ image of \eqref{eq:sinh_difference_equation_chi} even upon a suitable rescaling, as the last term would carry the coefficient
$-1$ instead of $\cos(\pi/b^2)$.

We will not use this equation to fix the multiplicative ambiguity, which is instead removed in the next section by the conditions that make the Hamiltonian self-adjoint, in the same spirit as in \cite{Griguolo:2026eoi}. The meaning of \eqref{eq:dual_difference_equation} will become clear in Section~\ref{sec:wave}, where we show that it is related, by a transformation commuting with \eqref{eq:sinh_difference_equation_chi}, to the dual Casimir equation of the modular double, see \eqref{eq:twisted_dual}.

\subsection{Uniqueness of the wavefunction}\label{subsec:uniqueness}
The solution \eqref{eq:chi_eigenfunction} was obtained from an ansatz, and the ambiguity
\eqref{eq:periodic_ambiguity} raises the question of whether it is the only acceptable solution of
\eqref{eq:sinh_difference_equation_chi}. For the finite cutoff JT wavefunction \eqref{eq:waveJT} the
question does not arise: the Schr\"odinger equation is a second-order differential equation in $L$, and a
boundary condition at $L=0$ together with boundedness at large $L$ fixes the solution up to
normalization. In sinh dilaton gravity, \eqref{eq:sinh_difference_equation_chi} is a difference equation
with step $2\pi ib^2$, and every periodic factor $g$ in \eqref{eq:periodic_ambiguity} is a function
of $e^{\Lambda/b^2}$, for instance $(e^{\Lambda/b^2}+a)/(e^{\Lambda/b^2}+c)$, $1+e^{\Lambda/b^2}$ or $1+e^{-\Lambda/b^2}$. The ambiguity is therefore a space
of functions rather than a finite set of integration constants. It disappears in the JT limit, where
$2 \pi b^2 \to 0$ and a function of $e^{\Lambda/b^2}$ either freezes to a constant or oscillates infinitely
fast, and it can be made invisible in the large cutoff limit $\nu\to\infty$ at fixed $L_\text{ren}$ as well.
The two limits therefore cannot decide which solution is the physical
one. We now show that the conditions under which \eqref{eq:sinh_quantum_hamiltonian2} is a well-defined
operator remove the ambiguity completely.

It is convenient to formulate the conditions for $\chi_s$ rather than $\psi_s$. Since $|\phi_b(x)|=1$, the prefactor in \eqref{eq:factor} is a pure phase, so $\psi_s$ and $\chi_s$ are unitarily related. The prefactor has neither zeros nor poles
in the strip $|\mathrm{Im}\,\Lambda|<\pi(1+b^2)$, so there $\psi_s$ and $\chi_s$ have the same analyticity and
growth. We require
\begin{itemize}
\item[(i)] $\chi_s(\Lambda)$ is holomorphic in the strip $|\mathrm{Im}\,\Lambda|<\eta$, with $\eta>2\pi b^2$,
\item[(ii)] $\chi_s(\Lambda)$ grows at most polynomially as $\mathrm{Re}\,\Lambda\to+\infty$,
\item[(iii)] $\chi_s(\Lambda)=c\,\mathcal E(\Lambda)\big(1+o(1)\big)$ as $\mathrm{Re}\,\Lambda\to-\infty$, with
$\mathcal E(\Lambda)\equiv e^{(\Lambda-2\pi b^2\nu)(\frac12+i\nu)}$,
\end{itemize}
where (ii) and (iii) hold uniformly inside the strip of (i). This makes the asymptotic data stable under
the imaginary translations performed by the Hamiltonian.

Condition (i) states that the wavefunction lies in the domain of the Hamiltonian. Acting with
$e^{a\hat\Pi}$, $a$ real, multiplies the momentum-space wavefunction by $e^{a\Pi}$, and the result is
normalizable only if the momentum distribution decays faster than $e^{-|a\Pi|}$. By the Paley--Wiener
theorem this means that $\psi_s(\Lambda)$ continues analytically to $|\mathrm{Im}\,\Lambda|<|a|$. This is
the standard domain of exponentials of the momentum and of the difference operators built from them; see
e.g.~\cite{FaddeevTakhtajan2015,PonsotTeschner2001}. The Hamiltonian
\eqref{eq:sinh_quantum_hamiltonian2} contains two successive shifts by $\pi b^2$, so $\hat H\psi_s$ is
defined only if $\psi_s$ is analytic for $|\mathrm{Im}\,\Lambda|\le2\pi b^2$: a state with a slower momentum
tail has infinite energy variance. The small margin $\eta>2\pi b^2$ is technical. As $b\to0$ the strip
collapses onto the real axis and (i) reduces to the usual smoothness of wavefunctions in the domain of
$\hat P^2$. Condition (i) is also where the restriction $b<1$ enters. The momentum space eigenfunction
\eqref{eq:fourier_solution} decays as $e^{-\pi(1+b^2)|\Pi|}$, so $\chi_s$ is analytic in
$|\mathrm{Im}\,\Lambda|<\pi(1+b^2)$ and has poles on the boundary of this strip. The shifts generated by the Hamiltonian stay inside it only if
$\pi(1+b^2)>2\pi b^2$, i.e. for $b<1$. The two strips coincide at the self-dual point $b=1$.

Condition (ii) simply states that we are considering scattering eigenfunctions. Their spectrum is continuous and they are $\delta$-normalizable as in \eqref{eq:orthogonality} if they are chosen to grow at most polynomially~\cite{GelfandVilenkin1964}. At large $\Lambda$, the terms
proportional to $e^{-\Lambda}$ in \eqref{eq:sinh_difference_equation_chi} drop out and the solutions are
superpositions of the plane waves $e^{\pm is\Lambda}$, whose amplitudes give the density of states
\eqref{eq:spectral_density_s}. The purpose of condition (ii) is to exclude periodic dressings such as $1+w$, which grow exponentially.

Finally, condition (iii) is the boundary condition at the point where the two boundaries touch and the length of the Cauchy slice goes to zero. As $\Lambda\to-\infty$, equation \eqref{eq:sinh_difference_equation_chi} has a regular singular point with
degenerate exponents: one solution behaves as $\mathcal E(\Lambda)$, i.e.\ $\psi_s(L)\sim L^{1/2}$, the other as
$\Lambda \mathcal E(\Lambda)$, i.e.\ $\psi_s(L)\sim L^{1/2}\log L$. As for the P\"oschl--Teller Hamiltonian of the JT
limit, both are square-integrable at $L=0$, and a boundary condition is needed to make the Hamiltonian
self-adjoint. Condition (iii) is the finite-$b$ analogue of the Friedrichs self-adjoint extension adopted for finite cutoff JT
gravity, reviewed in Section~\ref{sec:JT_review}~\cite{Griguolo:2026eoi}.

The result is that, for $0<b<1$, every solution of \eqref{eq:sinh_difference_equation_chi} satisfying
(i)--(iii) is proportional to \eqref{eq:chi_eigenfunction},
\begin{equation}\label{eq:uniqueness_result}
\chi(\Lambda)=N(s)\,\chi_s(\Lambda).
\end{equation}
The solution $\chi_s$ itself satisfies the three conditions, as follows from the asymptotics
\eqref{eq:eigenfunction_L=0} and \eqref{eq:eigenfunction_L=infty}.
We provide a proof of the uniqueness of the solution under the above assumptions in
Appendix~\ref{app:uniqueness}.

Remarkably, this statement interpolates between known results. As $b\to0$, (i) becomes a smoothness condition and (ii),
(iii) become the boundedness and Friedrichs self-adjoint extension that select \eqref{eq:waveJT}. As $\nu\to\infty$ at
fixed $L_\text{ren}$, the point $L=0$ is pushed to $L_\text{ren} \to -\infty$ and (iii)
is replaced by decay at the exponential barrier, which selects the Liouville gravity
wavefunction \eqref{eq:Liouville_limit2}.

\subsection{Partition function}
Having derived the wavefunctions \eqref{eq:chi_eigenfunction} and \eqref{eq:sinh_eigenfunction}, we are now in a position to compute the \rev{disk partition function} of the model. To do so, we briefly recall the prescription first presented in \corr{\cite{Yang:2018gdb,Saad:2019pqd}} and later adopted in several works on more general dilaton gravities \corr{\cite{Blommaert:2024ymv,Bossi:2024ffa,Blommaert:2025avl}} and JT at finite cutoff \corr{\cite{Griguolo:2026eoi}}. Within the canonical quantization framework at hand, the disk partition function of the model can be obtained as a transition amplitude between two states of definite vanishing length
\begin{equation}\label{eq:partition_function_prescription}
    Z(\beta) = \bra{L=0} e^{-\beta \hat{H}} \ket{L=0} \quad = \quad \vcenter
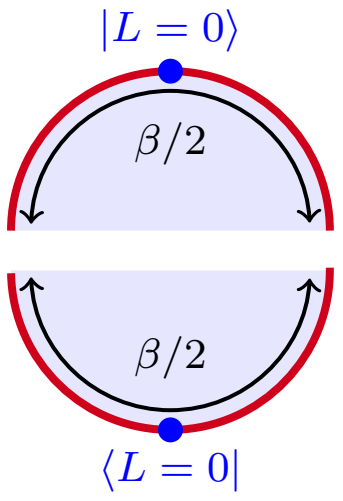
{\hbox{\begin{tikzpicture}[x=0.75pt,y=0.75pt,yscale=-1,xscale=1]

\draw  [draw opacity=0][fill={rgb, 255:red, 230; green, 230; blue, 255 }  ,fill opacity=1 ] (150,100) .. controls (150,100) and (150,100) .. (150,100) .. controls (150,122.09) and (132.09,140) .. (110,140) .. controls (87.91,140) and (70,122.09) .. (70,100) -- cycle ;
\draw  [draw opacity=0][line width=1.5]  (149.99,99.3) .. controls (150.38,121.39) and (132.79,139.61) .. (110.7,139.99) .. controls (88.61,140.38) and (70.39,122.79) .. (70.01,100.7) -- (110,100) -- cycle ; \draw  [color={rgb, 255:red, 208; green, 2; blue, 27 }  ,draw opacity=1 ][line width=1.5]  (149.99,99.3) .. controls (150.38,121.39) and (132.79,139.61) .. (110.7,139.99) .. controls (88.61,140.38) and (70.39,122.79) .. (70.01,100.7) ;  
\draw  [draw opacity=0][fill={rgb, 255:red, 230; green, 230; blue, 255 }  ,fill opacity=1 ] (70,90) .. controls (70,90) and (70,90) .. (70,90) .. controls (70,67.91) and (87.91,50) .. (110,50) .. controls (132.09,50) and (150,67.91) .. (150,90) -- cycle ;
\draw  [draw opacity=0][line width=1.5]  (70,90) .. controls (70,90) and (70,90) .. (70,90) .. controls (70,67.91) and (87.91,50) .. (110,50) .. controls (132.09,50) and (150,67.91) .. (150,90) -- (110,90) -- cycle ; \draw  [color={rgb, 255:red, 208; green, 2; blue, 27 }  ,draw opacity=1 ][line width=1.5]  (70,90) .. controls (70,90) and (70,90) .. (70,90) .. controls (70,67.91) and (87.91,50) .. (110,50) .. controls (132.09,50) and (150,67.91) .. (150,90) ;  
\draw  [color={rgb, 255:red, 0; green, 0; blue, 255 }  ,draw opacity=1 ][fill={rgb, 255:red, 0; green, 0; blue, 255 }  ,fill opacity=1 ] (107.37,140) .. controls (107.37,138.55) and (108.55,137.37) .. (110,137.37) .. controls (111.45,137.37) and (112.63,138.55) .. (112.63,140) .. controls (112.63,141.45) and (111.45,142.63) .. (110,142.63) .. controls (108.55,142.63) and (107.37,141.45) .. (107.37,140) -- cycle ;
\draw  [color={rgb, 255:red, 0; green, 0; blue, 255 }  ,draw opacity=1 ][fill={rgb, 255:red, 0; green, 0; blue, 255 }  ,fill opacity=1 ] (107.37,50) .. controls (107.37,48.55) and (108.55,47.37) .. (110,47.37) .. controls (111.45,47.37) and (112.63,48.55) .. (112.63,50) .. controls (112.63,51.45) and (111.45,52.63) .. (110,52.63) .. controls (108.55,52.63) and (107.37,51.45) .. (107.37,50) -- cycle ;
\draw[<->]  [draw opacity=0] (145,100) .. controls (145,119.36) and (129.33,135.06) .. (110,135.06) .. controls (90.67,135.06) and (75,119.36) .. (75,100) -- (110,100) -- cycle ; \draw[<->] (144.94,102.02) .. controls (143.9,120.44) and (128.65,135.06) .. (110,135.06) .. controls (91.25,135.06) and (75.94,120.29) .. (75.04,101.73) ; 
\draw[<->]  [draw opacity=0] (145,90) .. controls (145,70.64) and (129.33,54.94) .. (110,54.94) .. controls (90.67,54.94) and (75,70.64) .. (75,90) -- (110,90) -- cycle ; \draw[<->]    (144.94,87.98) .. controls (143.9,69.56) and (128.65,54.94) .. (110,54.94) .. controls (91.25,54.94) and (75.94,69.71) .. (75.04,88.27) ; 

\draw (110,143.4) node [anchor=north] [inner sep=0.75pt]  [font=\scriptsize]  {$\textcolor[rgb]{0,0,1}{\bra{L=0}}$};
\draw (110,46.6) node [anchor=south] [inner sep=0.75pt]  [font=\scriptsize]  {$\textcolor[rgb]{0,0,1}{\ket{L=0}}$};
\draw (110,128.6) node [anchor=south] [inner sep=0.75pt]  [font=\scriptsize]  {$\beta /2$};
\draw (110,61.4) node [anchor=north] [inner sep=0.75pt]  [font=\scriptsize]  {$\beta /2$};

\end{tikzpicture}}}.
\end{equation}
As discussed \corr{at} length in \corr{\cite{Griguolo:2026eoi}}, there are some caveats to be made about the prescription \eqref{eq:partition_function_prescription}.
First, we notice that there is nothing in Equation \eqref{eq:partition_function_prescription} preventing time evolution from leading the system towards states $\ket{s}$ with $s > \nu$. From Equation \corr{\eqref{quasilocal}} we know that these states correspond to \rev{complex} Hamiltonian eigenvalues and thus to a violation of unitarity. Two solutions to this problem are presented in \corr{\cite{Griguolo:2026eoi}} and we briefly recall them here: one can either replace the time evolution operator with its projection
\begin{equation}\label{eq:time_evolution_projected}
    e^{-\beta \hat{H}} \quad \to \quad \corrIV{\hat{\Pi}} e^{-\beta \hat{H}} \corrIV{\hat{\Pi}}, \qquad \corrIV{\hat{\Pi}} = \int_0^\nu \dd s \, \rho(s) \ket{s} \bra{s}
\end{equation}
or, alternatively, introduce a piecewise Hamiltonian that extends \corr{\eqref{quasilocal}} beyond the region $s < \nu$ while keeping its eigenvalues real. The consequences of both these choices in the context of JT gravity at finite cutoff are discussed \corr{at} length in \corr{\cite{Griguolo:2026eoi}} and they can be effortlessly extended to sinh dilaton gravity, thus we refer the interested reader to that work.

For concreteness, one can extend the Hamiltonian by continuing past $s=\nu$ on the branch that keeps the energy real,
\begin{equation}\label{eq:sinh_energy_piecewise}
E(s)=\frac{1}{2\pi b^2}
\begin{cases}
\sqrt{2\cosh(2\pi b^2\nu)-2}-\sqrt{2\cosh(2\pi b^2\nu)-2\cosh(2\pi b^2 s)}, & 0<s<\nu,\\[1mm]
\sqrt{2\cosh(2\pi b^2\nu)-2}+\sqrt{2\cosh(2\pi b^2 s)-2\cosh(2\pi b^2\nu)}, & s>\nu,
\end{cases}
\end{equation}
where the first branch is \eqref{quasilocal}.
The results that one would obtain by instead opting for \eqref{eq:time_evolution_projected} can be recovered by truncating every integral over energy labels at $s < \nu$.
Thus, inserting a resolution of the identity in terms of eigenstates of $\hat{H}$ into \eqref{eq:partition_function_prescription} yields the following expression for the partition function\footnote{Here we added an additional factor of $e^{2 \pi \nu}$ in the expression of $Z(\beta)$. In the JT case, this addition is motivated by the absence of a topological contribution that gets lost in the Hamiltonian approach. Its presence does not affect the physics, but we include it in sinh dilaton gravity as well in order to make the contact with JT smoother.}
\begin{equation}\label{eq:partition_function_2}
    Z(\beta) = e^{2 \pi \nu} \int_0^\infty \dd s \, \rho(s) \, \abs{\psi_s(0)}^2 e^{-\beta E(s)}.
\end{equation}
At this stage, we must acknowledge that the eigenfunctions \corr{\eqref{eq:sinh_eigenfunction}} we obtained in the previous section always vanish at $L = 0$, i.e. at $\Lambda \to -\infty$, hence suggesting that \eqref{eq:partition_function_2} should also vanish. Nevertheless, we can extract physically meaningful information from the expression \eqref{eq:partition_function_2} by formally replacing states $\ket{L=0}$ with their limit as\footnote{Recall that since $\psi_s(L) = \sqrt{\dv{\Lambda}{L}} \psi_s(\Lambda)$, $\lim_{L \to 0} \psi_s(L) = \lim_{\Lambda \to -\infty} e^{\Lambda/4 -\pi b^2\nu} = \lim_{L \to 0} \sqrt{L/2} \, e^{-\pi b^2 \nu}$.}
\begin{equation}\label{eq:L=0_prescription}
    \ket{L=0} \quad \to \quad \lim_{L \to 0} \frac{\ket{L}}{\psi_s(L)} = \lim_{L \to 0} \frac{\ket{L=0}}{\sqrt{L/2} \, e^{-\pi b^2 \nu}}.
\end{equation}
The reason why we are allowed to do so resides in the fact that physically relevant quantities are typically computed as
\begin{equation}
    \frac{\bra{L=0} e^{-(\beta-\tau) \hat{H}} \hat{\mathcal{O}} e^{-\tau \hat{H}} \ket{L=0}}{\bra{L=0} e^{-\beta \hat{H}} \ket{L=0}},
\end{equation}
hence both numerator and denominator vanish at the same rate.

Finally, combining the above prescriptions and using the asymptotics \eqref{eq:eigenfunction_L=0}, the partition function of the theory is given by
\begin{equation}\label{eq:partition_function}
\begin{split}
    Z(\beta) = e^{2 \pi \nu} \int_0^\infty \dd s \, \rho(s) \, e^{-\beta E(s)} = \int_0^\infty  \dd s \, \frac{e^{2 \pi \nu}}{\pi b^2} \frac{\sinh(2 \pi b^2 s)\,\sinh(2\pi s)}{\cosh(2\pi s)+\cosh(2\pi \nu)} 
    e^{-\beta E(s)},
\end{split}
\end{equation}
\corrIV{with $E(s)$ given by \eqref{eq:sinh_energy_piecewise}}.
Consistently, when $\nu \to \infty$ only the first branch survives and we recover the standard spectral density of Liouville gravity.

\section{Quantum group interpretation of the Hamiltonian}\label{sec:casimir}
The appearance of $q$-hypergeometric functions in \corr{\eqref{eq:sinh_eigenfunction}} is a strong hint that the finite cutoff problem of sinh dilaton gravity
carries a quantum group structure. In JT gravity the analogous statement is
precise and well understood: the finite cutoff Hamiltonian is the
$\mathfrak{sl}(2,\mathbb{R})$ Casimir reduced on functions of definite left and
right hyperbolic charge, and the wavefunctions of the universe are the hypergeometric functions solving the resulting radial equation. We now show that in our current setup this structure gets $q$-deformed following the same logic. Replacing $\mathfrak{sl}(2,\mathbb{R})$ by
$U_q(\mathfrak{sl}(2,\mathbb{R}))$ and the radial differential equation by a
finite-difference equation, an analogous reduction produces exactly the spectral
problem \corr{\eqref{eq:sinh_difference_equation_chi}} obtained from the bulk.

Throughout we set $q=e^{i\pi b^2}$ and take $U_q(\mathfrak{sl}(2,\mathbb{R}))$
to be generated by $K^{\pm1},E,F$ subject to
\begin{equation}\label{eq:Uq_relations}
KE=qEK,
\qquad
KF=q^{-1}FK,
\qquad
[E,F]=\frac{K^2-K^{-2}}{q-q^{-1}} ,
\end{equation}
\rev{where $K=q^{H}$, so that $[H,E]=E$ and $[H,F]=-F$,} with the coproduct
\begin{equation}
\Delta(K)=K\otimes K,
\qquad
\Delta(E)=E\otimes K^{-2}+1\otimes E,
\qquad
\Delta(F)=F\otimes1+K^2\otimes F .
\end{equation}
In these conventions the quadratic Casimir is
\begin{equation}
\label{eq:casimir}
\mathcal{C}
=
FE+\frac{qK^2+q^{-1}K^{-2}}{(q-q^{-1})^2} .
\end{equation}

To make contact with the usual $\mathfrak{sl}(2,\mathbb{R})$ story, we need to define the functions on the quantum group on which the generators of $U_q(\mathfrak{sl}(2,\mathbb{R}))$ can act and which will be the elements of the Hilbert space of our theory. To do that we
parametrize the coordinate algebra by the quantum analogue of the Gauss
coordinates,
\begin{equation}
g=
\begin{pmatrix}
A&B\\
C&D
\end{pmatrix}
=
\begin{pmatrix}
a&a\beta\\
\gamma a&\gamma a\beta+a^{-1}
\end{pmatrix},
\qquad
a\equiv e^\phi ,
\end{equation}
which obey
\begin{equation}
\label{eq:gauss-algebra}
a\gamma=q\gamma a,
\qquad
a\beta=q\beta a,
\qquad
[\beta,\gamma]=0 .
\end{equation}
The peculiarity of these coordinates is the fact that they do not commute, so every expression below must be understood with some ordering convention, we adopt the
normal order $(\gamma,a,\beta)$.

Functions on the quantum group carry two commuting actions of the algebra, denoted by $R_X$ and $L_X$,
induced by multiplying $g$ on the right and on the left respectively by an algebra element.
Both are defined through the Hopf
pairing between $U_q(\mathfrak{sl}(2,\mathbb{R}))$ and the coordinate algebra,
and their explicit form on ordered monomials has been worked out in \cite{Fan:2021bwt}, from which we take the expressions used below. Since the two
actions operate at opposite ends of $g$ they commute with one another, and since
$\mathcal{C}$ is central both commute with $R_{\mathcal{C}}$ as well. A function
on the quantum group can therefore be labelled at once by a Casimir eigenvalue
and by a pair of Cartan charges, one on each side, which resembles the structure found in Section \corr{\ref{sec:JT_review}} for the wave \corr{function} of the universe\corr{.}

What's more, on ordered monomials the two Cartan actions are diagonal,
\begin{equation}
\label{eq:cartan-monomials}
    L_{K^2}\!\left(\gamma^m a^\ell\beta^n\right)
    =
    q^{-\ell+2m}\,\gamma^m a^\ell\beta^n ,
    \qquad
    R_{K^2}\!\left(\gamma^m a^\ell\beta^n\right)
    =
    q^{\ell-2n}\,\gamma^m a^\ell\beta^n ,
\end{equation}
so that the three conditions
\begin{equation}
\label{eq:joint-diagonalization}
    L_{K^2}\Psi=q^{-2ih_L}\,\Psi ,
    \qquad
    R_{K^2}\Psi=q^{2ih_R}\,\Psi ,
    \qquad
    R_{\mathcal{C}}\,\Psi=c_q(s)\,\Psi
\end{equation}
can be easily imposed simultaneously. We denote the resulting function by
$\Psi^{s}_{h_L,h_R}$. The charges $h_L$ and $h_R$ are left free for
now and will be tied to the gravitational cutoff only after the reduction, while the
eigenvalue $c_q(s)$ will be fixed once we specify the representation in which
\eqref{eq:joint-diagonalization} is realized.

The first two conditions alone instruct us about the fact that any bi-invariant monomial must obey $\ell = 2m = 2n$, hence it can only be a power of $Z \equiv \gamma e^{2 \phi} \beta$. Consequently, every bi-invariant function can only depend on the three coordinates through the combination $\gamma e^{2 \phi} \beta$.

The same observation fixes the dependence of a function carrying non-vanishing
Cartan charges. With the convention of \eqref{eq:cartan-monomials}, the factors
$\gamma^{-ih_L}$ and $\beta^{-ih_R}$ carry respectively the required left
and right weights, so that the most general solution of the first two conditions
in \eqref{eq:joint-diagonalization} reads
\begin{equation}
\label{eq:radial-factorization}
    \Psi^{s}_{h_L,h_R}(g)
    =
    \gamma^{-ih_L} \, \Phi^{s}_{h_L,h_R}(Z) \, \beta^{-ih_R}.
\end{equation}

We can now impose the remaining condition in \eqref{eq:joint-diagonalization}.
On functions of the Gauss coordinates the right action of the Casimir
\eqref{eq:casimir} takes the form \corr{\cite{Fan:2021bwt}}
\begin{equation}
\label{eq:regular-casimir}
R_{\mathcal C}
=
\frac{
q\,T^\phi_{i\pi b^2}
+
q^{-1}T^\phi_{-i\pi b^2}
}{
(q-q^{-1})^2
}
+
D_{\gamma,q^{-2}}\,
e^{-2\phi}\,
D_{\beta,q^2},
\end{equation}
where
\begin{equation}
(T^\phi_c f)(\phi)=f(\phi+c),
\qquad
D_{x,\rev{p}}f(x)
\equiv
\frac{f(\rev{p}x)-f(x)}{(\rev{p}-1)x}.
\end{equation}
Once again, the ordering in the second term is part of the definition of the operator.
Since $R_{\mathcal C}$ commutes with both Cartan actions, it must map
\eqref{eq:radial-factorization} into a function of the same form, acting only on the radial part
$\Phi(Z)$. To deduce the form of the radially reduced operator, it is enough to consider its action on a monomial
$\Phi(Z)=Z^\omega$, for which \eqref{eq:radial-factorization} becomes
\begin{equation}
\Psi_\omega
=
\gamma^{-ih_L+\omega}
a^{2\omega}
\beta^{-ih_R+\omega}.
\end{equation}
The Cartan part of \eqref{eq:regular-casimir} translates the power of $a$,
\begin{equation}
T^\phi_{i\pi b^2}a^{2\omega}
=
q^{2\omega}a^{2\omega},
\end{equation}
while
\begin{equation}
D_{\gamma,q^{-2}}
\gamma^{-ih_L+\omega}
=
\frac{
q^{2ih_L-2\omega}-1
}{
q^{-2}-1
}
\gamma^{-ih_L+\omega-1},
\qquad
D_{\beta,q^{2}}
\beta^{-ih_R+\omega}
=
\frac{
q^{-2ih_R+2\omega}-1
}{
q^{2}-1
}
\beta^{-ih_R+\omega-1}.
\end{equation}
As a result, these contributions combine into
\begin{equation}
R_{\mathcal C}
\Psi^{(s)}_{h_L,h_R}
=
\gamma^{-ih_L}
\left(
\mathcal C^{\rm rad}_{h_L,h_R}
\Phi^{s}_{h_L,h_R}
\right)(Z)\,
\beta^{-ih_R},
\end{equation}
where the radial Casimir acts as
\begin{align}
\label{eq:radial-casimir}
\left(
\mathcal C^{\rm rad}_{h_L,h_R}\Phi
\right)(Z)
={}&
\frac{
q\,\Phi(q^2Z)
+
q^{-1}\Phi(q^{-2}Z)
}{
(q-q^{-1})^2
}
\nonumber\\
&+
\frac{
\left(1+q^{2i(h_L-h_R)}\right)\Phi(Z)
-q^{2ih_L}\Phi(q^{-2}Z)
-q^{-2ih_R}\Phi(q^2Z)
}{
Z\,(q^{-2}-1)(q^2-1)
}.
\end{align}
The Casimir problem on the three non-commuting Gauss coordinates has thus become
a second order finite-difference equation in the single radial coordinate $Z$,
for arbitrary values of the two Cartan charges.

To make contact with the gravitational \corr{Hamiltonian} \corrIX{\eqref{eq:sinh_quantum_hamiltonian2}} we now have to identify the sector describing the finite Dirichlet wall. This corresponds to the two boundaries carrying the same
hyperbolic charge,
\begin{equation}
\label{eq:equal-hyperbolic-charges}
h_L=h_R=\nu,
\end{equation}
identifying it with the boundary value of the dilaton introduced in
Section~\corr{\ref{subsec:sinh_quantization}}. It also remains to relate $Z$ to the
canonical variable $\Lambda$ used there, and it turns out that the correct identification is
\begin{equation}
\label{eq:Lambda-Z}
e^\Lambda=e^{2\pi b^2\nu}Z.
\end{equation}
Indeed, with this choice, a rescaling of Z corresponds to a shift in $\Lambda$, i.e. $q^{\pm2}Z \leftrightarrow \Lambda\pm2\pi i b^2$, which accounts for the finite translations appearing in the difference equation derived from the bulk.
However we notice that in \eqref{eq:radial-casimir} the shifted
arguments enter with unequal coefficients $q$ and $q^{-1}$, whereas the
gravitational equation is symmetric between the two shifts. Writing
\begin{equation}
\label{eq:Phi-chi}
\Phi(Z)=Z^{-1/2}\chi(\Lambda)
\end{equation}
produces exactly the compensating factors,
\begin{equation}
\Phi(q^2Z)
=
q^{-1}Z^{-1/2}
\chi(\Lambda+2\pi ib^2),
\qquad
\Phi(q^{-2}Z)
=
q\,Z^{-1/2}
\chi(\Lambda-2\pi ib^2),
\end{equation}
while \eqref{eq:Lambda-Z} gives $Z^{-1}=e^{2\pi b^2\nu}e^{-\Lambda}$. 
Substituting into the Casimir eigenvalue equation and removing the common factor \(Z^{-1/2}\), one obtains
\begin{align}
\label{eq:radial-casimir-Lambda}
(q-q^{-1})^2
c_q(s)
\chi
(\Lambda)
={}&
\left(1+q\,e^{-\Lambda}\right)
\chi(\Lambda-2\pi ib^2)
\nonumber\\
&+
\left(1+q^{-1}e^{4\pi b^2\nu}e^{-\Lambda}\right)
\chi(\Lambda+2\pi ib^2)
\nonumber\\
&-
2e^{2\pi b^2\nu}e^{-\Lambda}\chi(\Lambda).
\end{align}

The only missing ingredient is now the Casimir eigenvalue $c_q(s)$, which requires specifying the
representation. In the present case, the relevant one is the $q$-deformation of the
principal series carrying the asymptotic theory, realized on functions of a real
variable $t$ by \corr{\cite{Ponsot:1999uf,Fan:2021bwt}}
\begin{equation}
\label{eq:carrier}
K=T^t_{ib/2},
\qquad
E=
e^{2\pi bt}\,
\frac{
e^{i\pi b\lambda}T^t_{ib/2}
-
e^{-i\pi b\lambda}T^t_{-ib/2}
}{q-q^{-1}},
\qquad
F=
-e^{-2\pi bt}\,
\frac{
e^{-i\pi b\lambda}T^t_{ib/2}
-
e^{i\pi b\lambda}T^t_{-ib/2}
}{q-q^{-1}},
\end{equation}
with $\lambda=\frac b2+ibs$. We notice that this representation is the anti-hermitian version of the one used in \corr{\cite{Fan:2021bwt}}. Evaluating \eqref{eq:casimir} on this realization
gives
\begin{equation}
\label{eq:casimir-eigenvalue}
c_q(s)
=
-\frac{
\cosh(2\pi b^2s)
}{
2\sin^2(\pi b^2)
},
\end{equation}
so that, dividing \eqref{eq:radial-casimir-Lambda} by two, the reduced Casimir equation becomes
\begin{align}
\label{eq:casimir-gravity-match}
\cosh(2\pi b^2s)\,
\chi_s(\Lambda)
={}&
\frac12
\left(1+q\,e^{-\Lambda}\right)
\chi_s(\Lambda-2\pi ib^2)
\nonumber\\
&+
\frac12
\left(
1+q^{-1}e^{4\pi b^2\nu}e^{-\Lambda}
\right)
\chi_s(\Lambda+2\pi ib^2)
\nonumber\\
&-
e^{2\pi b^2\nu}e^{-\Lambda}\chi_s(\Lambda),
\end{align}
which is precisely the finite-difference equation obtained by canonical
quantization of sinh dilaton gravity in
Section~\corr{\ref{subsec:sinh_quantization}}. This agreement allows us to read the finite cutoff problem in group-theoretic terms.

\section{Modular double and the regular wavefunction}\label{sec:wave}

The amplitudes of sinh dilaton gravity at large cutoff, i.e.\ of Liouville gravity, are governed by the modular double of $SL_q(2,\mathbb{R})$ \cite{Mertens:2020hbs,Fan:2021bwt}. This is due to the fact that sinh dilaton gravity can be recast as a timelike and a spacelike Liouville theory \cite{Mertens:2020hbs}, and the $b \to 1/b$ self-duality of Liouville theory is closely tied to the modular double \cite{Ponsot:1999uf,PonsotTeschner2001}.
In this section, we discuss to what extent the modular double structure survives in the finite cutoff setup.

We follow the JT discussion at the end of Section~\ref{sec:JT_review}, where the wavefunction is the combination of the hyperbolic matrix elements $K_{++}$ and $K_{--}$ regular at $L=0$. First, we construct the hyperbolic matrix element of the modular double that reduces to $K_{++}$ in the JT limit, and show that its radial part is the solution \corrIX{$\chi^-_s(\Lambda)$} behaving as $e^{-i\Lambda s}$ at $\Lambda\to\infty$. Second, we show that \corrIX{$\chi^-_s(\Lambda)$} also satisfies a difference equation with shifts $\Lambda\to\Lambda\pm2\pi i$: the dual Casimir equation, continued in the representation label. \rev{The two equations together leave no periodic factor in front of \corrIX{$\chi^-_s(\Lambda)$}, and the dual one distinguishes it from} its partner \corrIX{$\chi^+_s(\Lambda)$}, obtained by $s\to-s$ and that reduces to $K_{--}$ in the JT limit. Finally, we show that the wavefunction $\chi_s(\Lambda)$ of Section~\ref{subsec:sinh_quantization} is the unique combination of $\chi^\pm_s(\Lambda)$ \rev{with constant coefficients that is} regular at $\Lambda\to-\infty$. 

\subsection{An explicit realization of the modular double}

Besides $E,F,H$, we consider the generators $\tilde E,\tilde F,\tilde H$ of the dual algebra \rev{$U_{\tilde q}(\mathfrak{sl}(2,\mathbb R))$}, with $\tilde q=e^{i\pi/b^2}$. The full algebra is the tensor product $U_q(\mathfrak{sl}(2,\mathbb{R})) \otimes U_{\tilde{q}}(\mathfrak{sl}(2,\mathbb{R}))$\corrIX{, up to signs that we specify below}.

Our construction of the modular double is slightly non-standard. Its representations are usually labeled by $\alpha=Q/2+iP$, with $Q=b+b^{-1}$ and $P\in\mathbb R^+$ \cite{Ponsot:1999uf}. This label manifestly respects the $b\to1/b$ symmetry, but, as is generally the case for this symmetry, it is incompatible with the semiclassical limit $b\to0$ \cite{Ponsot:1999uf}: these representations do not naturally reduce to the principal series of $SL(2,\mathbb R)$, labeled by $j=-1/2+is$ \cite{Blommaert:2018oro,Blommaert:2018iqz}. In gravity, however, $b\to0$ is the JT limit, which we need to preserve. We therefore replace $\alpha$  with the label  $\lambda=b/2+ibs$, the natural $q$-deformation of the JT principal series label\corrIX{, $\lambda=b(j+1)$} \cite{Fan:2021bwt}. This choice breaks the $b\to1/b$ symmetry of the label, but not the modular double structure itself\chg{, in a sense that we make precise below.}

Moreover, we consider the following realization of the $U_q(\mathfrak{sl}(2,\mathbb{R})) \otimes U_{\tilde{q}}(\mathfrak{sl}(2,\mathbb{R}))$ quantum algebra on functions of $x>0$:
\begin{equation}\label{eq:algebra1}
K=\mathcal{D}^{x}_{q}, \quad
E=-i x\frac{e^{\pi i b \lambda}\mathcal{D}^{x}_{q}-e^{-\pi i b \lambda}\mathcal{D}^{x}_{q^{-1}}}{2\sin(\pi b^2)}, \quad
F=i x^{-1} \frac{e^{-\pi i b \lambda} \mathcal{D}^{x}_{q}-e^{\pi i b \lambda} \mathcal{D}^{x}_{q^{-1}}}{2 \sin (\pi b^2)},
\end{equation}
and
\begin{equation}\label{eq:algebra2}
\tilde{K}=\mathcal{D}^{x}_{e^{i\pi}}, \quad
\tilde{E}=-i x^{\frac{1}{b^2}}\frac{e^{\frac{\pi i \lambda}{b}}\mathcal{D}^{x}_{e^{i\pi}}-e^{-\frac{\pi i \lambda}{b}}\mathcal{D}^{x}_{e^{-i\pi}}}{2\sin(\pi/b^2)}, \quad
\tilde{F}=i x^{-\frac{1}{b^2}} \frac{e^{-\frac{\pi i \lambda}{b}} \mathcal{D}^{x}_{e^{i\pi}}-e^{\frac{\pi i \lambda}{b}} \mathcal{D}^{x}_{e^{-i\pi}}}{2 \sin (\pi/ b^2)}
\end{equation}
where we defined the dilatation operator
\begin{equation}
\mathcal{D}^{x}_{a} f(x)=f(ax), \qquad \mathcal{D}^{x}_{a}=e^{\log a\, x\partial_x}.
\end{equation}
\chg{The first set is the realization \eqref{eq:carrier} of Section~\ref{sec:casimir} written in the variable $x=e^{2\pi bt}$, so $K,E,F$ satisfy the relations \eqref{eq:Uq_relations} of $U_q(\mathfrak{sl}(2,\mathbb R))$, while the second set satisfies the same relations with $q$ replaced by $\tilde q$. We use $x$ rather than $t$ because it is the coordinate in which the $\mathfrak{sl}(2,\mathbb{R})$ generators are represented in the JT limit, see Appendix \ref{app:JTsteps}.} \corrIX{As is standard for the modular double, the two sets of generators commute only up to signs. Since $\mathcal{D}^{x}_{q}\,x^{\pm1/b^2}=-x^{\pm1/b^2}\,\mathcal{D}^{x}_{q}$ and $\mathcal{D}^{x}_{e^{i\pi}}\,x^{\pm1}=-x^{\pm1}\,\mathcal{D}^{x}_{e^{i\pi}}$, $K$ anticommutes with $\tilde E$ and $\tilde F$, $\tilde K$ anticommutes with $E$ and $F$, and all other pairs of generators commute. In particular $K^2$, $\tilde K^2$ and the two Casimir operators commute with all the generators, which is all that is used below.} 

Evaluating the Casimir operator \eqref{eq:casimir} on the realization \eqref{eq:algebra1} gives the eigenvalue \eqref{eq:casimir-eigenvalue}. Similarly, the dual Casimir $\tilde c_{\tilde q}$, obtained from \eqref{eq:casimir} by replacing $q$ with $\tilde q$ and $E,F,K$ with $\tilde E,\tilde F,\tilde K$, takes on the realization \eqref{eq:algebra2} the value
\begin{equation}\label{eq:two_casimirs}
\tilde{c}_{\tilde q}(\lambda)=\frac{\cosh\left(2\pi s+i\pi/b^2\right)}{2\sin^2(\pi/b^2)} .
\end{equation}
This follows from \eqref{eq:casimir-eigenvalue} by $b\to1/b$ at fixed $\lambda$, and will play a central role below.

\chg{Notice that these are the Casimir eigenvalues of the standard representations labelled by $\alpha$ once we perform an analytic continuation of their expression. Indeed for $\alpha=Q/2+iP$ the two Casimirs are proportional to $\cosh(2\pi bP)$ and $\cosh(2\pi P/b)$, and recalling that $\lambda=Q/2+iP$ with $P=bs+\frac{i}{2b}$, one finds}
\begin{equation}\label{eq:casimirs_continued}
\cosh(2\pi b P)=-\cosh(2\pi b^2 s),\qquad \cosh(2\pi P/b)=\cosh\left(2\pi s+\frac{i\pi}{b^2}\right),
\end{equation}
\chg{which are \eqref{eq:casimir-eigenvalue} and \eqref{eq:two_casimirs} up to normalization. This is the precise sense in which the choice of label breaks the $b\to1/b$ symmetry while preserving the modular double structure: the two Casimir equations are those of the modular double, but, since the label itself depends on $b$, the dual one is not the $b\to1/b$ image of the first when written in terms of $s$ and $\nu$.}

\chg{The carrier space is equipped with the scalar product $\displaystyle\bra{f_1}\ket{f_2}=\int_0^{\infty}\frac{dx}{x}f_1^*(x)f_2(x)$, discussed in Appendix~\ref{app:hyperbolic}, with respect to which $K^{\dagger}=K$, $E^{\dagger}=-E$ and $F^{\dagger}=-F$. The} generators of the dual algebra are related to the original ones via the transcendental property \cite{Mertens:2022aou}
\begin{equation}\label{eq:transcendental}
\begin{split}
\tilde{e}&=2 \sin (\pi/b^2)\tilde{E}=e^{1/b^2}=\left(2 \sin(\pi b^2)E\right)^{1/b^2} \\
\tilde{f}&=2 \sin (\pi/b^2)\tilde{F}=f^{1/b^2}=\left(2 \sin(\pi b^2)F\right)^{1/b^2}.
\end{split}
\end{equation}

\paragraph{Group element}
Since we do not have at our disposal a hyperbolic decomposition of the modular double quantum group analogous to \eqref{eq:JT_hyperbolic_decomposition_review}, we construct a $q$-deformation of the Gauss--Euler decomposition
\begin{equation}\label{eq:gausseuler}
g=e^{\gamma F} e^{2\phi H} e^{\beta E}
\end{equation}
where $\left(\phi,\beta,\gamma\right)$ are the Gauss coordinates on the group manifold already encountered above. Specifically, we use
\begin{equation}\label{eq:groupdecomp}
g=g_{b}\left(2i \sin (\pi b^2)\gamma F\right)e^{2\phi H}g_{b}^{*}\left(-2 i\sin (\pi b^2)\beta E\right) =g_{b}\left(i\gamma f\right)e^{2\phi H}g_{b}^{*}(-i\beta e),
\end{equation}
which is almost identical to the one introduced in \cite{Mertens:2022aou}.\footnote{The group decomposition of the modular double proposed in \cite{Mertens:2022aou} is $g=g_{b}\left(\gamma f\right)e^{2\phi H}g_{b}^{*}(\beta e)$. However, this does not reduce to the Gauss--Euler decomposition as $b\rightarrow 0$, which is instead a relevant property for our purposes.} Here $g_{b}(z)=1/\phi_b\left(\frac{\log z}{2 \pi b}\right)$, see \eqref{eq:gb_def}, is Faddeev's quantum dilogarithm in the standard normalization of \cite{Ip2013QuantumPlane}, its properties are collected in Appendix \ref{app:gb}.\footnote{This is also the normalization for which $g_b(z)$ is bounded on the branches of $\log z$ selected by the transcendental relation \eqref{eq:transcendental} at negative arguments.} The proposal generalizes the decomposition
\begin{equation}\label{eq:Fronsdal}
g=e_{q^{-2}}^{\gamma F} e^{2\phi H} e_{q^{2}}^{\beta E}
\end{equation}
of Fronsdal and Galindo, valid for a single quantum group \cite{FronsdalGalindo1993,Mertens:2022aou}, where $e_{q^2}(z)$ is the $q$-exponential defined in \eqref{eq:qexponential}. Thanks to the modular properties of the quantum dilogarithm, $g_b(z)$ factorizes into a product of a $q$-exponential and a $\tilde q$-exponential, see Appendix \ref{app:gb}. Together with \eqref{eq:transcendental}, this makes \eqref{eq:groupdecomp} the natural candidate for a modular double decomposition. Using the small-$b$ expansion \eqref{eq:dilog_b=0}, one finds $g_b(z)\sim\exp\left(\frac{1}{2\pi i b^2}\mathrm{Li}_2(-z)\right)$, so that
\begin{equation}
g_b\left(2i \sin (\pi b^2)\gamma F\right)\sim e^{-\gamma F},\qquad g_b^{*}\left(-2i \sin (\pi b^2)\beta E\right)\sim e^{-\beta E}.
\end{equation}
Therefore \eqref{eq:groupdecomp} reduces in the JT limit to the Gauss--Euler decomposition \eqref{eq:gausseuler} with $(\beta,\gamma)\to(-\beta,-\gamma)$. This sign is immaterial for what follows, since the matrix elements depend on $\beta$ and $\gamma$ only through $\beta\gamma$ and the invariant $Z=\beta e^{2\phi}\gamma$.

\subsection{The hyperbolic matrix element}

We now compute the hyperbolic matrix element
\begin{equation}\label{eq:amplitude}
\bra{h}g\ket{h} \qquad h\equiv\nu ,
\end{equation}
where $\ket{h}$ is a hyperbolic state (an eigenvector of $K$) with hyperbolic charge fixed by the boundary cutoff, as dictated by the previous Section. In the JT limit this becomes the $K_{++}$ element \eqref{eq:K++JT}, which provides one of the two basis solutions at $L=\infty$. As in \cite{Fan:2021bwt}, we insert two complete sets of parabolic states: $\ket{\mu}$ diagonalizing $E$ and $\tilde E$, and $\ket{\nu}$ diagonalizing $F$ and $\tilde F$,\footnote{On the half-line the parabolic states are not mutually orthogonal, but they still provide a resolution of the identity, which is all that is needed here.}
\begin{equation}\label{eq:eigenvalue1}
e \ket{\mu}=-2\pi i b^2 \mu \ket{\mu}, \qquad \tilde{e}\ket{\mu}=\left(-2\pi i b^2 \mu\right)^{1/b^2}\ket{\mu},
\end{equation}
\begin{equation}\label{eq:eigenvalue2}
f \ket{\nu}=-2\pi i b^2 \nu \ket{\nu}, \qquad \tilde{f}\ket{\nu}=\left(-2\pi i b^2 \nu\right)^{1/b^2}\ket{\nu},
\end{equation}
consistently with $E$ and $F$ being antihermitian and with \eqref{eq:transcendental}. Inserting \eqref{eq:groupdecomp}, the functions $g_b$ and $g_b^*$ act diagonally on the parabolic states, and one obtains
\begin{equation}\label{eq:amplitude2}
\bra{h}g\ket{h}=\int_{-\infty}^{+\infty} \frac{\mathrm{d} \mu}{2\pi} \int_{-\infty}^{+\infty} \frac{\mathrm{d} \nu}{2\pi}\, \bra{h}\ket{\nu}\bra{\mu}\ket{h} \, g_{b}\left(2 \pi  b^2 \gamma \nu \right)\, g_{b}^{*}\left(-2 \pi  b^2 \beta \mu\right) \psi_{\mu \nu}(\phi),
\end{equation}
where $\psi_{\mu \nu}(\phi)=\bra{\nu}e^{2\phi H}\ket{\mu}$ is the Whittaker function \cite{Fan:2021bwt}. \chg{The hyperbolic states are the eigenvectors of the dilatation $K=e^{i\pi b^2 x \partial_x}$, the same as in the JT limit,}
\begin{equation}\label{eq:hyperbolic}
\bra{x}\ket{h}=x^{i h},
\end{equation}
with $K\bra{x}\ket{h}=e^{-\pi b^2 h}\bra{x}\ket{h}$ and $\tilde{K}\bra{x}\ket{h}=e^{-\pi h}\bra{x}\ket{h}$, consistently with the hyperbolic charges in \eqref{eq:joint-diagonalization}.

\chg{The evaluation of \eqref{eq:amplitude2} is carried out in Appendix~\ref{app:hyperbolic}}, where using the identification
\begin{equation}\label{eq:ipelement}
Z=\beta e^{2\phi}\gamma=e^{\Lambda-2\pi b^2 h},
\end{equation}
which is the same as \eqref{eq:Lambda-Z}, we find
\begin{equation}\label{eq:K+++}
\boxed{
\begin{aligned}
\bra{h}g(\phi,\beta,\gamma)\ket{h}
&=
2\pi b\,
\frac{S_{b}\left(\frac{b}{2}+ibs-ibh\right)S_{b}\left(\frac{b}{2}+ibs+ibh\right)}
{S_{b}\left(b+2ibs\right)}\,e^{2i\phi h}
\\[1mm]
&\quad\times e^{-\left(\Lambda-2\pi b^2 h \right)(\frac12+ih+is)}
\\[1mm]
&\quad\times F_{b}\left(\frac{b}{2}+i b h+ibs,\frac{b}{2}+i b h+ibs, b+2ibs; -e^{-\Lambda}e^{4\pi b^2 h}\right)
\end{aligned}
}
\end{equation}
In normal-ordered form, $e^{2i\phi h}e^{-(\Lambda-2\pi b^2 h)(\frac12+ih)}=\gamma^{-ih}Z^{-1/2}\beta^{-ih}$. So \eqref{eq:K+++} has exactly the radial factorization \eqref{eq:radial-factorization}, \eqref{eq:Phi-chi} derived in the previous Section,
\begin{equation}\label{eq:Kpp_radial}
\bra{h}g\ket{h}\propto\gamma^{-ih}\,Z^{-1/2}\,\corrIX{\chi^{-}_s(\Lambda)}\,\beta^{-ih},
\end{equation}
with radial function
\begin{equation}\label{eq:sol_sinh_ininifty}
\corrIX{\chi^{-}_s(\Lambda)}=e^{-i \Lambda s}F_{b}\left(\lambda+ib\nu,\lambda+ib\nu,2\lambda; -e^{-\Lambda+4\pi b^2 \nu}\right),
\end{equation}
where we have used $h=\nu$ and recognized the representation label $\lambda=b/2+ibs$. In the JT limit \corrIX{$\chi^-_s(\Lambda)\to\psi^-_s(L)$}, and \eqref{eq:K+++} reduces to the $SL(2,\mathbb{R})$ element $K_{++}$ in \eqref{eq:K++JT}. If one replaces $\lambda$ with the standard label $\alpha=Q/2+iP$, \eqref{eq:sol_sinh_ininifty} is the hyperbolic matrix element of the modular double computed in \cite{Ip2013QuantumPlane},\footnote{See equation 7.35 of \cite{Ip2013QuantumPlane}.} with the hyperbolic labels identified with the cutoff $\nu$.

The function \corrIX{$\chi^-_s(\Lambda)$} solves the gravitational difference equation \eqref{eq:sinh_difference_equation_chi}, as it must, since it is a radial $U_q(\mathfrak{sl}(2,\mathbb{R}))$ Casimir eigenfunction. At $\Lambda\to\infty$ its argument vanishes, $F_b\to1$, and \corrIX{$\chi^-_s(\Lambda)$}$\simeq e^{-i\Lambda s}$: it is the $q$-deformation of the solution \corrIX{$\psi^-_s(L)$} around $L=\infty$.

\subsection{The dual Casimir equation and the basis at infinity}\label{sec:dual_equation}

Since \eqref{eq:K+++} is a representation matrix element of the modular double, and not only of $U_q(\mathfrak{sl}(2,\mathbb R))$, it must be an eigenfunction of both Casimir operators. The $U_q$ Casimir, radially reduced as in Section \ref{sec:casimir}, gives the gravitational difference equation \eqref{eq:sinh_difference_equation_chi}, with shifts $\Lambda\to\Lambda\pm2\pi i b^2$. The dual Casimir must then give a second difference equation for the same radial function \corrIX{$\chi^-_s(\Lambda)$}, with shifts $\Lambda\to\Lambda\pm2\pi i$, generated by $\tilde K$. This is indeed the case. The second equation can be read off directly from the integral representation of \corrIX{$\chi^-_s(\Lambda)$},\footnote{Write \eqref{eq:b_hypergeometric} in the variable $bt$. At fixed labels $(\alpha,\beta,\gamma)$ the integrand depends on $b$ only through the double sine functions, which are invariant under $b\to1/b$. The gravitational equation \eqref{eq:sinh_difference_equation_chi} follows from the shift $bt\to bt+b$ of the integration variable together with $S_b(x+b)=2\sin(\pi b x)S_b(x)$. The shift $bt\to bt+1/b$ together with $S_b(x+1/b)=2\sin(\pi x/b)S_b(x)$ then yields the equation obtained from the first by $b\to1/b$ at fixed labels, which is \eqref{eq:dual_eq}.} and reads
\begin{equation}\label{eq:dual_eq}
\begin{split}
\cosh\!\left(\! 2\pi s+\frac{i\pi}{b^2}\right)\corrIX{\chi^-_s(\Lambda)}
={}&\frac12\left(\tilde q^{-1}+e^{-\Lambda/b^2}\right)\corrIX{\chi^-_s(\Lambda-2\pi i)}
+\frac12\left(\tilde q+e^{4\pi\nu}e^{-\Lambda/b^2}\right)\corrIX{\chi^-_s(\Lambda+2\pi i)}\\
&+e^{2\pi\nu}e^{-\Lambda/b^2}\corrIX{\chi^-_s(\Lambda)}.
\end{split}
\end{equation}

The eigenvalue in \eqref{eq:dual_eq} is the dual Casimir $\tilde c_{\tilde q}(\lambda)$ in \eqref{eq:two_casimirs}, up to normalization. This identifies \eqref{eq:dual_eq} as the dual Casimir equation of the modular double\chg{, in the analytically continued form discussed after \eqref{eq:two_casimirs}.}

\paragraph{The second solution}
The gravitational difference equation \eqref{eq:sinh_difference_equation_chi} is invariant under $s\to-s$, and so is the $U_q$ Casimir \rev{\eqref{eq:casimir-eigenvalue}}: $c_q(\lambda)=c_q(\bar\lambda)$, with $\bar\lambda=b/2-ibs$. Repeating the construction of the previous subsection in the realization \eqref{eq:algebra1}, \eqref{eq:algebra2} with $\lambda$ replaced by $\bar\lambda$ gives the second solution
\begin{equation}\label{eq:chi_minus}
\corrIX{\chi^{+}_s(\Lambda)=\chi^{-}_{-s}(\Lambda)}=e^{i \Lambda s}F_{b}\left(\bar{\lambda}+ib\nu,\bar{\lambda}+ib\nu,2\bar{\lambda}; -e^{-\Lambda+4\pi b^2 \nu}\right).
\end{equation}
It behaves as $e^{i\Lambda s}$ at $\Lambda\to\infty$ and reduces to \corrIX{$\psi^+_s(L)$} in the JT limit, that is, to the radial part of $K_{--}$. It satisfies the dual equation \eqref{eq:dual_eq} with $s\to-s$: the operator on the right-hand side is the same, while the eigenvalue becomes $\cosh\left(2\pi s-i\pi/b^2\right)$, which is $\tilde c_{\tilde q}(\bar\lambda)$ up to normalization. Notice that $\lambda$ and $\bar\lambda$ share the $U_q$ Casimir, hence the gravitational energy, but not the dual one. \corrIX{Since $\lambda/b=j+1$ and $\bar\lambda/b=-j=(-1-j)+1$, in the JT limit the reflection $\lambda\to\bar\lambda$ is the shadow reflection $j\to-1-j$, which preserves the $\mathfrak{sl}(2,\mathbb R)$ Casimir $j(j+1)$.} In the modular double, by contrast, the reflection that preserves both Casimirs is $\lambda\to Q-\lambda$, and $\bar\lambda=b-\lambda$ is a genuinely different label.

\chg{For $b^2\notin\mathbb{Q}$ the two Casimir equations leave no periodic \rev{factor}: a meromorphic factor $p(\Lambda)$ such that $p(\Lambda)\chi^\pm_s(\Lambda)$ still solves both must be periodic under the two incommensurate shifts $\Lambda\to\Lambda+2\pi ib^2$ and $\Lambda\to\Lambda+2\pi i$, and is therefore constant, as already noted below \eqref{eq:dual_difference_equation}.}

\subsection{The regular solution}

At fixed energy, the modular double has thus supplied two \rev{distinguished} eigenfunctions of the gravitational Hamiltonian, $\chi^\pm_s(\Lambda)$, behaving as \corrIX{$e^{\pm i\Lambda s}$} at $\Lambda\to\infty$. As in JT gravity, the physical wavefunction should be the combination of them that is regular at the origin $\Lambda\to-\infty$. We now show what is the combination that coincides with $\chi_s(\Lambda)$ in \eqref{eq:chi_eigenfunction}.

The key observation is that the three functions $\chi_s(\Lambda)$ and $\chi^\pm_s(\Lambda)$ are the same Mellin--Barnes integral, up to a trigonometric factor. Defining
\begin{equation}
P(t)\equiv S_b\left(\tfrac b2+ib\nu\pm ibs+bt\right)S_b^2(-bt),
\end{equation}
one finds\footnote{In \eqref{eq:chi_eigenfunction} we rewrite $1/S_b(b+bt)=S_b(1/b-bt)=-2\sin(\pi t)\,S_b(-bt)$, using $S_b(x)S_b(Q-x)=1$ and the shift property. \corrIX{For $\chi^\pm_s$, eqs.~\eqref{eq:chi_minus} and \eqref{eq:sol_sinh_ininifty}, we rewrite in the same way $1/S_b(b\mp2ibs+bt)$ and change variables $t\to-t-\frac12-i\nu\pm is$.} The contour still separates the same sequences of poles, and the double sine functions recombine exactly into $P(t)$.}
\begin{align}\label{eq:three_integrals}
\chi_s(\Lambda)&=-\frac{4\pi b^2}{S_b(\frac b2+ib\nu\pm ibs)}\,\mathcal{E}(\Lambda) \int_{\mathcal C_t}\frac{\dd t}{2\pi i}\,e^{(\Lambda-2\pi b^2\nu)t}\,P(t)\, \sin(\pi t),\\
\chi^\pm_s(\Lambda)&=\frac{4\pi b\,S_b(b\corrIX{\mp}2ibs)}{S_b^2(\frac b2+ib\nu\corrIX{\mp} ibs)}\,e^{\corrIX{\pm}2\pi i b^2\nu s}\,\mathcal{E}(\Lambda)
\int_{\mathcal C_t}\frac{\dd t}{2\pi i}\,e^{(\Lambda-2\pi b^2\nu)t}\,P(t)\,\cos\big(\pi(t+i\nu\corrIX{\pm} is)\big).\notag
\end{align}

At $\Lambda\to-\infty$ we close the contour to the right. There $P(t)$ has double poles at $t=n\in\mathbb N$, coming from $S_b^2(-bt)$. A double pole produces, through the derivative of $e^{(\Lambda-2\pi b^2\nu)t}$, a term linear in $\Lambda$. The leading one, at $t=0$, gives
\begin{equation}
\chi^\pm_s(\Lambda)\overset{\Lambda\to-\infty}{\sim}\Lambda\, e^{\Lambda(\frac12+i\nu)},
\end{equation}
which is the $q$-analogue of the logarithmic branch-cut singularity of $\psi^\pm_s(L)$ at $L=0$ in JT gravity. \chg{In a combination $A\chi^+_s+B\chi^-_s$ with constant coefficients the integrand is $P(t)$ times a combination of $\cos(\pi t)$ and $\sin(\pi t)$, and the coefficient of the leading logarithmic term is the value of this factor at $t=0$. Requiring it to vanish leaves a factor proportional to $\sin(\pi t)$, which vanishes at \emph{every} integer and therefore turns all the double poles at $t=n$ into simple ones at once.}\footnote{The poles at $t=n+mb^{-2}$ with $m\geq1$ are double for every combination, and give contributions suppressed by $e^{m\Lambda/b^2}$, which are invisible in the JT limit and do not affect the behavior at the origin.} \chg{By \eqref{eq:three_integrals} the resulting function is precisely $\chi_s(\Lambda)$, which behaves as $e^{\Lambda(\frac12+i\nu)}$ at $\Lambda\to-\infty$, in agreement with \eqref{eq:eigenfunction_L=0}. \rev{The restriction to constant coefficients is not an assumption: by the corollary of Appendix~\ref{app:uniq_two}, combinations of $\chi^\pm_s$ with constant coefficients exhaust the solutions that satisfy conditions (i) and (ii) of Section~\ref{subsec:uniqueness} and are at most logarithmic at the origin. Together with the present result, this reproduces the uniqueness statement \eqref{eq:uniqueness_result}; \corrIX{$\chi^-_s$} is the logarithmic solution used in Appendix~\ref{app:uniqueness}.}}

\paragraph{Connection formula}
The coefficients follow from the elementary identity
\begin{equation}
\sin(\pi t)=-\frac{i\cosh\pi(\nu+s)}{\sinh 2\pi s}\cos\big(\pi(t+i\nu-is)\big)+\frac{i\cosh\pi(\nu-s)}{\sinh 2\pi s}\cos\big(\pi(t+i\nu+is)\big),
\end{equation}
which can be checked by expanding the cosines. Inserting it into \eqref{eq:three_integrals} and simplifying the prefactors,\footnote{We use \eqref{eq:double_sine_norm} and $S_b(-2ibs)S_b(b+2ibs)=i/(2\sinh2\pi s)$, both of which follow from the reflection and shift properties of $S_b$.} we obtain
\begin{equation}\label{eq:sinh_connection}
\boxed{\;
\begin{aligned}
&\chi_s(\Lambda)=\corrIX{C_s\,e^{-2\pi i b^2\nu s}\,\chi^+_s(\Lambda)+C_s^*\,e^{2\pi i b^2\nu s}\,\chi^-_s(\Lambda)},\\[1mm]
&C_s=\frac{b\,S_b(2ibs)}{S_b(\frac{b}{2}+ib\nu+ibs)\,S_b(\frac{b}{2}-ib\nu+ibs)}
\end{aligned}\;}
\end{equation}
This is the $q$-deformation of the Gauss connection formula
\eqref{eq:connection}. The asymptotic relation for $b\to 0,$
$S_b(bx)\simeq b^{2x-1}(2\pi)^{x-1}\Gamma(x)$,
then shows that the coefficients reduce to
\begin{equation}
C_s\to\frac{\Gamma(2is)}{\Gamma\left(\frac12+i\nu+is\right)\Gamma\left(\frac12-i\nu+is\right)},
\end{equation}
so that \eqref{eq:sinh_connection} reduces exactly to \eqref{eq:connection}. Moreover, since $\chi^\pm_s(\Lambda)\simeq e^{\corrIX{\pm} i\Lambda s}$ at $\Lambda\to\infty$, \eqref{eq:sinh_connection} reproduces the plane-wave asymptotics \eqref{eq:eigenfunction_L=infty} of the wavefunction, with the same coefficients $C_s$, and hence the spectral density \eqref{eq:spectral_density_s}. In this sense the subleading terms of the asymptotic expansion \eqref{eq:eigenfunction_L=infty} are resummed exactly by $\chi^\pm_s(\Lambda)$.

\paragraph{The dual equation of the wavefunction}
Since $\chi_s(\Lambda)$ combines matrix elements of both labels, it is not an eigenfunction of the dual Casimir of either $\lambda$ or $\bar\lambda$. Nevertheless it satisfies the dual equation \eqref{eq:dual_difference_equation}, and the present construction explains why. By \eqref{eq:three_integrals}, the integrands of $\chi_s$ and \corrIX{$\chi_s^-$} differ by the factor $\sin\pi t/\cos\pi(t+i\nu-is)$. Since the dual Cartan shift $\mathcal T:\Lambda\to\Lambda+2\pi i$ multiplies the integrand by $-e^{-2\pi\nu}e^{2\pi it}$, this factor is a ratio of operators linear in $\mathcal T$:
\begin{equation}\label{eq:twisted_dual}
(e^{2\pi s}\mathcal T-1)\,\chi_s\propto(e^{2\pi\nu}\mathcal T+1)\,\corrIX{\chi^-_s} .
\end{equation}
These operators commute with \eqref{eq:sinh_difference_equation_chi}, whose coefficients depend only on $e^{-\Lambda}$, but not with \eqref{eq:dual_eq}. Transporting \eqref{eq:dual_eq} through them reproduces \eqref{eq:dual_difference_equation}, including the coefficient $\cos(\pi/b^2)$ that distinguishes it from the $b\to1/b$ image of \eqref{eq:sinh_difference_equation_chi}. The dual equation of the wavefunction is therefore the dual Casimir equation of \corrIX{$\chi_s^-$}, twisted by an operator that commutes with $U_q(\mathfrak{sl}(2,\mathbb R))$.

\chg{To summarize, \rev{the analytic conditions select a two-dimensional space of solutions; in it,} the modular double \rev{identifies} the basis at infinity, $\chi^\pm_s(\Lambda)$, and regularity at the origin fixes the combination, which is the wavefunction $\chi_s(\Lambda)$ found by canonical quantization, with coefficients equal to its asymptotic amplitudes. This is the structure of JT gravity reviewed in Section \ref{sec:JT_review}, where the regular wavefunction is the combination of $K_{++}$ and $K_{--}$ that is regular at $L=0$.}

\section{Concluding remarks and outlook}\label{sec:conclusions}

In this paper we canonically quantized sinh dilaton gravity with finite Dirichlet boundary conditions, extending the bulk analysis of finite cutoff JT gravity \corr{\cite{Griguolo:2025kpi,Griguolo:2026eoi}} to a theory with an underlying quantum group structure. The main lesson is that the two deformations of asymptotic JT gravity, the finite cutoff controlled by $\Phi_b$ and the quantum deformation controlled by $b$, can be implemented at the same time while keeping full analytic control.
 
At the classical level, we described the reduced phase space by the length of the ERB, measured in an auxiliary AdS$_2$ geometry, and its conjugate momentum. The resulting Hamiltonian is a two-parameter deformation of Liouville quantum mechanics. At the quantum level, we obtained a finite-difference eigenvalue equation for the gravitational Hamiltonian and solved it exactly in terms of $q$-hypergeometric functions. From these we derived the finite cutoff spectral density, which interpolates between finite cutoff JT gravity and asymptotic Liouville gravity, and we checked that the JT and large cutoff limits commute. \corr{The difference equation alone fixes the wavefunction only up to periodic factors. We showed that this ambiguity is removed by the conditions that make the Hamiltonian well defined, namely analyticity in a strip, tempered growth at large length and regularity at $L=0$, which select a unique wavefunction.} Finally, we showed that the bulk equation is the radially reduced Casimir of $U_q(\mathfrak{sl}(2,\mathbb R))$ with equal left and right hyperbolic charges set by the cutoff\corr{, and that the dictionary extends to the wavefunctions once $U_q(\mathfrak{sl}(2,\mathbb R))$ is completed to the modular double. The two plane-wave solutions are hyperbolic matrix elements of the modular double with labels $\lambda$ and $\bar\lambda$, \rev{each singled out by the dual Casimir of its label within the two-dimensional space of solutions allowed by the analytic conditions}, and the wavefunction is their unique regular combination, with the asymptotic amplitudes as coefficients}. This is the $q$-deformed version of the dictionary between finite cutoff JT gravity and $SL(2,\mathbb R)$ hyperbolic matrix elements.
 
We conclude with some comments and open directions.
 
\paragraph{The auxiliary geometry as a Liouville saddle.}
A central ingredient in our construction is the Weyl rescaling that maps the sinh dilaton black hole to an auxiliary AdS$_2$ black hole, in which the ERB length is computed. We fixed the residual freedom in this map by requiring that the asymptotic boundaries of the two geometries be identified. With this choice the Weyl factor is $\omega^2=e^{2\pi b^2\Phi}$, and the radial map satisfies $\Phi'(\rho)=e^{-2\pi b^2\Phi(\rho)}$. This choice has a natural interpretation in the Liouville formulation of sinh dilaton gravity \corr{\cite{Mertens:2020hbs}}. There, sinh dilaton gravity is rewritten as a pair of Liouville theories, one spacelike and one timelike, whose fields are linear combinations of the metric conformal factor and the dilaton. In conformal gauge, $\dd s^2=e^{2\varphi}\dd z\dd\bar z$, the auxiliary metric reads
\begin{equation}
\dd s^2_{\rm AdS}=e^{2\pi b^2\Phi}\,\dd s^2=e^{2\corrIV{\sigma}}\,\dd z\,\dd\bar z,
\qquad
\corrIV{\sigma}\equiv\varphi+\pi b^2\Phi .
\end{equation}
Since the auxiliary metric has constant negative curvature, $\corrIV{\sigma}$ solves the classical Liouville equation. Up to normalization conventions, $\corrIV{\sigma}$ is precisely one of the two Liouville fields of \corr{\cite{Mertens:2020hbs}}. The classical solution in which we computed the ERB length is therefore the classical saddle of one of the two Liouville fields. The auxiliary AdS$_2$ is not an ad hoc device: it is the geometry naturally seen by this Liouville sector, and the canonical variable $L$ is a geodesic length in the Liouville metric $e^{2\corrIV{\sigma}}\dd z\dd\bar z$.
 
This observation suggests a CFT interpretation of our results. In this picture, the finite cutoff corresponds to a deformation of the standard FZZT and ZZ boundary conditions of the two Liouville sectors \corr{\cite{Mertens:2020hbs}}. A CFT interpretation of the results presented here will appear in a forthcoming paper \cite{newpaper}.
 
\paragraph{Further aspects of the sinh model.}
Several questions remain open within sinh dilaton gravity itself. The first concerns the group-theoretic construction. At $b=0$ we have an explicit map between the finite cutoff chart $(L,u,v)$ and the Gauss--Euler coordinates. At finite $b$ we only identified the invariant combination $Z=\gamma e^{2\phi}\beta$ with $e^{\Lambda-2\pi b^2\nu}$\corr{, and computed the matrix elements with a $q$-deformed Gauss--Euler decomposition}. An explicit hyperbolic decomposition of the \corr{modular double}, analogous to $g=e^{uH}e^{\frac L2(E+F)}e^{vH}$, \corr{would give $L$ a direct geometric meaning at finite $b$. It would also allow one to construct the analogue of $K_{+-}$, which in JT gravity is the regular wavefunction itself, rather than selecting the regular combination by its behavior at $L=0$.} It would also clarify the role of the Faddeev quantum dilogarithm in the similarity transformation \corr{\eqref{eq:factor}}. A second question concerns our choice of representation label $\lambda=b/2+ibs$, made to keep the JT limit manifest. This choice breaks the $b \to 1/b$ self-duality of Liouville theory at the level of representations\corr{: the reflection $\lambda\to\bar\lambda$ relating the two plane-wave solutions is a symmetry of $U_q(\mathfrak{sl}(2,\mathbb R))$ but not of $U_{\tilde q}(\mathfrak{sl}(2,\mathbb R))$, and the dual equation \eqref{eq:dual_difference_equation} is not the $b\to 1/b$ image of \eqref{eq:sinh_difference_equation_chi}.} It would be interesting to understand whether the finite cutoff is compatible with this self-duality, or whether it breaks it in a physically meaningful way. Finally, the finite cutoff partition function deserves a more detailed analysis:  the fate of states with $s>\nu$ and its relation to a $T\bar T$-like deformation of the asymptotic Liouville gravity answer \corr{\cite{Gross:2019ach,Iliesiu:2020zld,Griguolo:2026eoi}}. 
 
\paragraph{Sine dilaton gravity and DSSYK.}
The most pressing extension is to sine dilaton gravity, the bulk dual of DSSYK \corr{\cite{Blommaert:2024ymv,Blommaert:2024whf,Blommaert:2025avl,Bossi:2024ffa}}. Formally, sine dilaton gravity is obtained from the sinh model by continuing $b^2$ to imaginary values, and many of our formulas extend directly. However, the location of the asymptotic boundary is more subtle in this case. In the sine dilaton hologram of DSSYK \corr{\cite{Blommaert:2024ymv}}, the holographic screen is not at a large real value of the dilaton. It sits at a \emph{complexified} location, with the real part of the dilaton fixed and the imaginary part sent to infinity. This complexification is essential to reproduce the DSSYK transfer matrix and the discreteness of the chord-number basis. Consequently, a finite Dirichlet boundary in sine dilaton gravity should be placed along a complexified contour for the boundary value of the dilaton. As the cutoff is removed along this contour, the boundary must approach the asymptotic DSSYK point, rather than moving out along the real dilaton axis. Defining the reduced phase space, the operator ordering, and the Hilbert space along such a contour raises new conceptual and technical issues. Among them are the reality of the spectrum, the periodicity of the effective dilaton potential, and the fate of the quantum group structure at $|q|<1$. We leave a more detailed investigation of this problem to future work, while acknowledging that proposals for such deformations have already been put forward in the literature from the boundary perspective \cite{Aguilar-Gutierrez:2026ogo,Aguilar-Gutierrez:2026qqe}. A successful resolution would provide a finite cutoff deformation of DSSYK holography, and, through the recent connections with three-dimensional de Sitter gravity \corr{\cite{Tietto:2025oxn,Blommaert:2026ofx}}, a new handle on the description of observers \cite{Freivogel:2026loz} and finite regions in quantum de Sitter space \cite{Buchmuller:2026jbr}.

\section*{Acknowledgments}
We thank Andreas Belaey, Andreas Blommaert and Thomas Mertens for useful discussions. LG, LR, DS, AT have been supported in part by the Italian Ministero
dell’Università e della Ricerca (MUR), and by Istituto Nazionale di Fisica Nucleare (INFN)
through the ``Gauge and String Theory'' (GAST) research project. JP acknowledges financial support from the European Research Council (grant BHHQG-101040024). Funded
by the European Union. Views and opinions expressed are however those of the author(s)
only and do not necessarily reflect those of the European Union or the European Research
Council. Neither the European Union nor the granting authority can be held responsible
for them.
JP acknowledges fruitful discussions during the workshop ``Observers, wormholes and complex saddles in cosmology'', organized at the Bernoulli Center for Fundamental Studies (EPFL, Lausanne) from 18--22 May 2026.

\appendix

\section{$b$-formulae}\label{app:bformulae}

\subsection{Faddeev's quantum dilogarithm}\label{app:faddeev}
The Faddeev's quantum dilogarithm is defined as
\begin{equation}\label{eq:quantum_dilog_def}
    \phi_b(z) \equiv \exp\!\left( - \frac{1}{4} \int_{-\infty}^{+\infty} \frac{e^{-2 i z t}}{\sinh(t b) \sinh(t/b)} \frac{\dd t}{t} \right),
\end{equation}
where the integration contour is understood to circle the singularity at $t=0$ from above. It satisfies the following shift properties
\begin{equation}\label{eq:dilog_shift_app}
    \phi_b(z \pm i b) = (1 + e^{2 \pi b z}e^{\pm i \pi b^2})^{\pm 1} \phi_b(z).
\end{equation}
Its large-argument asymptotics is given by \cite{Kashaev:2011se}
\begin{equation}\label{eq:dilog_asymptotics}
    \phi_b(z) \sim
    \begin{cases}
        \corrIV{e^{-\frac{i\pi}{12}(b^2+b^{-2})}\,e^{-i\pi z^2}}, &\quad \text{Re}{z} \to +\infty,\\
        1, &\quad \text{Re}{z} \to -\infty. 
    \end{cases}
\end{equation}
\corrIV{Its semiclassical limit at fixed $z$ is}
\begin{equation}\label{eq:dilog_b=0}
    \phi_b\!\left(\frac{z}{2 \pi b}\right) \overset{b \to 0}{\sim} \exp\!\left( \frac{i}{2 \pi b^2} \text{Li}_2(-e^z) \right).
\end{equation}
\corrIX{It} can be rewritten as a ratio of \corr{$q$-Pochhammer} symbols as \cite{Kashaev:2001,Garoufalidis:2020pax}
\begin{equation}\label{eq:phib_product}
\corr{\phi_{b}(z)}=\frac{\left(e^{\frac{2\pi}{b} \left(z-i Q/2\right)};\tilde{q}^{-2}\right)_{\infty}}{\left(e^{2\pi b \left(z+i Q/2\right)};q^2\right)_{\infty}},
\end{equation}
where $q=e^{i \pi b^2}$ and $\tilde{q} = e^{i \pi/b^2}$.
\rev{With our conventions, and for real $b$, $|\phi_b(z)|=1$ for real $z$, and $\phi_b(z)$ has zeros at $z=i\left(\frac Q2+nb+mb^{-1}\right)$ and poles at $z=-i\left(\frac Q2+nb+mb^{-1}\right)$, with $n,m\geq0$.}

\subsection{Double sine function}
The double sine function is related to the  Faddeev's quantum dilogarithm as
\begin{equation}
    S_{b}(x)=\corr{e^{\frac{i\pi}{2}x\left(Q-x\right)}\,e^{-\frac{i\pi}{12}\left(Q^2+1\right)}\,\phi_{b}\left(i x-i Q/2\right)}
\end{equation}
\rev{Besides the shift relation \eqref{eq:double_sine_shift}, the double sine function satisfies the dual shift and the reflection relations}
\begin{equation}\label{eq:double_sine_dual_shift}
    \rev{S_b(x+b^{-1}) = 2\sin(\pi x/b)\, S_b(x),}
\end{equation}
\begin{equation}\label{eq:double_sine_reflection}
    \rev{S_b(x)\,S_b(Q-x)=1,}
\end{equation}
\rev{and it has a simple pole at the origin,}
\begin{equation}\label{eq:double_sine_residue}
    \rev{S_b(x)\simeq\frac{1}{2\pi x},\qquad x\to0 .}
\end{equation}
\rev{Together with \eqref{eq:double_sine_shift}, the last relation gives $S_b(b)=b$, which is used to identify \eqref{eq:chi_eigenfunction} with \eqref{eq:sinh_eigenfunction} and in \eqref{eq:eigenfunction_L=infty}.}
The double sine function \rev{also} satisfies
\begin{equation}\label{eq:double_sine_norm}
    \abs{S_b\!\left(\frac{b}{2}+ibx\right)}^2 = \frac{1}{2 \cosh(\pi x)}, \qquad \abs{S_b(ibx)}^2 = \frac{1}{4 \sinh(\pi b^2 x) \sinh(\pi x)}.
\end{equation}
The double sine function reduces to the ordinary Gamma function according to
\begin{equation}\label{eq:double_sine_b=0}
    S_b(bx)
    \overset{b\to0}{\sim}
    \frac{1}{\sqrt{2\pi}}
    \left(2\pi b^2\right)^{x-\frac{1}{2}} \Gamma(x).
\end{equation}
The double sine function $S_b(bz)$ has a set of poles ($\times$) and zeros ($\circ$) located at
\begin{equation}\label{eq:poleszeros}
    z_\times = - n - m b^{-2}, \qquad z_\circ = (n+1) + (m+1) b^{-2}.
\end{equation}

\subsection{The function \texorpdfstring{$g_b$}{g\_b}}\label{app:gb}
The group element \eqref{eq:groupdecomp} of Section~\ref{sec:wave} is built from
\begin{equation}\label{eq:gb_def}
g_b(z)=\frac{1}{\phi_b\big(\frac{\log z}{2\pi b}\big)},
\end{equation}
which is Faddeev's quantum dilogarithm in the normalization of \cite{Ip2013QuantumPlane}. For $z>0$ one
has $|g_b(z)|=1$, and $g_b(z)\to1$ as $z\to0$. The shift relation \eqref{eq:dilog_shift_app} and its
$b\to1/b$ image give the two functional equations
\begin{equation}\label{eq:gb_functional}
g_b(q^{-1}z)=(1+z)\,g_b(qz),\qquad g_b(e^{-i\pi}z)=\big(1+z^{1/b^2}\big)\,g_b(e^{i\pi}z).
\end{equation}

By \eqref{eq:phib_product}, $g_b(z)=(-qz;q^2)_\infty/(-\tilde q^{-1}z^{1/b^2};\tilde q^{-2})_\infty$. With
the $q$-exponential
\begin{equation}\label{eq:qexponential}
e_{q^2}(x)=\sum_{n=0}^{\infty}\frac{x^n}{[n]_{q^2}!}=\frac{1}{\big(x(1-q^2);q^2\big)_\infty},\qquad
e_{q^2}(x)\,e_{q^{-2}}(-x)=1,
\end{equation}
this becomes a product of a $q$- and a $\tilde q$-exponential,
\begin{equation}
g_b(z)=e_{q^{-2}}\!\left(\frac{iz}{2\sin\pi b^2}\right)e_{\tilde q^{-2}}\!\left(\frac{iz^{1/b^2}}{2\sin(\pi/b^2)}\right).
\end{equation}
For $z=i\gamma f$, the transcendental relation \eqref{eq:transcendental} gives
\begin{equation}
g_b(i\gamma f)=e_{q^{-2}}(-\gamma F)\;e_{\tilde q^{-2}}\big(i(i\gamma)^{1/b^2}\tilde F\big),
\end{equation}
and similarly for $g_b^*(-i\beta e)$. The group element \eqref{eq:groupdecomp} is therefore the product of
a $U_q$ and a $U_{\tilde q}$ factor, and it generalizes the Gauss--Euler decomposition of a single
quantum group in terms of $q$-exponentials, due to Fronsdal and Galindo.

The integrals over the parabolic labels in Section~\ref{sec:wave} are Mellin transforms of $g_b$. For
real $\alpha\neq0$,
\begin{equation}\label{eq:identity_both}
\begin{split}
\int_0^{\infty}\!dx\,x^{s-1}g_b(2\pi b^2\alpha x)&=2\pi b\,(2\pi b^2|\alpha|)^{-s}\,
e^{\frac{i\pi}2\mathrm{sgn}(\alpha)s}\,e^{\frac{i\pi b^2}{2}(s^2+s)}\,S_b(bs),\\
\int_0^{\infty}\!dx\,x^{s-1}g_b^*(2\pi b^2\alpha x)&=2\pi b\,(2\pi b^2|\alpha|)^{-s}\,
e^{-\frac{i\pi}2\mathrm{sgn}(\alpha)s}\,e^{-\frac{i\pi b^2}{2}(s^2+s)}\,S_b(bs),
\end{split}
\end{equation}
where $g_b^*(z)\equiv\overline{g_b(\bar z)}$, and the integrals at large $x$ are defined by analytic
continuation in $s$. For $\alpha<0$ the argument is $e^{i\pi}|2\pi b^2\alpha x|$ in $g_b$ and
$e^{-i\pi}|2\pi b^2\alpha x|$ in $g_b^*$. These are the branches selected by \eqref{eq:transcendental}
on the parabolic states with principal branches of the eigenvalues, and on them both functions are
bounded.

The identities follow from the Fourier transform of $g_b$, eq.~(2.30) of \cite{Ip:2021wnr}, whose conventions
agree with ours, with $G_b(x)=e^{\frac{i\pi}{2}x(x-Q)}S_b(x)$. Setting $x=e^{2\pi br}$ and inverting it gives, for $2\pi b^2\alpha=1$,
\begin{equation}
\int_0^\infty\dd x\,x^{s-1}g_b(x)=\frac{2\pi b\,e^{i\pi b^2s^2}}{G_b(Q-bs)}
=2\pi b\,e^{\frac{i\pi}{2}s}\,e^{\frac{i\pi b^2}{2}(s^2+s)}\,S_b(bs),
\end{equation}
where we used \eqref{eq:double_sine_reflection}. The factor $(2\pi b^2|\alpha|)^{-s}$ follows by rescaling $x$.
For $\alpha<0$, rotating the contour by $e^{-i\pi}$, which is allowed since $g_b(e^{i\theta}y)$ is regular for
$0\le\theta\le\pi$, multiplies the result by $e^{-i\pi s}$. The second identity is the complex conjugate of the first.
As $b\to0$, both reduce to \eqref{eq:ide}.

\section{Proof of $q$-hypergeometric orthogonality}\label{appendix:overlap}
In this Appendix we provide a proof of the orthogonality relation
\begin{equation}
    \bra{s'}\ket{s} = \int_{-\infty}^{+\infty} \dd\Lambda \, \psi^*_{s'}(\Lambda) \psi_s(\Lambda) = \frac{\delta(s-s')}{\rho(s)},
\end{equation}
with $\rho(s)$ given in \eqref{eq:spectral_density_s}, using the asymptotic behaviors \eqref{eq:eigenfunction_L=0} and \eqref{eq:eigenfunction_L=infty} computed in the main text.
We start by recalling that the eigenfunction \eqref{eq:sinh_eigenfunction} solves the difference equation
\begin{equation}\label{eq:diff_equation_appendix}
    D_s \psi_s(\Lambda) = 0, \quad D_s \equiv \frac{1}{2} \left( e^{-i \pi b^2 \partial_\Lambda} f(\Lambda) e^{-i \pi b^2 \partial_\Lambda} + e^{i \pi b^2 \partial_\Lambda} f(\Lambda) e^{i \pi b^2 \partial_\Lambda} \right) + V(\Lambda) - \cosh\!{(2 \pi b^2 s)},
\end{equation}
hence we can write
\begin{equation}
    \begin{split}
        0
        &= \int_{-\infty}^{+\infty} \dd\Lambda \bigg( \psi^*_{s'}(\Lambda) \, D_s \psi_s(\Lambda) - \psi_{s}(\Lambda) \, D_{s'} \psi^*_{s'}(\Lambda) \bigg).
    \end{split}
\end{equation}
The potential terms cancel and we are left with
\begin{equation}\label{eq:orthogonality_computation}
    \int_{-\infty}^{+\infty} \dd\Lambda \, \psi^*_{s'}(\Lambda) \psi_s(\Lambda) = \frac{I_{b}(s',s) - I^*_{b}(s,s')}{\cosh\!{(2 \pi b^2 s)} - \cosh\!{(2 \pi b^2 s')}},
\end{equation}
where we introduced the quantity\footnote{Here the shift operators are expanded in Taylor series and integrated by parts term by term. This expansion is formal, since $\psi_s$ \rev{is singular} at $|\mathrm{Im}\,\Lambda|=\pi(1+b^2)$.}
\begin{equation}\label{eq:orthogonality_integral}
\begin{split}
    I_b(s',s)
    &\equiv \frac{1}{2} \int_{-\infty}^{+\infty} \dd\Lambda \, \psi^*_{s'}(\Lambda) \, e^{i \pi b^2 \partial_\Lambda} \, f(\Lambda) \, e^{i \pi b^2 \partial_\Lambda} \, \psi_s(\Lambda) + (b^2 \to -b^2)\\
    &= \sum_{n=0}^{\infty} \left( i \pi b^2 \right)^{2n} \sum_{k=0}^{2n} \frac{1}{k! (2n-k)!} \int_{-\infty}^{+\infty} \dd\Lambda \, \psi^*_{s'} \, \partial_\Lambda^k\!\left[ f(\Lambda) \partial_\Lambda^{2n-k} \psi_s  \right].
\end{split}
\end{equation}
Performing $k$ partial integrations within the integral term in \eqref{eq:orthogonality_integral}, we are left with
\begin{equation}
    \begin{split}
        I_b(s',s)
        &= \sum_{n=0}^{\infty} \left( i \pi b^2 \right)^{2n} \sum_{k=0}^{2n} \frac{1}{k! (2n-k)!} \Bigg( (-1)^k \int_{-\infty}^{+\infty} \dd\Lambda \, f(\Lambda) \, \partial_\Lambda^k \psi^*_{s'} \, \partial_\Lambda^{2n-k} \psi_{s} + \\
        &\quad + \sum_{j=1}^k (-1)^{j-1} \left[ \partial_\Lambda^{j-1} \psi^*_{s'} \cdot \sum_{p=0}^{k-j} \binom{k}{p} \partial_\Lambda^p f(\Lambda) \, \partial_\Lambda^{2n-j-p} \psi_s \right]_{\Lambda \to -\infty}^{\Lambda \to +\infty} \Bigg).
    \end{split}
\end{equation}
The bulk term will cancel with its counterpart coming from $I^*_b(s,s')$.
Next, we focus on the $\Lambda \to -\infty$ contribution. From the earlier result \eqref{eq:eigenfunction_L=0}, we know that in this regime the eigenfunction is real and independent of the energy parameter $s$. This means that every contribution from $\Lambda \to -\infty$ will cancel between the two terms appearing at the \corr{numerator} in \eqref{eq:orthogonality_computation}.

The $\Lambda \to +\infty$ contribution, instead, involves derivatives of $f(\Lambda)$. From its definition \corr{\eqref{eq:sinh_f_function}}, we know that this function approaches $1$ in this regime and thus all its derivatives vanish. The sum over $p$ thus collapses to $p=0$ and we have
\begin{equation}
    I_b(s',s) = \sum_{n=0}^{\infty} \left( i \pi b^2 \right)^{2n} \sum_{k=0}^{2n} \frac{1}{k! (2n-k)!} \sum_{j=1}^k (-1)^{j-1} \lim_{\Lambda \to \infty} \left[ \partial_\Lambda^{j-1} \psi^*_{s'} \cdot \, \partial_\Lambda^{2n-j} \psi_s \right] + \text{bulk term}.
\end{equation}
Making use \corr{of} the explicit asymptotic expression of the eigenfunction \eqref{eq:eigenfunction_L=infty}\corrIV{, in which we absorb the phases $e^{\mp2\pi ib^2\nu s}$ into $C_s$ and $C_s^*$ (since only $|C_s|$ enters the result)}, we obtain
\begin{equation}
\begin{split}
    I_b(s',s) &- \text{bulk term}
    = -i \sum_{n=0}^{\infty} \left( \pi b^2 \right)^{2n} \sum_{k=0}^{2n} \frac{1}{k! (2n-k)!} \times \\
    &\quad\times \sum_{j=1}^k s'^{j-1} s^{2n-j} \lim_{\Lambda \to \infty} \left[ (-1)^{j-1} \left(  C_{s'} C_s e^{i \Lambda (s+s')} - \text{c.c.} \right) + \left( C^*_{s'} C_s e^{i \Lambda (s-s')} - \text{c.c.} \right) \right].
\end{split}
\end{equation}
Remarkably, the complex exponentials $\lim_{\Lambda \to \infty} e^{\pm i\Lambda (s \pm s')}$ vanish in the distributional sense unless $s \pm s' = 0$. This condition is met only when $s = s'$ since we have $s,s' \geq 0$. We are thus interested in the case $s-s' \to 0$, for which the numerator in \eqref{eq:orthogonality_computation} simplifies to
\begin{equation}\label{eq:numerator_computation}
    I_b(s',s) - I^*_b(s,s') = 2 \pi \abs{C_s}^2 \delta(s-s') \times \sum_{n=1}^\infty \frac{(2 \pi b^2)^{2n}}{(2n)!} 2 n (s-s') s^{2n-1}.
\end{equation}
Here we used the identity
\begin{equation}
    \sum_{k=1}^{2n} \frac{1}{(k-1)! (2n-k)!} = \frac{2^{2n-1}}{(2n-1)!}
\end{equation}
and the Dirac delta expression
\begin{equation}
    \lim_{\Lambda \to \infty} \frac{\sin((s-s') \Lambda)}{s-s'} =  \pi \delta(s-s').
\end{equation}
Finally, the denominator in \eqref{eq:orthogonality_computation} reduces, for $s-s' \to 0$, to
\begin{equation}
    \cosh\!{(2 \pi b^2 s)} - \cosh\!{(2 \pi b^2 s')} = \sum_{n=1}^\infty \frac{(2 \pi b^2)^{2n}}{(2n)!} 2n (s-s') s^{2n-1}.
\end{equation}
This quantity precisely cancels the infinite sum appearing in \eqref{eq:numerator_computation}. Hence, from Equation \eqref{eq:orthogonality_computation}, we conclude that the overlap between two eigenfunctions reads
\begin{equation}
    \int_{-\infty}^{+\infty} \dd\Lambda \, \psi^*_{s'}(\Lambda) \psi_s(\Lambda) = 2 \pi \abs{C_s}^2 \delta(s-s').
\end{equation}
Finally, using the definition of $C_s$ \corr{\eqref{eq:eigenfunction_L=infty}} together with Equations \eqref{eq:double_sine_norm}, we can read off
\begin{equation}
    \rho(s) = \frac{1}{2 \pi \abs{C_s}^2} = \frac{1}{\pi b^2} \frac{\sinh\!{(2 \pi b^2 s)} \, \sinh(2 \pi s)}{\cosh(2 \pi \nu) + \cosh(2 \pi s)}.
\end{equation}
which indeed corresponds to the spectral density \eqref{eq:spectral_density_s} given in the main text.

\section{Proof of the uniqueness of the wavefunction}\label{app:uniqueness}

In this appendix we prove the statement of Section~\ref{subsec:uniqueness}: every solution $\chi$ of
\eqref{eq:sinh_difference_equation_chi} satisfying conditions (i)--(iii) is a constant multiple of
$\chi_s$. The argument parallels the one for a second-order differential equation, where any solution
is a combination of two independent ones, with coefficients given by Wronskians and fixed by the
boundary conditions. For a difference equation the Wronskian is replaced by the Casoratian, which
vanishes whenever the ratio of two solutions is periodic with the period $\Omega=2\pi ib^2$ of the
equation, not only when it is constant. The coefficients of the combination are therefore periodic
functions of $\Lambda$ rather than constants: this is the ambiguity \eqref{eq:periodic_ambiguity}.
The argument still works because suitably dressed Casoratians of solutions holomorphic in a strip are
holomorphic functions of $w=e^{\Lambda/b^2}$ on the punctured plane, so the periodic coefficients are
fixed by Liouville's theorem once their behavior at $w\to0$ and $w\to\infty$ is known, which is what
conditions (i)--(iii) provide.

As reference solutions we use $\chi_s$ and the plane-wave solution \corrIX{$\chi_s^-$} of
Section~\ref{sec:wave}, eq.~\eqref{eq:sol_sinh_ininifty}. The latter is the $q$-deformation of the JT
matrix element $K_{++}$, the solution with the logarithmic singularity at $L=0$: it behaves as
$\Lambda\,\mathcal E(\Lambda)$ there, and is precisely what condition (iii) excludes.
Section~\ref{app:uniq_ref} recalls the properties of the two solutions that the proof uses, which are
derived in Sections~\ref{sec:sinh_box} and~\ref{sec:wave} or follow from the integral representations
given there. Section~\ref{app:uniq_per} introduces the dressed Casoratian, which is periodic, and
shows that it is a nonzero constant for $\chi_s$ and \corrIX{$\chi_s^-$}, and that it vanishes for $\chi_s$ and
any solution $\chi$ satisfying (i)--(iii). Section~\ref{app:uniq_const} concludes: $\chi=g\chi_s$ with
$g$ periodic and equal to a ratio of two dressed Casoratians, hence holomorphic on the punctured
plane, bounded, and constant.

Throughout, $0<b<1$, $s>0$ and $\nu\in\mathbb R$, and we use the shorthand
\begin{equation}
q=e^{i\pi b^2},\qquad \Omega=2\pi ib^2,\qquad w=e^{\Lambda/b^2},\qquad
\mathcal E(\Lambda)=e^{(\Lambda-2\pi b^2\nu)(\frac12+i\nu)} .
\end{equation}
The step $\Omega$ and the variable $w$ are related by $e^{\Omega/b^2}=1$, so a function of $\Lambda$ with
period $\Omega$ is a function of $w$. We write \eqref{eq:sinh_difference_equation_chi} in the form
\begin{equation}\label{uq:ACD}
A(\Lambda)\chi(\Lambda+\Omega)+C(\Lambda)\chi(\Lambda-\Omega)=D(\Lambda)\chi(\Lambda),
\end{equation}
with
\begin{equation}
A(\Lambda)=\tfrac12\big(1+q^{-1}e^{4\pi b^2\nu}e^{-\Lambda}\big),\quad
C(\Lambda)=\tfrac12\big(1+q\,e^{-\Lambda}\big),\quad
D(\Lambda)=\cosh(2\pi b^2s)+e^{2\pi b^2\nu}e^{-\Lambda}.
\end{equation}
Since $\Omega$ is purely imaginary, the equation relates the values of $\chi$ at three points on a
vertical line. A solution holomorphic in the strip $S_\eta=\{|\mathrm{Im}\,\Lambda|<\eta\}$ thus
satisfies \eqref{uq:ACD} for $|\mathrm{Im}\,\Lambda|<\eta-2\pi b^2$, a nonempty strip since
$\eta>2\pi b^2$. All asymptotic estimates below, for $\mathrm{Re}\,\Lambda\to\pm\infty$, are understood
to hold uniformly in $\mathrm{Im}\,\Lambda$ for $|\mathrm{Im}\,\Lambda|\le2\pi b^2$, a closed strip
contained in $S_\eta$, which is where they will be used.

We shall repeatedly use two facts about periodic functions. First, let $f$ be holomorphic on a
horizontal strip of height larger than $2\pi b^2$, with $f(\Lambda+\Omega)=f(\Lambda)$ wherever both
points lie in the strip. Then $f(\Lambda)=\hat F(w)$, where $\hat F$ is holomorphic on the
punctured plane $\mathbb C^*$, and the limits $|w|\to\infty$ and $|w|\to0$ correspond to
$\mathrm{Re}\,\Lambda\to+\infty$ and $\mathrm{Re}\,\Lambda\to-\infty$. Second,
a polynomial bound in $\Lambda$ is a logarithmic bound in $w$, and
a function $\hat F$ that is holomorphic for $|w|>R$ and satisfies
$|\hat F(w)|\le C(\log|w|)^N$ is automatically bounded at infinity.\footnote{Since
$\hat F(w)/w\to0$ as $|w|\to\infty$, Riemann's theorem on removable singularities implies that the
point at infinity is a removable singularity, so $\hat F$ has a finite limit there.} Below, a hat
on a function indicates that it is regarded as a function of $w$.

The properties of $\phi_b$ and $S_b$ that enter the proof, namely the shift relation and the
asymptotics of $\phi_b$ and the location of the zeros and poles of both functions, are collected in
Appendix~\ref{app:bformulae}.

\subsection{Properties of the reference solutions \texorpdfstring{$\chi_s$ and \corrIX{$\chi_s^-$}}{chi\_s and chi\_s-}}\label{app:uniq_ref}

The proof uses only the following properties of the two reference solutions $\chi_s$ and
\corrIX{$\chi_s^-$},\footnote{They are independent, since their Casoratian does not vanish, as shown in
Section~\ref{app:uniq_per}.} which are derived in Sections~\ref{sec:sinh_box} and~\ref{sec:wave} or
follow directly from the integral representations given there.

The solution $\chi_s$ is holomorphic for $|\mathrm{Im}\,\Lambda|<\pi(1+b^2)$, because its Fourier
kernel \eqref{eq:fourier_solution} decays as $e^{-\pi(1+b^2)|\mathrm{Re}\,\Pi|}$, see
Section~\ref{subsec:uniqueness}. Its asymptotics are given by the residue expansions that lead to
\eqref{eq:eigenfunction_L=0} and \eqref{eq:eigenfunction_L=infty}, which for $\chi_s$ read
\begin{align}
\chi_s(\Lambda)&=\mathcal E(\Lambda)\big(1+O(e^{\mathrm{Re}\,\Lambda})\big),&&\mathrm{Re}\,\Lambda\to-\infty,\label{uq:asym-}\\
\chi_s(\Lambda)&=C_s^*e^{2\pi ib^2\nu s}\,e^{-is\Lambda}+C_s\,e^{-2\pi ib^2\nu s}\,e^{is\Lambda}+O(e^{-\mathrm{Re}\,\Lambda}),&&\mathrm{Re}\,\Lambda\to+\infty,\label{uq:asym+}
\end{align}
with $C_s$ defined in \eqref{eq:eigenfunction_L=infty}; it is finite and nonzero for $s>0$, since
$S_b$ has neither zeros nor poles at the arguments involved (Appendix~\ref{app:bformulae}). Hence
$\chi_s$ satisfies (i)--(iii) for every $\eta\in(2\pi b^2,\pi(1+b^2))$, an interval that is nonempty
because $b<1$. From now on we take $\eta$ in this interval, which is no restriction on $\chi$.

The solution \corrIX{$\chi_s^-$}, eq.~\eqref{eq:sol_sinh_ininifty}, solves \eqref{uq:ACD}, as shown in
Section~\ref{sec:wave} (see also the footnote to \eqref{eq:dual_eq}). By \eqref{eq:three_integrals}
its Mellin--Barnes integrand differs from that of $\chi_s$ only by a trigonometric factor with the
same growth, so \corrIX{$\chi_s^-$} is holomorphic in the same strip. Its asymptotics, derived in
Section~\ref{sec:wave}, are
\begin{align}
\corrIX{\chi^-_s(\Lambda)}&=e^{-is\Lambda}\big(1+O(e^{-\mathrm{Re}\,\Lambda})\big),&&\mathrm{Re}\,\Lambda\to+\infty,\label{uq:asymp+}\\
\corrIX{\chi^-_s(\Lambda)}&=\mathcal E(\Lambda)\big(\alpha\Lambda+\beta+O(\Lambda\,e^{\mathrm{Re}\,\Lambda})\big),&&\mathrm{Re}\,\Lambda\to-\infty,\label{uq:asymp-}
\end{align}
with constants $\alpha\neq0$ and $\beta$. Thus \corrIX{$\chi_s^-$} is the logarithmic solution of
Section~\ref{subsec:uniqueness}: it satisfies (i) and (ii) but not (iii).

All four estimates hold uniformly in $\mathrm{Im}\,\Lambda$ on closed sub-strips of
$|\mathrm{Im}\,\Lambda|<\pi(1+b^2)$, and in particular for $|\mathrm{Im}\,\Lambda|\le2\pi b^2$, since
they are obtained by shifting the contour of a Fourier integral whose kernel decays as above on
horizontal bands.

\subsection{The dressed Casoratian and uniqueness}\label{app:uniq_per}

For two solutions $\chi_1$, $\chi_2$ of \eqref{uq:ACD}, holomorphic in $S_\eta$, the Casoratian
\begin{equation}\label{uq:W}
K[\chi_1,\chi_2](\Lambda)=\chi_1(\Lambda)\chi_2(\Lambda+\Omega)-\chi_2(\Lambda)\chi_1(\Lambda+\Omega)
\end{equation}
is the discrete analogue of the Wronskian: it vanishes identically if and only if $\chi_2/\chi_1$ is
$\Omega$-periodic. For a second-order differential equation without first-derivative term the
Wronskian would be constant. Here the Casoratian is not, because the coefficients $A$ and $C$ of the
two shifted terms in \eqref{uq:ACD} are different, but it comes close: eliminating
$\chi_i(\Lambda-\Omega)$ from \eqref{uq:W} with the help of \eqref{uq:ACD} gives
\begin{equation}
C(\Lambda)K[\chi_1,\chi_2](\Lambda-\Omega)=A(\Lambda)K[\chi_1,\chi_2](\Lambda),
\end{equation}
so that $K$ changes under a shift by $\Omega$ only by the explicit factor $A/C$. Such a factor can be
compensated as follows. The function
\begin{equation}
M(\Lambda)=\frac{\phi_b\big(-\frac{\Lambda}{2\pi b}\big)}{\phi_b\big(-\frac{\Lambda}{2\pi b}+2b\nu\big)}
\end{equation}
is built so that, by the shift relation \eqref{eq:dilog_shift} with $e^\Omega=q^2$,
\begin{equation}
\frac{M(\Lambda)}{M(\Lambda-\Omega)}=\frac{1+q\,e^{4\pi b^2\nu}e^{-\Lambda}}{1+q\,e^{-\Lambda}}
=\frac{A(\Lambda-\Omega)}{C(\Lambda)} ,
\end{equation}
and therefore the dressed Casoratian
\begin{equation}\label{uq:Wper}
W[\chi_1,\chi_2]\equiv A\,M\,K[\chi_1,\chi_2]\qquad\text{satisfies}\qquad W(\Lambda)=W(\Lambda-\Omega),
\end{equation}
that is, it is $\Omega$-periodic.
\rev{The origin of the dressing is simple: $M$ is the square of the prefactor in \eqref{eq:factor}, and in terms of the functions $\psi_i$ related to $\chi_i$ by \eqref{eq:factor} one finds $W[\chi_1,\chi_2]=\frac12 f\big(\Lambda+\tfrac{\Omega}{2}\big)\big[\psi_1(\Lambda)\psi_2(\Lambda+\Omega)-\psi_2(\Lambda)\psi_1(\Lambda+\Omega)\big]$, with $f$ given in \eqref{eq:sinh_f_function}. The dressed Casoratian is thus the Casoratian of the symmetric form \eqref{eq:Schroedinger_equation} of the equation, in which the coefficients of the two shifted terms are related by a shift, and this is why it is periodic.}

Let us see where $W$ is defined and how it behaves. The Casoratian $K[\chi_1,\chi_2]$ involves
$\chi_i(\Lambda)$ and $\chi_i(\Lambda+\Omega)$, so it is holomorphic for
$-\eta<\mathrm{Im}\,\Lambda<\eta-2\pi b^2$, a strip of height $2\eta-2\pi b^2$, which is larger than
$2\pi b^2$ because $\eta>2\pi b^2$. The factor $A$ is entire, and $M$ has neither zeros nor poles for
$|\mathrm{Im}\,\Lambda|<\pi(1+b^2)$, a region that contains $S_\eta$. Thus $W$ is holomorphic on a
strip taller than one period, and by the first of the two facts recalled above it is a holomorphic
function $\hat W(w)$ on all of $\mathbb C^*$.

To determine the behavior of $W$ we also need that of $A$ and $M$ at large $|\mathrm{Re}\,\Lambda|$.
As $\mathrm{Re}\,\Lambda\to+\infty$, $A\to\frac12$ and $M\to1$, since both quantum dilogarithms tend
to $1$. As $\mathrm{Re}\,\Lambda\to-\infty$, $A\simeq\frac12q^{-1}e^{4\pi b^2\nu}e^{-\Lambda}$, while the
Gaussian asymptotics \eqref{eq:dilog_asymptotics} of the two quantum dilogarithms combine into
$M\simeq e^{4\pi ib^2\nu^2}e^{-2i\nu\Lambda}$; both estimates hold up to exponentially small
corrections. Together with
$\mathcal E(\Lambda)\mathcal E(\Lambda+\Omega)=q\,e^{-4\pi b^2\nu-4\pi ib^2\nu^2}e^{(1+2i\nu)\Lambda}$,
this gives
\begin{equation}\label{uq:AME}
A(\Lambda)M(\Lambda)\,\mathcal E(\Lambda)\mathcal E(\Lambda+\Omega)\longrightarrow\frac12,\qquad\mathrm{Re}\,\Lambda\to-\infty .
\end{equation}
Hence, at the two ends, the dressed Casoratian of two solutions reduces to
\begin{equation}\label{uq:Wends}
W[\chi_1,\chi_2]\simeq
\begin{cases}
\frac12\big[\chi_1(\Lambda)\chi_2(\Lambda+\Omega)-\chi_2(\Lambda)\chi_1(\Lambda+\Omega)\big], &\mathrm{Re}\,\Lambda\to+\infty,\\[1mm]
\frac12\big[u_1(\Lambda)u_2(\Lambda+\Omega)-u_2(\Lambda)u_1(\Lambda+\Omega)\big],\quad u_i\equiv\chi_i/\mathcal E, &\mathrm{Re}\,\Lambda\to-\infty,
\end{cases}
\end{equation}
up to a factor $1+o(1)$.

To proceed, we need three dressed Casoratians. The first is that of the two reference solutions. As
$\mathrm{Re}\,\Lambda\to+\infty$, inserting \eqref{uq:asym+} and \eqref{uq:asymp+} in \eqref{uq:Wends}, the
$e^{-is\Lambda}$ terms cancel and
\begin{equation}
W[\chi_s,\corrIX{\chi^-_s}]\longrightarrow\tfrac12\,C_s\,e^{-2\pi ib^2\nu s}\big(e^{-is\Omega}-e^{is\Omega}\big)
=C_s\,e^{-2\pi ib^2\nu s}\sinh(2\pi b^2s).
\end{equation}
As $\mathrm{Re}\,\Lambda\to-\infty$, \eqref{uq:asym-} and \eqref{uq:asymp-} give, for
$u_s\equiv\chi_s/\mathcal E$ and \corrIX{$u_-\equiv\chi^-_s/\mathcal E$}, $u_s=1+o(1)$ and
\corrIX{$u_-$}$=\alpha\Lambda+\beta+o(1)$, with exponentially small corrections, so
\begin{equation}
W[\chi_s,\corrIX{\chi^-_s}]\longrightarrow\tfrac12\big[\big(\alpha(\Lambda+\Omega)+\beta\big)-\big(\alpha\Lambda+\beta\big)\big]=\frac{\alpha\Omega}{2}.
\end{equation}
Thus $\hat W[\chi_s,\corrIX{\chi^-_s}]$ is holomorphic on $\mathbb C^*$ with finite limits at $w\to\infty$ and
$w\to0$: it extends to a bounded entire function and is constant by Liouville's theorem,
\begin{equation}\label{uq:W1}
W[\chi_s,\corrIX{\chi^-_s}]=C_s\,e^{-2\pi ib^2\nu s}\sinh(2\pi b^2s)\ne0 .
\end{equation}
In particular the two limits must coincide, $\alpha\Omega/2=C_s\,e^{-2\pi ib^2\nu s}\sinh(2\pi b^2s)$, which
confirms that $\alpha\neq0$.

The second is that of $\chi_s$ with a solution $\chi$ satisfying (i)--(iii). As
$\mathrm{Re}\,\Lambda\to+\infty$, $\chi$ grows at most polynomially by (ii) and $\chi_s$ is bounded, so by
\eqref{uq:Wends} $|\hat W(w)|\le C(\log|w|)^N$, and $\hat W$ is bounded at infinity by the second
fact. As $\mathrm{Re}\,\Lambda\to-\infty$, \eqref{uq:asym-} and condition (iii) give $u_s=1+\epsilon_1$ and
$u\equiv\chi/\mathcal E=c+\epsilon_2$ with $\epsilon_{1,2}\to0$, and the leading terms cancel:
\begin{equation}
W[\chi_s,\chi]\simeq\tfrac12\Big[\big(1+\epsilon_1(\Lambda)\big)\big(c+\epsilon_2(\Lambda+\Omega)\big)
-\big(c+\epsilon_2(\Lambda)\big)\big(1+\epsilon_1(\Lambda+\Omega)\big)\Big]\longrightarrow0 .
\end{equation}
So $\hat W[\chi_s,\chi]$ is a bounded entire function vanishing at the origin, and by Liouville's
theorem
\begin{equation}\label{uq:W0}
W[\chi_s,\chi]\equiv0 .
\end{equation}

The third is that of $\chi$ with \corrIX{$\chi^-_s$}. As $\mathrm{Re}\,\Lambda\to+\infty$ the same argument shows
that $\hat W[\chi,\corrIX{\chi^-_s}]$ is bounded at infinity; its behavior at the origin will not be needed.

\paragraph{Conclusion}\label{app:uniq_const}

By \eqref{uq:W0}, and since $A$ and $M$ have no zeros in the strip where $W$ is defined,
$K[\chi_s,\chi]\equiv0$, which means that
\begin{equation}\label{uq:g}
\chi=g\,\chi_s,\qquad g(\Lambda+\Omega)=g(\Lambda),
\end{equation}
that is, $\chi$ has the form \eqref{eq:periodic_ambiguity}. The remaining task is to show that $g$ is
constant. Here the second reference solution enters. Since $g$ is periodic, it passes through the
Casoratian, $K[g\chi_s,\corrIX{\chi^-_s}]=g\,K[\chi_s,\corrIX{\chi^-_s}]$, and therefore
\begin{equation}
g=\frac{W[\chi,\corrIX{\chi^-_s}]}{W[\chi_s,\corrIX{\chi^-_s}]}=\frac{\hat W[\chi,\corrIX{\chi^-_s}](w)}{W[\chi_s,\corrIX{\chi^-_s}]} .
\end{equation}
The denominator is the nonzero constant \eqref{uq:W1} and the numerator is holomorphic on $\mathbb C^*$.
So $g$, which a priori was only meromorphic, with possible poles at the zeros of $\chi_s$, is in fact a
holomorphic function $\hat G(w)$ on the whole punctured plane: the Casoratian representation shows that
$\chi$ vanishes wherever $\chi_s$ does, to at least the same order. At infinity $\hat G$ is bounded,
because $\hat W[\chi,\corrIX{\chi^-_s}]$ is. At the origin we use \eqref{uq:g} directly: by condition (iii) and
\eqref{uq:asym-}, $\hat G=\chi/\chi_s\to c$ as $w\to0$, so the origin is a removable singularity. Hence
$\hat G$ is a bounded entire function, a constant by Liouville's theorem, and
\begin{equation}
\chi=N(s)\,\chi_s,\qquad N(s)=c,
\end{equation}
which is the statement \eqref{eq:uniqueness_result}.

\subsection{A corollary: the two-dimensional space of solutions}\label{app:uniq_two}

\rev{The same argument determines all the solutions that satisfy (i) and (ii) but are allowed to be logarithmic at the origin. Let us replace (iii) by the weaker condition}
\begin{itemize}
\item[\rev{(iii$'$)}] \rev{$|\chi(\Lambda)|\le C_0\,(1+|\Lambda|)^N\,|\mathcal E(\Lambda)|$ as $\mathrm{Re}\,\Lambda\to-\infty$, for some constants $C_0$ and $N$, uniformly inside the strip of (i).}
\end{itemize}
\rev{Then every solution of \eqref{eq:sinh_difference_equation_chi} satisfying (i), (ii) and (iii$'$) is a combination of the two reference solutions with constant coefficients,
\begin{equation}\label{uq:two}
\chi(\Lambda)=a_1\,\chi_s(\Lambda)+a_2\,\corrIX{\chi^-_s(\Lambda)},\qquad a_1,a_2\in\mathbb C .
\end{equation}}

\rev{Indeed, let $\chi$ be any solution holomorphic in $S_\eta$. Solving the two relations $\chi(\Lambda)=a_1\chi_s(\Lambda)+a_2\corrIX{\chi^-_s(\Lambda)}$ and $\chi(\Lambda+\Omega)=a_1\chi_s(\Lambda+\Omega)+a_2\corrIX{\chi^-_s(\Lambda+\Omega)}$ for $a_1$ and $a_2$ at each $\Lambda$ gives
\begin{equation}\label{uq:cramer}
a_1=\frac{K[\chi,\corrIX{\chi^-_s}]}{K[\chi_s,\corrIX{\chi^-_s}]}=\frac{W[\chi,\corrIX{\chi^-_s}]}{W[\chi_s,\corrIX{\chi^-_s}]},\qquad
a_2=\frac{K[\chi_s,\chi]}{K[\chi_s,\corrIX{\chi^-_s}]}=\frac{W[\chi_s,\chi]}{W[\chi_s,\corrIX{\chi^-_s}]} ,
\end{equation}
which are $\Omega$-periodic: this is the general solution, with periodic coefficients. The denominator is the nonzero constant \eqref{uq:W1}, and the numerators are holomorphic functions of $w$ on $\mathbb C^*$. As $\mathrm{Re}\,\Lambda\to+\infty$, condition (ii) and the boundedness of $\chi_s$ and \corrIX{$\chi^-_s$} give, by \eqref{uq:Wends}, $|\hat W(w)|\le C(\log|w|)^N$ for both numerators, which are therefore bounded at infinity. As $\mathrm{Re}\,\Lambda\to-\infty$, \eqref{uq:Wends} together with \eqref{uq:asym-}, \eqref{uq:asymp-} and (iii$'$) gives $|\hat W(w)|\le C\,\big(\log(1/|w|)\big)^{N+1}$, so that $w\,\hat W(w)\to0$ and the origin is a removable singularity. Both numerators are thus bounded entire functions of $w$, hence constants by Liouville's theorem, and so are $a_1$ and $a_2$.}

\rev{Therefore, under (i), (ii) and (iii$'$) the space of solutions of \eqref{eq:sinh_difference_equation_chi} is two-dimensional, as for a second-order differential equation: analyticity in the strip and the growth conditions at the two ends remove the periodic ambiguity completely, and leave two integration constants. By the connection formula \eqref{eq:sinh_connection}, with $C_s\neq0$, the space is equivalently spanned by $\chi^+_s$ and $\chi^-_s$, both of which satisfy the three conditions. Condition (iii) then requires $a_2=0$, because $\alpha\neq0$ in \eqref{uq:asymp-}, and one recovers \eqref{eq:uniqueness_result}.}

\rev{Finally, every solution holomorphic in $S_\eta$ extends meromorphically to the whole $\Lambda$ plane by iterating \eqref{uq:ACD}. On this extension the dual Casimir equations of Section~\ref{sec:dual_equation} can be imposed, and they distinguish $\chi^+_s$ from $\chi^-_s$ within the two-dimensional space, as discussed there.}

\section{Large cutoff limit}\label{appendix:large_cutoff}
We now consider the large cutoff behavior of the operator \ref{eq:sinh_quantum_hamiltonian2}, i.e. we send $\Phi_b \to \infty$ while keeping the renormalized length
\begin{equation}\label{eq:renorm}
    \hat{L}_\text{ren} \equiv \hat{L} - 2 \log( \frac{2 e^{2 \pi b^2 \Phi_b}}{2 \pi b^2})
\end{equation}
finite. In the limit \eqref{eq:renorm}, we therefore have 
\begin{equation}
    \begin{split}
        \sinh\!{(\hat{L}/2)} \overset{\Phi_b \to \infty}{\longrightarrow} \frac{e^{2 \pi b^2 \Phi_b}}{2 \pi b^2} \, e^{\hat{L}_\text{ren}/2},
        \qquad \cosh\!{(\hat{L}/2)}  \overset{\Phi_b \to \infty}{\longrightarrow} \frac{e^{2 \pi b^2 \Phi_b}}{2 \pi b^2} \, e^{\hat{L}_\text{ren}/2},
        \qquad \tanh\!{(\hat{L}/2)}  \overset{\Phi_b \to \infty}{\longrightarrow} 1,
    \end{split}
\end{equation}
thus the operator $f(\hat{L})$ defined in \corr{\eqref{eq:sinh_f_function}} reduces to
\begin{equation}
    f(\hat{L}) \overset{\Phi_b \to \infty}{\longrightarrow} \sqrt{1 + (2 \pi b^2)^2 \, e^{-\hat{L}_\text{ren}}},
\end{equation}
whereas the momentum reduces to
\begin{equation}
    \hat{\Pi} \overset{\Phi_b \to \infty}{\longrightarrow} \hat{P},
\end{equation}
and the potential term $e^{2 \pi b^2 \Phi_b}/\sinh^2(\hat{L}/2)$ is subleading. In summary, the prepotential operator at large cutoff reads
\begin{equation}\label{eq:sinh_prepotential_large_cutoff}
    \cosh(2 \pi b^2 \hat{\Phi}_h) = \frac{1}{2} e^{\pi b^2 \hat{P}} \sqrt{1 + (2 \pi b^2)^2 \, e^{-\hat{L}_\text{ren}}} e^{\pi b^2 \hat{P}} + \frac{1}{2} e^{-\pi b^2 \hat{P}} \sqrt{1 + (2 \pi b^2)^2 \, e^{-\hat{L}_\text{ren}}} e^{- \pi b^2 \hat{P}},
\end{equation}
and we shall compare it to the results presented in Section 5 of \cite{Mertens:2020hbs}. There, two difference equations are presented (see Equation 5.6 and 5.7), which correspond to the Schr\"odinger equations produced by the difference operators
\begin{equation}\label{eq:sinh_prepotential_non_symmetric}
    \begin{split}
        \cosh_{-}(2 \pi b^2 \hat{\Phi}_h) &= \frac{1}{2} e^{+\pi b^2 \hat{P}} \left( 1 + (2 \pi b^2)^2 e^{-\hat{L}_\text{ren}} \right) e^{+\pi b^2 \hat{P}} + \frac{1}{2} \rev{e^{-2 \pi b^2 \hat{P}}}, \\
        \cosh_{+}(2 \pi b^2 \hat{\Phi}_h) &= \frac{1}{2} \rev{e^{+2 \pi b^2 \hat{P}}} + \frac{1}{2} e^{-\pi b^2 \hat{P}} \left( 1 + (2 \pi b^2)^2 e^{-\hat{L}_\text{ren}} \right) e^{- \pi b^2 \hat{P}},
    \end{split}
\end{equation}
respectively.
Neither operator in Equation \eqref{eq:sinh_prepotential_non_symmetric} matches our large cutoff operator \eqref{eq:sinh_prepotential_large_cutoff} as it stands. Nevertheless, we can show that they share the same spectrum. To see this, we now relate them by a similarity transformation realized by some operator(s) $S_{\pm}(\hat{L}_\text{ren})$ as follows
\begin{equation}\label{eq:similarity_relation}
    S^{-1}_{\pm}(\hat{L}_\text{ren}) \cosh{(2 \pi b^2 \hat{\Phi}_h)} S_{\pm}(\hat{L}_\text{ren}) = \cosh_{\pm}(2 \pi b^2 \hat{\Phi}_h).
\end{equation}
To understand the form of the operator $S_{\pm}(\hat{L}_\text{ren})$ to choose, we take advantage of the following operator \corr{identities}: given two canonically conjugate operators $\hat{L}_\text{ren}$ and $\hat{P}$,
\begin{equation}
    [\hat{L}_\text{ren}, \hat{P}] = i,
\end{equation}
one can show that
\begin{equation}\label{eq:commutation_shift}
    e^{\pm \pi b^2 \hat{P}} f(\hat{L}_\text{ren}) = f(\hat{L}_\text{ren} \mp i \pi b^2) e^{\pm \pi b^2 \hat{P}}.
\end{equation}
If we sandwich the prepotential operator \eqref{eq:sinh_prepotential_large_cutoff} between $S^{-1}_{\pm}(L_\text{ren})$ and $S_{\pm}(L_\text{ren})$ and employ (\ref{eq:commutation_shift}) twice, we end up with
\begin{equation}
    \begin{split}
        S^{-1}_{\pm}&(\hat{L}_\text{ren}) \cosh{(2 \pi b^2 \hat{\Phi}_h)} S_{\pm}(\hat{L}_\text{ren}) = \\
        &= \frac{1}{2} e^{\pi b^2 \hat{P}} S^{-1}_{\pm}(\hat{L}_\text{ren} + i \pi b^2) \sqrt{1 + (2 \pi b^2)^2 \, e^{-\hat{L}_\text{ren}}} S_{\pm}(\hat{L}_\text{ren} - i \pi b^2) e^{\pi b^2 \hat{P}} + (b^2 \to -b^2).
    \end{split}
\end{equation}
Hence, requiring the relation (\ref{eq:similarity_relation}) to hold translates into the following condition on $S_{\pm}(\hat{L}_\text{ren})$:
\begin{equation}
    \begin{split}
        S_{\pm}(\hat{L}_\text{ren} + i \pi b^2) = \left( 1 + (2 \pi b^2)^2 \, e^{-\hat{L}_\text{ren}} \right)^{\pm 1/2} S_{\pm}(\hat{L}_\text{ren} - i \pi b^2),
    \end{split}
\end{equation}
which is readily satisfied if we \corr{relate} the operator $S_{\pm}(\hat{L}_\text{ren})$ to Faddeev's quantum dilogarithm as
\begin{equation}\label{eq:operator}
    S_{\pm}(\hat{L}_\text{ren}) = \phi_b\!\left(-\frac{\hat{L}_\text{ren} - 2 \log(2 \pi b^2)}{2 \pi b} \right)^{\mp 1/2}.
\end{equation}
Since $|\phi_b|=1$ on the real axis, $S_\pm$ is unitary. Thus, we conclude that our result at large cutoff is consistent with those presented in \cite{Mertens:2020hbs}.

We conclude the discussion about the large cutoff limit of sinh dilaton gravity by verifying that the $b \to 0$ limit of (\ref{eq:sinh_prepotential_large_cutoff}) reproduces the Liouville Hamiltonian (\corr{\ref{eq:JT_Liouville_H_review}}) discussed in Section \corr{\ref{sec:JT_review}}. Indeed, if we expand both sides as $b \to 0$ and keep only terms up to order $\sim b^2$, then the left-hand-side reads
\begin{equation}
    \lim_{b \to 0} \cosh(2 \pi b^2 \hat{\Phi}_h) = 1 + \frac{(2 \pi b^2)^2}{2} \hat{\Phi}_h^2,
\end{equation}
whereas the right-hand-side is
\begin{equation}
    \lim_{b \to 0} \frac{1}{2} e^{\pi b^2 \hat{P}} \sqrt{1 + (2 \pi b^2)^2 \, e^{-\hat{L}_\text{ren}}} e^{\pi b^2 \hat{P}} + (b^2 \to -b^2)
    = 1 + \frac{(2 \pi b^2)^2}{2} \left( \hat{P}^2 + e^{-\hat{L}_\text{ren}} \right), 
\end{equation}
hence confirming the expected behavior.

\section{Hyperbolic matrix elements}\label{app:hyperbolic}

In this Appendix we collect the steps leading to the hyperbolic matrix element \eqref{eq:K+++} of Section \ref{sec:wave}: the parabolic vectors, the parabolic-hyperbolic overlaps, the Whittaker function, the evaluation of the matrix element, and the JT limit of each step. \chg{We use the shift and reflection relations of the double sine function, \eqref{eq:double_sine_shift}, \eqref{eq:double_sine_dual_shift} and \eqref{eq:double_sine_reflection}, and its behavior \eqref{eq:double_sine_residue} at the origin, all collected in Appendix~\ref{app:bformulae}.}

\paragraph{Carrier space and scalar product}
\chg{The realization \eqref{eq:algebra1}, \eqref{eq:algebra2} acts on functions of $x>0$. Its commutation relations follow from}
\begin{equation}
e^{i \pi b^2 x \partial_x}x^{\pm 1} e^{-i\pi b^2 x \partial_x}=q^{\pm 1} x^{\pm 1}, \qquad
e^{i \pi x \partial_x}x^{\pm 1/b^2} e^{-i\pi x \partial_x}=\tilde{q}^{\pm 1} x^{\pm 1/b^2},
\end{equation}
\chg{which are a consequence of} $\left[x\partial_x,x^{\pm 1}\right]=\pm x^{\pm 1}$ and $\left[x\partial_x,x^{\pm 1/b^2}\right]=\pm \frac{1}{b^2}x^{\pm 1/b^2}$. Inner products are defined as
\begin{equation}\label{eq:scalar}
\bra{f_1}\ket{f_2}=\int_{0}^{+\infty}\frac{dx}{x}f_{1}^{*}(x)f_{2}(x).
\end{equation}
The measure $dx/x$ is the one that renders the scalar product invariant, $\bra{f_1}\ket{f_2}=\bra{g \circ f_1}\ket{g \circ f_2}$, in the JT limit, where $g$ is the finite action of the group.\footnote{The latter is different from the standard $SL(2,\mathbb{R})$ one because we are exponentiating the ``symmetric'' form of the generators. See the discussion around equations 2.22 and 2.23 of \cite{Fan:2021bwt}.} We assume this measure is the correct one also in the $b$-deformed case. With respect to \eqref{eq:scalar}, one has $K^{\dagger}=K$, $E^{\dagger}=-E$ and $F^{\dagger}=-F$.

\subsection{Parabolic vectors and overlaps}\label{app:overlaps}

Let us find the simultaneous eigenvector $\phi_{\mu}(x)=\bra{x}\ket{\mu}$ of $e$ and $\tilde e$, that is, the solution of
\begin{equation}\label{eq:paraboliceq}
\begin{split}
e\, \phi_{\mu}(x)&=-i x \left[e^{\pi i b \lambda}\phi_\mu(e^{i\pi b^2}x)-e^{-\pi i b \lambda}\phi_\mu(e^{-i\pi b^2}x)\right]= -2\pi i b^2 \mu\, \phi_{\mu}(x),\\
\tilde{e}\, \phi_{\mu}(x)&=-i x^{1/b^2}\left[ e^{\pi i \lambda/b} \phi_\mu\left(e^{i \pi }x\right)-e^{-\pi i \lambda/b} \phi_\mu\left(e^{-i \pi }x\right)\right]=\left(-2\pi i b^2 \mu\right)^{1/b^2} \phi_{\mu}(x),
\end{split}
\end{equation}
for $x>0$. With the ansatz $\phi_{\mu}(x)=x^{-\lambda/b} f_{\mu}(x)$, and since $(e^{\pm i\pi b^2}x)^{-\lambda/b}=e^{\mp i\pi b\lambda}x^{-\lambda/b}$, the first equation becomes
\begin{equation}\label{eq:reducedeq}
(-ix)\left(f_{\mu}(e^{i\pi b^2}x)-f_{\mu}(e^{-i\pi b^2}x)\right)=-2 \pi i b^2 \mu \, f_{\mu}(x).
\end{equation}
The solution is given by the Mellin--Barnes integral
\begin{equation}\label{eq:mellinf}
f_{\mu}(x)=\frac{2\pi b}{2\pi i}\int_{-i \infty}^{+i\infty} d \xi \ \abs{\mu}^{-\xi} \   S_{b}(b \xi)\,x^{\xi}(2\pi b^2)^{-\xi}\, e^{i \frac{\pi}{2}\text{sign}(\mu)\xi},
\end{equation}
where the contour runs to the right of the poles of $S_b(b\xi)$ at $\xi=-n-mb^{-2}$. Indeed, acting with the left-hand side of \eqref{eq:reducedeq} multiplies the integrand by $(-ix)\,2i\sin(\pi b^2\xi)$, and $2\sin(\pi b^2\xi)S_b(b\xi)=S_b(b(\xi+1))$ by \eqref{eq:double_sine_shift}. After the change of variables $\xi+1=\zeta$ we get
\begin{equation}
(-ix)\left(f_{\mu}(e^{i\pi b^2}x)-f_{\mu}(e^{-i\pi b^2}x)\right)=(2\pi b^2)\, e^{-i \frac{\pi}{2}\text{sign}(\mu)} \abs{\mu}\,  f_{\mu}(x)=-2 \pi i b^2 \mu\,  f_{\mu}(x),
\end{equation}
as it should. For the second equation, the factor $e^{\pm i\pi\lambda/b}$ again cancels against $(e^{\pm i\pi}x)^{-\lambda/b}$. The difference of the two shifted terms multiplies the integrand by $2i\sin(\pi\xi)$, and $2\sin(\pi\xi)S_b(b\xi)=S_b(b(\xi+1/b^2))$. The shift $\xi+1/b^2=\zeta$ then gives the eigenvalue $\left(2\pi b^2\abs{\mu}e^{-i\frac{\pi}{2}\text{sign}(\mu)}\right)^{1/b^2}=\left(-2\pi i b^2 \mu\right)^{1/b^2}$. Here $f_\mu(e^{\pm i\pi}x)$ is understood as the analytic continuation of $f_\mu(x)$, which extends beyond the region of convergence of \eqref{eq:mellinf} by means of \eqref{eq:reducedeq}. Similarly, the vector diagonalizing $f$ and $\tilde{f}$ is
\begin{equation}\label{eq:eigenvectorF}
\bra{x}\ket{\nu}=\phi_{\nu}(x)=x^{\lambda/b} \frac{2\pi b}{2\pi i}\int_{-i \infty}^{+i\infty} d \xi \ \abs{\nu}^{-\xi} \   S_{b}(b \xi)\,x^{-\xi}(2\pi b^2)^{-\xi}\, e^{i \frac{\pi}{2}\text{sign}(\nu)\xi},
\end{equation}
with eigenvalues $-2\pi i b^2 \nu$ and $\left(-2\pi i b^2 \nu\right)^{1/b^2}$.

\paragraph{Resolution of the identity}
On the whole real line, with the measure $dx/\abs{x}$ and the relative phase between the branches $x\gtrless0$ fixed as in the JT case, the parabolic vectors satisfy $\bra{\mu}\ket{\mu'}=2\pi\delta(\mu-\mu')$ for any $b$. Restricting the corresponding completeness relation to $x,x'>0$ gives the resolution of the identity on the half-line,
\begin{equation}\label{eq:completeness}
    1=\int_{-\infty}^{+\infty}\frac{d\mu}{2\pi}\,\ket{\mu}\bra{\mu}=\int_{-\infty}^{+\infty}\frac{d\nu}{2\pi}\,\ket{\nu}\bra{\nu},
\end{equation}
which is what is used in \eqref{eq:amplitude2}.

\paragraph{Overlaps}
The overlaps with the hyperbolic vectors \eqref{eq:hyperbolic} are
\begin{equation}
\bra{h}\ket{\mu}=\frac{2\pi b}{2\pi i}\int_{-i \infty}^{+i\infty} d \xi \ \abs{\mu}^{-\xi} \   S_{b}(b \xi)(2\pi b^2)^{-\xi} e^{i \frac{\pi}{2}\text{sign}(\mu)\xi}\int_{0}^{+\infty} \frac{dx}{x}\, x^{-ih}\, x^{\xi}\, x^{-\frac12-is}.
\end{equation}
By Mellin inversion, the integral over $x$ sets $\xi=\frac12+is+ih$ and removes the factor $1/(2\pi i)$. Proceeding in the same way for $\bra{h}\ket{\nu}$, we obtain
\begin{equation}\label{eq:boverlaps_app}
\begin{split}
\bra{h}\ket{\mu}&=2\pi b\,S_b \left(b\left(\tfrac12+is+ih\right)\right) \abs{\mu}^{-\frac12-ih-is} \left(2 \pi b^2\right)^{-\frac12-ih-is} e^{i \frac{\pi}{2}\text{sign}(\mu)\left(\frac12+is+ih\right)}, \\
\bra{h}\ket{\nu}&=2\pi b\,S_b \left(b\left(\tfrac12+is-ih\right)\right) \abs{\nu}^{-\frac12+ih-is} \left(2 \pi b^2\right)^{-\frac12+ih-is} e^{i \frac{\pi}{2}\text{sign}(\nu)\left(\frac12+is-ih\right)},
\end{split}
\end{equation}
Using \eqref{eq:double_sine_b=0}, one checks that $2\pi b\,(2\pi b^2)^{-x}S_b(bx)\to\Gamma(x)$, so that \eqref{eq:boverlaps_app} reduces to the $SL(2,\mathbb R)$ overlaps of Appendix \ref{app:JTsteps}.

\subsection{Mixed-parabolic Whittaker function}\label{app:witta}

We have to compute
\begin{equation}
\psi_{\mu\nu}(\phi)=\bra{\nu}e^{2\phi H}\ket{\mu}=\int_{0}^{+\infty} \frac{dx}{x}\,\phi_{\nu}^{*}(x)\,\phi_{\mu}(e^{2\phi}x).
\end{equation}
The two functions involved are
\begin{equation}
\begin{split}
\phi_{\nu}^{*}(x)&=x^{\frac12-is} \frac{2\pi b}{2\pi i}\int_{-i \infty}^{+i\infty} d \xi \ \abs{\nu}^{-\xi} \   S_{b}(b \xi)\,x^{-\xi}(2\pi b^2)^{-\xi} e^{-i \frac{\pi}{2}\text{sign}(\nu)\xi}, \\
\phi_{\mu}(e^{2\phi}x)&=(x e^{2\phi})^{-\frac12-is} \frac{2\pi b}{2\pi i}\int_{-i \infty}^{+i\infty} d \xi' \ e^{2\phi \xi'}\abs{\mu}^{-\xi'} \   S_{b}(b \xi')\,x^{\xi'}(2\pi b^2)^{-\xi'} e^{i \frac{\pi}{2}\text{sign}(\mu)\xi'},
\end{split}
\end{equation}
where in the first line we used $\overline{S_b(z)}=S_b(\bar z)$ and relabeled $\bar\xi\to\xi$. The integral over $x$, $\int_0^{+\infty}\frac{dx}{x}x^{\xi'-\xi-2is}$, sets $\xi'=\xi+2is$ by Mellin inversion and removes one factor $1/(2\pi i)$. Performing the shift $\xi \rightarrow \xi-is$ we obtain
\begin{equation}\label{eq:wittaker_app}
\begin{split}
\psi_{\mu\nu}(\phi)=e^{-\phi}\frac{(2 \pi b)^2}{2 \pi i} \int_{-i \infty}^{+i \infty} d \xi\, &e^{2\phi \xi}\abs{\mu}^{-\xi-is} \abs{\nu}^{-\xi+is} S_{b}\left(b \left(\xi\pm is\right)\right) (2\pi b^2)^{-2 \xi}\\
&\times e^{i\frac{\pi}{2}\text{sign}(\mu)(\xi+is)}e^{-i\frac{\pi}{2}\text{sign}(\nu)(\xi-is)},
\end{split}
\end{equation}
As $b\to0$, using \eqref{eq:double_sine_b=0}, this becomes
\begin{equation}\label{eq:truemellinbarnes}
\psi_{\mu\nu}(\phi)\to e^{-\phi}\frac{1}{2\pi i}\int_{-i\infty}^{+i \infty} d \xi\, e^{2\phi \xi}\abs{\mu}^{-\xi-is} \abs{\nu}^{-\xi+is}\, \Gamma(\xi \pm i s)\, e^{i \frac{\pi}{2}\text{sign}(\mu)(\xi+is)}e^{-i \frac{\pi}{2}\text{sign}(\nu)(\xi-is)},
\end{equation}
which for $\mu,\nu>0$ is $2e^{-\phi}e^{-\pi s}(\mu/\nu)^{-is}K_{2is}(2\sqrt{\mu\nu}\,e^{-\phi})$.

\subsection{Evaluation of $K_{++}$}\label{app:K++}

\chg{Inserting \eqref{eq:completeness} and \eqref{eq:boverlaps_app} into \eqref{eq:amplitude2}, the matrix element becomes}
\begin{equation}\label{eq:doubleK++}
\begin{split}
\bra{h}g(\phi,\beta,\gamma)\ket{h}=&\ \frac{(2\pi b^2)^{2i h}}{2\pi}\,S_b\left(b\left(\tfrac{1}{2}-is-ih\right)\right)S_b\left(b\left(\tfrac{1}{2}+is-i h\right)\right)\\
& \times\int_{-\infty}^{+\infty} \mathrm{d}\mu\,
|\mu|^{ih+i s-\frac{1}{2}}
\exp\left[-\tfrac{i\pi}{2}\operatorname{sgn}(\mu)\left(\tfrac{1}{2}-ih-i s\right)\right] g_{b}^{*}\left(-2 \pi  b^2 \beta \mu\right)  \\
& \times \int_{-\infty}^{+\infty} \mathrm{d} \nu\,
|\nu|^{ih-is-\frac{1}{2}}
\exp\left[\tfrac{i\pi}{2}\operatorname{sgn}(\nu)\left(\tfrac{1}{2}-ih+i s\right)\right] g_{b}\left(2 \pi  b^2 \gamma \nu \right) \psi_{\mu \nu}(\phi).
\end{split}
\end{equation}
\chg{After inserting the Mellin--Barnes representation \eqref{eq:wittaker_app} of $\psi_{\mu\nu}(\phi)$,} we exchange the order of integration and perform the integrals over $\mu$ and $\nu$ first. Both integrals contain the same power, $\abs{\mu}^{A-1}$ and $\abs{\nu}^{A-1}$, with
\begin{equation}
A\equiv\tfrac12+ih-\xi .
\end{equation}
The integral over $\mu$ is
\begin{equation}
\int_{-\infty}^{+\infty} \mathrm{d}\mu \ \abs{\mu}^{A-1} e^{-i \frac{\pi}{2}\text{sign}(\mu)\left(\frac12-i h-2is-\xi\right)} g_{b}^{*}\left(-2 \pi  b^2 \beta \mu\right).
\end{equation}
For $\beta>0$, the argument of $g_b^*$ has sign $-\text{sign}(\mu)$. The second identity in \eqref{eq:identity_both} then gives the phase $e^{i\frac{\pi}{2}\text{sign}(\mu)A}$, and the sum over the two signs of $\mu$ produces $2\cos\left(\frac{\pi}{2}(-2ih-2is)\right)=2\cosh\pi(s+h)$:
\begin{equation}
\int_{-\infty}^{+\infty} \mathrm{d}\mu \ \ldots=2 (2\pi b) \left(2 \pi b^2\right)^{-A}\beta^{-A} S_{b}\left(bA\right) \cosh\left(\pi(s+h)\right) e^{-\frac{\pi i b^2}{2}(A^2+A)}.
\end{equation}
The integral over $\nu$ is
\begin{equation}
\int_{-\infty}^{+\infty} \mathrm{d}\nu \ \abs{\nu}^{A-1} e^{i \frac{\pi}{2}\text{sign}(\nu)\left(\frac12-i h+2is-\xi\right)} g_{b}\left(2 \pi  b^2 \gamma \nu\right).
\end{equation}
For $\gamma>0$, the first identity in \eqref{eq:identity_both} gives the phase $e^{i\frac{\pi}{2}\text{sign}(\nu)A}$, and the sum over the two signs of $\nu$ produces $2\cos\left(\frac{\pi}{2}(1+2is-2\xi)\right)=2\sin\pi(\xi-is)$:
\begin{equation}
\int_{-\infty}^{+\infty} \mathrm{d}\nu \ \ldots=2 (2\pi b) \left(2 \pi b^2\right)^{-A}\gamma^{-A} S_{b}\left(bA\right) \sin\left(\pi(\xi-is)\right) e^{\frac{\pi i b^2}{2}(A^2+A)}.
\end{equation}
The Gaussian factors cancel in the product. Collecting all factors of $2\pi b^2$, the dependence on $\xi$ cancels and one is left with
\begin{equation}
\begin{split}
\bra{h}g(\phi,\beta,\gamma)\ket{h}=&-8\pi i b^2\, S_b\left(b\left(\tfrac{1}{2}-is-ih\right)\right)S_b\left(b\left(\tfrac{1}{2}+is-i h\right)\right)\cosh\left(\pi(s+h)\right)\left(\beta \gamma\right)^{-ih-\frac12}e^{-\phi} \\
&\times \int_{-i \infty}^{+i \infty} d \xi \left(\beta e^{2\phi } \gamma\right)^{\xi}S_{b}\left(b \left(\xi\pm is\right)\right)  S_{b}^2\left(b\left(\tfrac12+ih-\xi\right)\right) \sin\left(\pi(\xi-is)\right).
\end{split}
\end{equation}

\paragraph{Identification with $F_b$}
By \eqref{eq:double_sine_dual_shift} and \eqref{eq:double_sine_reflection}, $\sin(\pi w)S_b(bw)=\frac12 S_b(bw+1/b)=1/\big(2S_b(b(1-w))\big)$. With $w=\xi-is$, the integrand becomes
\begin{equation}
\left(\beta e^{2\phi } \gamma\right)^{\xi}\,\frac{S_{b}\left(b \left(\xi+is\right)\right)  S_{b}^2\left(b\left(\tfrac12+ih-\xi\right)\right)}{2\,S_b\left(b(1+is-\xi)\right)} .
\end{equation}
Changing variables $t=-\xi-is$ and using $\beta e^{2\phi}\gamma=e^{\Lambda-2\pi b^2h}$, this is exactly the integrand of \eqref{eq:b_hypergeometric} with $\alpha=\beta=\frac b2+ibh+ibs$, $\gamma=b+2ibs$ and $z=-e^{-\Lambda}e^{4\pi b^2h}$. Indeed $(-z)^te^{i\pi bt(\alpha+\beta-\gamma)}=e^{(-\Lambda+2\pi b^2h)t}=\left(\beta e^{2\phi}\gamma\right)^{\xi+is}$. Therefore
\begin{multline}
\int_{-i \infty}^{+i \infty} d \xi \ldots=\frac{2\pi i}{4\pi b}\,e^{-i\left(\Lambda-2\pi b^2 h \right)s}\,\frac{S_b^2\left(\frac b2+ibh+ibs\right)}{S_b(b+2ibs)}\\
\times F_{b}\left(\tfrac{b}{2}+i b h+ibs,\tfrac{b}{2}+i b h+ibs, b+2ibs; -e^{-\Lambda}e^{4\pi b^2 h}\right).
\end{multline}
Finally, $S_b\left(b(\tfrac12-is-ih)\right)S_b\left(\tfrac b2+ibh+ibs\right)\cosh\pi(s+h)=\frac12$ by \eqref{eq:double_sine_norm}, and $(\beta\gamma)^{-ih-\frac12}e^{-\phi}=e^{2i\phi h}e^{-(\Lambda-2\pi b^2h)(\frac12+ih)}$. Putting everything together gives the result \eqref{eq:K+++} of the main text. As $b\to0$, \chg{its} prefactor tends to $\Gamma(\frac12+is-ih)\Gamma(\frac12+is+ih)/\Gamma(1+2is)$, the coefficient of \eqref{eq:K++JT}.

\subsection{JT limit}\label{app:JTsteps}

The $b \rightarrow 0$ limit of the algebra \eqref{eq:algebra1} is
\begin{equation}
H=x \partial_x,\qquad
E= x^2 \partial_x+\left(\tfrac12+is\right)x,\qquad
F=-\partial_x+\frac{\frac12+is}{x}\corrIX{,}
\end{equation}
\corrIX{with $\frac12+is=j+1$. On this realization the Casimir is $FE+H^2+H=j(j+1)$.}
The parabolic vectors, obtained by diagonalizing $E$ and $F$, are
\begin{equation}
\bra{x}\ket{\mu}=x^{-\frac12-is}e^{i \mu/x}, \qquad \bra{x}\ket{\nu}=x^{\frac12+is}e^{i \nu x},
\end{equation}
while the hyperbolic one, obtained by diagonalizing $H$, is $\bra{x}\ket{h}=x^{ih}$. Since all generators are antihermitian, they are diagonalized with imaginary eigenvalues. The parabolic-hyperbolic overlaps follow from the identity
\begin{equation}\label{eq:ide}
\int_{0}^{+\infty} \mathrm{d}z \  z^{\delta-1}e^{i b z}=\Gamma(\delta) \abs{b}^{-\delta} \exp \left(i\frac{\pi}{2} \text{sign}(b)\delta\right),
\end{equation}
and are
\[
\langle h|\mu\rangle=\Gamma\left(\tfrac{1}{2}+is+i h\right)|\mu|^{-ih-i s-\frac{1}{2}}e^{\frac{i\pi}{2}\operatorname{sgn}(\mu)\left(\frac12+ih+i s\right)},\]
\[
\langle h|\nu\rangle=\Gamma\left(\tfrac{1}{2}+is-ih\right)|\nu|^{ih-is-\frac{1}{2}}e^{\frac{i\pi}{2}\operatorname{sgn}(\nu)\left(\frac{1}{2}-ih+i s\right)}.
\]
These are precisely the $b \rightarrow 0$ limit of \eqref{eq:boverlaps_app}, and \eqref{eq:truemellinbarnes} is the $b\to0$ limit of the Whittaker function. Since $g_b(2\pi b^2w)\to e^{iw}$ and $g_b^*(2\pi b^2w)\to e^{-iw}$ as $b\to0$, the group element \eqref{eq:groupdecomp} becomes $e^{-\gamma F}e^{2\phi H}e^{-\beta E}$, with
\begin{equation}
\bra{\nu}e^{-\gamma F}e^{2\phi H}e^{-\beta E}\ket{\mu}=e^{i\gamma\nu}\,e^{i\beta\mu}\,\psi_{\mu\nu}(\phi),
\end{equation}
and the Mellin identities \eqref{eq:identity_both} reduce to \eqref{eq:ide}. Every step of Appendix \ref{app:K++} then has a smooth limit. The product $\Gamma(\xi\pm is)\Gamma^2(\frac12+ih-\xi)\sin\pi(\xi-is)$ is resummed by the Mellin--Barnes representation of the hypergeometric function,
\begin{equation}
{}_2F_1(a,b,c;z)=\frac{1}{2\pi i } \frac{\Gamma(c)}{\Gamma(a)\Gamma(b)} \int_{c-i\infty}^{c+i \infty} ds \frac{\Gamma(a+s)\Gamma(b+s)\Gamma(-s)}{\Gamma(c+s)} (-z)^{s},
\end{equation}
and one recovers \eqref{eq:K++JT} with unit coefficient. Equivalently, in terms of the radial function, \corrIX{$\chi^-_s(\Lambda)\to\psi^-_s(L)$}.

\paragraph{The $K_{+-}$ element}
In JT gravity one can also compute directly the element that connects the two branches of the carrier space, which is the regular wavefunction of the universe \eqref{eq:K+-JT}. To this end we extend the carrier space to the whole real line, with measure $dx/\abs{x}$, and use the parabolic vectors
\begin{equation}
\bra{x}\ket{\mu}=\abs{x}^{-\frac12-is}e^{i \mu/x}, \qquad \bra{x}\ket{\nu}=\abs{x}^{\frac12+is}e^{i \nu x},
\end{equation}
which on the whole line satisfy $\bra{\mu}\ket{\mu'}=2\pi\delta(\mu-\mu')$, so that \eqref{eq:completeness} still holds. The hyperbolic vectors now come in two branches,
\begin{equation}\label{eq:hyperbolicJT}
\bra{x}\ket{h,+}=\theta(x)x^{i h} \qquad \bra{x}\ket{h,-}=\theta(-x)(-x)^{i h},
\end{equation}
and the overlaps follow from \eqref{eq:ide} as
\begin{equation}
\begin{split}
\langle h,\pm|\mu\rangle&=\Gamma\left(\tfrac{1}{2}+is+i h\right)|\mu|^{-ih-i s-\frac{1}{2}}\exp\left[\pm\tfrac{i\pi}{2}\operatorname{sgn}(\mu)\left(\tfrac12+ih+i s\right)\right],\\
\langle h,\pm|\nu\rangle&=\Gamma\left(\tfrac{1}{2}+is-ih\right)|\nu|^{ih-is-\frac{1}{2}}\exp\left[\pm\tfrac{i\pi}{2}\operatorname{sgn}(\nu)\left(\tfrac{1}{2}-ih+i s\right)\right].
\end{split}
\end{equation}
Splitting the integral over $x$ in the positive and negative range, the Whittaker function becomes the sum of two terms, $\psi_{\mu\nu}(\phi)=I^{+}_{\mu\nu}+I^{-}_{\mu\nu}$, with
\begin{equation}\label{eq:truemellinbarnes2}
I_{\mu \nu}^{\pm}=e^{-\phi}\frac{1}{2 \pi i}\int_{-i\infty}^{+i \infty} d \xi\, e^{2\phi \xi}\abs{\mu}^{-\xi-is} \abs{\nu}^{-\xi+is}\, \Gamma(\xi \pm i s)\, e^{\pm i \frac{\pi}{2}\text{sign}(\mu)(\xi+is)}e^{\mp i \frac{\pi}{2}\text{sign}(\nu)(\xi-is)}.
\end{equation}
Here $I^+_{\mu\nu}$ comes from $x>0$ and is \eqref{eq:truemellinbarnes}, while $I^-_{\mu\nu}$ comes from $x<0$.

The goal is to compute the $+-$ sector of the hyperbolic matrix element as a double integral transform of the mixed parabolic matrix element,
\begin{equation}\label{eq:K+-}
\begin{split}
\bra{h,-}g\ket{h,+}&=\int_{-\infty}^{+\infty} \frac{\mathrm{d}\mu}{2\pi} \int_{-\infty}^{+\infty} \frac{\mathrm{d} \nu}{2\pi}\, \bra{h,-}\ket{\nu}\bra{\nu} g\ket{\mu} \bra{\mu}\ket{h,+},\\
\bra{\nu}g\ket{\mu}&=e^{i\gamma\nu}\,e^{i\beta\mu}\,\psi_{\mu\nu}(\phi),
\end{split}
\end{equation}
where $g=e^{-\gamma F}e^{2\phi H}e^{-\beta E}$ is the JT limit of \eqref{eq:groupdecomp}. Inserting the overlaps given above, we obtain
\begin{equation}\label{eq:hyperbolic1}
\begin{split}
\bra{h,-}g\ket{h,+}=&\ \frac{1}{4\pi^2}\,\Gamma\left(\tfrac{1}{2}+is-ih\right)\Gamma\left(\tfrac{1}{2}-is-i h\right)\\
&\times\int_{-\infty}^{+\infty} \mathrm{d}\mu\,
|\mu|^{ih+i s-\frac{1}{2}}
\exp\left[-\tfrac{i\pi}{2}\operatorname{sgn}(\mu)\left(\tfrac{1}{2}-ih-i s\right)\right] e^{i \beta \mu}\\
& \times \int_{-\infty}^{+\infty} \mathrm{d} \nu\,
|\nu|^{ih-is-\frac{1}{2}}
\exp\left[-\tfrac{i\pi}{2}\operatorname{sgn}(\nu)\left(\tfrac{1}{2}-ih+i s\right)\right] e^{i \gamma \nu} \left(I_{\mu \nu}^{+}+I_{\mu \nu}^{-}\right).
\end{split}
\end{equation}
We exchange the order of integration and perform the integrals over $\mu$ and $\nu$ first, using \eqref{eq:ide}. With $A=\frac12+ih-\xi$ as before, for $\beta,\gamma>0$, the term $I^+_{\mu\nu}$ gives $2\Gamma(A)\beta^{-A}\cosh\pi(s+h)$ from the integral over $\mu$ and $2\Gamma(A)\gamma^{-A}\cos\pi(\xi-ih)$ from the integral over $\nu$. The term $I^-_{\mu\nu}$ gives $2\Gamma(A)\beta^{-A}\cos\pi(\xi-ih)$ and $2\Gamma(A)\gamma^{-A}\cosh\pi(s-h)$. Using $\Gamma(A)\cos\pi(\xi-ih)=\pi/\Gamma(\frac12-ih+\xi)$, the two contributions combine into
\begin{equation}
\begin{split}
&\frac{4 \pi\,(\beta\gamma)^{-ih+\xi-\frac12}\,\Gamma \left(\frac12+ih-\xi\right)}{\Gamma \left(\frac12-ih+\xi\right)} \Big(\cosh\pi(s-h)+\cosh\pi(s+h)\Big)\\
&\qquad=\frac{8 \pi\,(\beta\gamma)^{-ih+\xi-\frac12}\,\Gamma \left(\frac12+ih-\xi\right)}{\Gamma \left(\frac12-ih+\xi\right)} \cosh(\pi s)\cosh(\pi h).
\end{split}
\end{equation}
Plugging this result into the remaining integral over $\xi$, we end up with
\begin{equation}
\begin{split}
\bra{h,-}g\ket{h,+}=&\ \frac{2}{\pi}\,\Gamma\left(\tfrac{1}{2}-is-ih\right)\Gamma\left(\tfrac{1}{2}+is-i h\right) \cosh(\pi s)\cosh(\pi h)\, Z^{-\frac12} \left(\beta \gamma\right)^{-ih}\\
&\times \frac{1}{2 \pi i}\int_{-i \infty}^{+i \infty} d \xi \ Z^{\xi} \ \Gamma \left(\xi \pm i s\right) \frac{\Gamma \left(\frac12+i h-\xi\right)}{\Gamma \left(\frac12-i h+\xi\right)},
\end{split}
\end{equation}
with $Z=\beta e^{2\phi}\gamma$. The remaining integral is the Mellin--Barnes representation of the hypergeometric function given above, with $a=\frac12+ih+is$, $b=\frac12+ih-is$ and $c=1$, and equals $Z^{\frac12+ih}\,\Gamma(\frac12+ih+is)\Gamma(\frac12+ih-is)\,{}_2F_1(a,b;1;-Z)$. Using $\Gamma(\frac12+x)\Gamma(\frac12-x)=\pi/\cos(\pi x)$ twice, we obtain
\begin{equation}
\bra{h,-}g\ket{h,+}=4\pi\,\frac{\cosh(\pi s)\cosh(\pi h)}{\cosh(2\pi s)+\cosh(2\pi h)}\,(\beta\gamma)^{-ih}Z^{ih}\,{}_2F_1\left(\tfrac12+ih+is,\tfrac12+ih-is;1;-Z\right).
\end{equation}
Finally, in the finite cutoff chart $g=e^{uH}e^{\frac{L}{2}(E+F)}e^{vH}$, one has $\beta\gamma=e^{-(u+v)}\tanh^2(L/2)$ and $Z=\sinh^2(L/2)$, so that
\begin{equation}
\begin{split}
\bra{h,-}g(L,u,v)\ket{h,+}={}&  4 \pi\,\frac{\cosh(\pi s)\cosh(\pi h)}{\cosh(2\pi s)+\cosh(2\pi h)}\,e^{i h(u+v)}\left(\cosh\tfrac L2\right)^{2ih}\\
&\times{}_2F_1\left(\tfrac12+ih+is,\tfrac12+ih-is;1;-\sinh^2\tfrac L2\right),
\end{split}
\end{equation}
which reproduces \eqref{eq:K+-JT}, up to the normalization of the hyperbolic states and the factor $e^{ih(u+v)}$. The same computation for $\bra{h,+}g\ket{h,-}$ gives zero: for $\beta,\gamma>0$, the action of $g$ on functions restricted to $x>0$ only involves their values at $x>0$. At finite $b$, the analogue of this computation would require the dual generators on the negative branch, which is why in Section \ref{sec:wave} the regular wavefunction is obtained instead from the connection formula \eqref{eq:sinh_connection}.

\bibliographystyle{JHEP}
\bibliography{biblio}

\end{document}